\documentclass[10pt]{article}
\pdfoutput=1
\usepackage{graphicx}
\usepackage{amsmath}
\usepackage{amssymb}
\usepackage{amsthm}
\usepackage{ytableau}
\usepackage{fullpage}
\usepackage{hyperref}

\newcommand{\beq}{\begin{equation}}
\newcommand{\eeq}{\end{equation}}
\newcommand{\bal}{\begin{align}}
\newcommand{\eal}{\end{align}}
\newcommand{\CR}{\nonumber \\}

\newcommand{\gq}{\mathfrak{q}}
\newcommand{\gd}{\mathfrak{d}}
\newcommand{\deltabar}{\overline{\delta}}
\newcommand{\alphabar}{\overline{\alpha}}

\newcommand{\ch}{\mathrm{ch}}
\newcommand{\Tr}{\mathrm{Tr}}

\makeatletter
 
 \@addtoreset{equation}{section}
\makeatother

\begin{document}

\title{
\Large{ \bf $(q,t)$-KZ equations for
quantum toroidal algebra \\
and Nekrasov partition functions on ALE spaces
}}

\author{
{\bf Hidetoshi Awata$^a$}\footnote{awata@math.nagoya-u.ac.jp},
\ {\bf Hiroaki Kanno$^{a,b}$}\footnote{kanno@math.nagoya-u.ac.jp},
\ {\bf Andrei Mironov$^{c,d,e,f}$}\footnote{mironov@lpi.ru; mironov@itep.ru},\\
\ {\bf Alexei Morozov$^{d,e,f}$}\thanks{morozov@itep.ru},
\ {\bf Kazuma Suetake$^a$}\footnote{m14020z@math.nagoya-u.ac.jp},
\ \ and \ {\bf Yegor Zenkevich$^{d,g,h}$}\thanks{yegor.zenkevich@gmail.com}
\date{ }
}

\maketitle

\vspace{-6.2cm}

\begin{center}
\hfill FIAN/TD-30/17\\
\hfill IITP/TH-24/17\\
\hfill ITEP/TH-41/17
\end{center}

\vspace{4cm}

\begin{center}
$^a$ {\small {\it Graduate School of Mathematics, Nagoya University,
Nagoya, 464-8602, Japan}}\\
$^b$ {\small {\it KMI, Nagoya University,
Nagoya, 464-8602, Japan}}\\
$^c$ {\small {\it Lebedev Physics Institute, Moscow 119991, Russia}}\\
$^d$ {\small {\it ITEP, Moscow 117218, Russia}}\\
$^e$ {\small {\it Institute for Information Transmission Problems, Moscow 127994, Russia}}\\
$^f$ {\small {\it National Research Nuclear University MEPhI, Moscow 115409, Russia }}\\
$^g$ {\small {\it Dipartimento di Fisica, Universit\`a di Milano-Bicocca,
Piazza della Scienza 3, I-20126 Milano, Italy}}\\
$^h$ {\small {\it INFN, sezione di Milano-Bicocca,
I-20126 Milano, Italy
  }}
\end{center}

\vspace{.5cm}

\begin{abstract}
We describe the general strategy for lifting the
  Wess-Zumino-Witten model from the level of one-loop Kac-Moody
  $U_q(\widehat{\mathfrak{g}})_k$ to generic quantum toroidal algebras. A nearly exhaustive presentation is given for the two series
  $U_{q,t}(\widehat{\widehat{\mathfrak{gl}}}_1)$ and
  $U_{q,t}(\widehat{\widehat{\mathfrak{gl}}}_n)$, when
  screenings do not exist and thus all the correlators are purely
  algebraic, i.e.\ do not include additional hypergeometric type
  integrations/summations.

  Generalizing the construction of the intertwiner (refined
  topological vertex) of the Ding-Iohara-Miki (DIM) algebra, we obtain
  the intertwining operators of the Fock representations of the quantum toroidal
  algebra of type $A_n$. The correlation functions of these operators
  satisfy the $(q,t)$-Knizhnik-Zamolodchikov (KZ) equation, which features the
  ${\cal R}$-matrix. The matching with the Nekrasov function for the
  instanton counting on the ALE space is worked out explicitly.

 We also present an important application of the DIM formalism to the
 study of $6d$ gauge theories described by the double elliptic integrable
 systems. We show that the modular and periodicity properties of the
 gauge theories are neatly explained by the network matrix models
 providing solutions to the elliptic $(q,t)$-KZ
 equations.
\end{abstract}


\tableofcontents

\section{Introduction and outline}

Conformal field theories \cite{CFT1,CFT2,CFT3,CFT4} (CFT) are connected by the AGT relations
\cite{AGT1,AGT2,AGT3}\footnote{For various AGT-related issues, see \cite{AGT5d1}-\cite{AGT5d16}.} to the low-energy supersymmetric Yang-Mills theories
\cite{SW1,SW2}\footnote{Integrability behind these theories was discovered in \cite{GKMMM1} and studied in \cite{GKMMM2,GKMMM3,GKMMM4}, see also \cite{GM} for a review and \cite{Nek1,Nek2,Nek3} for these theories in the $\Omega$-background.} and therefore are once again at the center of
attention in modern theoretical and mathematical physics.  One of the
immediate results of this is a new interest in various extensions
and deformations, needed to match higher-dimensional
generalizations on the Yang-Mills side.  In conformal field theory, the
central personage is the Wess-Zumino-Witten theory \cite{WZW1}-\cite{WZW5} (WZW), which is the theory with an
extended Kac-Moody symmetry, of which all other important (if not all) models,
including Liouville and Toda theories, are various reductions. This model and
its reductions are straightforwardly handled by various versions of
the free-field methods \cite{DF1}-\cite{GMMOS4}.  Nowadays problem is the
lifting of this model to the level of adequately extended toroidal
algebras (which corresponds to lifting from 4d to 6d on the Yang-Mills side
of the AGT relations) and development of an efficient generalization of
the free-field formalism to describe the resulting ``network matrix models'' \cite{MMSh1}-\cite{Mironov:2016yue}.
This paper is a continuation of our study of this problem in
\cite{Awata:2016riz,Awata:2016mxc,Awata:2016bdm,Awata:2017cnz}\footnote{It goes in parallel with other important
efforts in the same direction, see \cite{Matsuo1}-\cite{O3}.}.  Mathematically the problem
is that of the full-fledged representation theory of the
Ding-Iohara-Miki algebra \cite{DI,Miki} and its various generalizations \cite{DIM1}-\cite{DIMl}, which we also refer
to as DIM. In this paper, we focus on a small corner of this very broad
area and describe two generalizations and one application of the
$(q,t)$-KZ equations introduced in~\cite{Awata:2017cnz}.

The plan of the paper is the following: in sec.2, we propose a generalization of the $(q,t)$-KZ for   $U_{q,t}(\widehat{\widehat{\mathfrak{gl}}}_1)$ with arbitrary central charge.
In sec.3, we describe the quantum toroidal algebra and its vertical and horizontal (Fock) representations that we deal with in the paper. In sec.4, we construct the operators that intertwine these representations. In sec.5, we derive the (level one) Knizhnik-Zamolodchikov equation for the correlation functions of these intertwining operators, Eqs.(\ref{KZ1})-(\ref{KZ2}). In sec.6, we use solutions to the elliptic KZ equations to obtain modular properties of the $6d$ $U(N)$ gauge
theories with adjoint hypermultiplet of mass $m$ compactified on torus
$T^2$ derived in \cite{Gleb} basing on the description in terms of double elliptic integrable
systems.  Concluding remarks in sec.7 are followed by Appendices A-E that contain various technical details.

With the help of all the technical exercises, we would like to
demonstrate a simple idea: that the DIM intertwiner formalism is not just
an interesting toy but an important tool, which can find its use
in gauge theories, as well as in other related fields.

In the remaining part of the introduction, we first describe
these three objectives (sec.~\ref{sec:strt-object}) and then give some
general description of the methods by which we are going to achieve
them (sec.~\ref{sec:strat-comp}).

\subsection{Strategic objectives}
\label{sec:strt-object}
\subsubsection{Abelian $(q,t)$-KZ  for general central charge}
\label{sec:abelian-q-t}
The $(q,t)$-KZ equation was introduced
in~\cite{Awata:2017cnz}\footnote{See \cite{KZ}-\cite{Sun} for the standard KZ and $q$KZ equations and \cite{O1,O2,O3} for their extensions.}.
Its first generalization relaxes the condition on the central charge of the
``horizontal'' representation, i.e.\ we no longer require this space
to be the Fock space, but instead assume that it has a general central
charge $(k,N)$. The vertical representations are still assumed to be
Fock spaces with central charge $(0,1)$. In this case, the modification of the KZ
equation is not hard to guess and the solution to the
equations can also be explicitly obtained. The solution is algebraic,
i.e.\ no integrals of screening charges appear in the answer\footnote{For solutions of various KZ equations, see original papers \cite{SchV1}-\cite{FV1} and a review in \cite{EFK}.}.


\subsubsection{Non-Abelian $(q,t)$-KZ equation for unit central charge and its
algebraic solutions}
\label{sec:nonabelian-q-t}

The second generalization of the $(q,t)$-KZ equation is the much
sought non-Abelian version, that for the algebra
$U_{q,t}(\widehat{\widehat{\mathfrak{gl}}}_n)$.
In order to derive the $(q,t)$-KZ equation, we first construct
the intertwiners for the horizontal and vertical Fock representations
with unit central charges, i.e.\ $(1,N)$ and $(0,1)$.
The intertwining relations for the intertwiners are determined by the coproduct
structure of the quantum toroidal algebra $U_{q,t}(\widehat{\widehat{\mathfrak{gl}}}_n)$.
The same strategy as in the $\widehat{\widehat{\mathfrak{gl}}}_1$ case \cite{AFS}
can be used, and the $\lambda$-component of intertwiner $\Phi_\lambda(v)$
can be expressed as the normal ordered product of the currents $E_i(z)$
over the boxes of the Young diagram $\lambda$, where the argument $z$ is
shifted according to the position of the boxes. One of the important
differences with the $\widehat{\widehat{\mathfrak{gl}}}_1$ case is
the appearance of the zero mode factor in the free field realization of
the horizontal Fock representation. The zero modes are group algebra valued
and their commutation relation is crucial for obtaining the correct intertwining relations.
Another new aspect of $U_{q,t}(\widehat{\widehat{\mathfrak{gl}}}_n)$ is the \lq\lq color
selection\rq\rq\ rules. Some combinatorial arguments for such rules are required,
especially when we establish the relation to the Nekrasov partition function
for gauge theories on the ALE space $ALE_n$ of type $A_n$, which is a resolution of
the orbifold $\mathbb{C}^2/\mathbb{Z}_n$.

Once we obtain the intertwiners, we can introduce the $\mathcal{T}$-operator
and the ${\cal R}$-matrix in a similar way to \cite{Awata:2016mxc,Awata:2016bdm} and
write down the $(q,t)$-KZ equations \cite{Awata:2017cnz}, where the ${\cal R}$-matrix
is featured as the connection matrix for $\gq$-shift of the argument of the intertwiner.
The ${\cal R}$-matrix can be identified with $\gq$-difference of the operator product expansion (OPE) factor
of the intertwiners and is essentially diagonal.
Since the OPE factor of the intertwiners agrees with the Nekrasov factor
(the bi-fundamental contribution to the partition function),
we have a fundamental relationship between the ${\cal R}$-matrix and the Nekrasov
partition function. Basing on this relation, we can find explicit solutions to our
$(q,t)$-KZ equations, which turn out to be the Nekrasov functions
for 5d gauge theories on $ALE_n \times S^1$.
Since we consider only the setup with
unit central charges, all the solutions to the KZ equations are still algebraic.

Moreover, for the unrefined case with unit central charges,
we actually demonstrate that the intertwiners of
$U_{q,q}(\widehat{\widehat{\mathfrak{gl}}}_n)$ essentially factorize
into products of noninteracting
$U_{q^n,q^n}(\widehat{\widehat{\mathfrak{gl}}}_1)$ intertwiners. The
most complicated and interesting case of representations with general
central charges in non-Abelian DIM algebra is left for the future.


\subsubsection{Modular and periodic properties of $6d$ gauge theories. }
\label{sec:modul-peri-prop}
\begin{figure}[h]
  \centering
  \includegraphics[width=4cm]{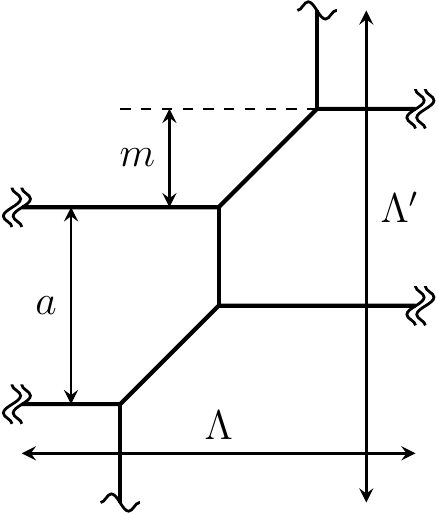}
  \caption{Compactified $(p,q)$-brane web corresponding to the $6d$
    $U(2)$ gauge theory with adjoint hypermultiplet compactified on
    $T^2$. The wavy (double wavy) lines are understood to be
    identified with each other. The parameters of the gauge theory are
    encoded in the distances between the branes: $a$ is the Coulomb
    modulus, $m$ is the mass of the adjoint field, $\Lambda$ is the
    exponentiated complexified coupling, and $\Lambda'$ is the
    exponentiated complex structure modulus of the compactification
    torus $T^2$.}
  \label{fig:SW}
\end{figure}

\begin{figure}[h]
  \centering
  \includegraphics[width=9cm]{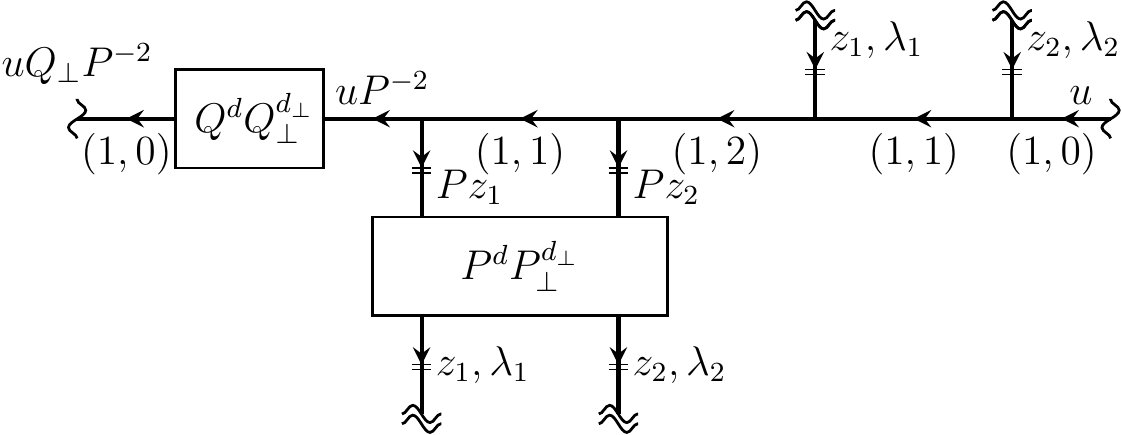}
  \caption{Compactified network of intertwiners corresponding to the
    same $6d$ $U(2)$ gauge theory with adjoint hypermultiplet, as in
    Fig.~\ref{fig:SW}. The grading operators $Q^d
    Q_{\perp}^{d_{\perp}}$ and $P^d P_{\perp}^{d_{\perp}}$ count the
    states of the Fock representations in the same way as the Boltzman
    factor $e^{-\beta H}$ counts the states in a quantum mechanical
    partition function.}
  \label{fig:ellqKZ}
\end{figure}

To demonstrate the effectiveness of DIM formalism, we are going to
describe an important application of network matrix models to $6d$
gauge theories with adjoint matter compactified on the torus
$T^2$. These theories are, in a certain sense, the highest step in the
hierarchy of gauge theories with eight supercharges, for which the
Seiberg-Witten and Nekrasov solutions are available. Within the
Seiberg-Witten paradigm, they are described by the double
elliptic integrable systems with both coordinates and momenta entering
the Hamiltonians through elliptic functions \cite{Delli1,Delli2,Delli3}. Despite the recent important
progress~\cite{Gleb1,Gleb2,Gleb3}, these systems are still quite mysterious and
require more explicit description. In particular, the behavior of the
$6d$ gauge theories and double elliptic systems under the $S$-duality and
the modular transformations of the compactification torus is quite
peculiar. It turns out that the modular transformation rule
\emph{mixes} the complex structure of the compactification torus with
the complexified coupling in a very specific way~\cite{Gleb}.  We will
explain this behavior using the network matrix model of the Abelian
DIM algebra $U_{q,t} (\widehat{\widehat{\mathfrak{gl}}}_1)$
corresponding to the gauge theories in question. Adding the
adjoint matter in the gauge theory corresponds to the
\emph{compactification} of the network diagram in the horizontal
direction, and an extra sixth dimension also implies the
compactification of the vertical direction. Thus, we get the ``doubly
compactified'' network, which geometrically corresponds to a CY
three-fold with an elliptic fibration. In the algebraic language, the
double compactification corresponds to taking the trace of the product
of the intertwiners over both the vertical and the horizontal
representations. The network of intertwiners is modeled after the
Seiberg-Witten Type IIB $(p,q)$-brane diagram associated with the
gauge theory. The brane diagram for the case of $U(2)$ gauge theory is
shown in Fig.~\ref{fig:SW}, and the network of intertwiners, in
Fig.~\ref{fig:ellqKZ}.

The picture of DIM intertwiners corresponding to the double elliptic
system is given by the intersection of one horizontal and $N$
vertical lines. Notice that the picture of intertwiners is rotated by
$\frac{\pi}{2}$ with respect to the Seiberg-Witten $(p,q)$-brane
diagram (Fig.~\ref{fig:SW}) usually given in the literature. Of course,
this does not change the answer since NS5 and D5 branes of Type IIB
string theory are $S$-dual. The ends of the lines should be
identified with each other so that the picture is essentially drawn on a
two-dimensional torus.

The ``fugacities'' $Q$, $Q_{\perp}$ can be also understood as the
twisting parameters of the fibration giving the background for the
M-theory, which hints at possible duality between them and
equivariant parameters $q$ and $t$ of the $\Omega$-background. Notice
that we set the preferred direction (determining the coproduct
structure and thus the intertwiners) to be vertical. However, the
final answer for the character/partition function is independent of
the preferred direction. In these conventions, the trace over the
horizontal representation can be taken straightforwardly giving a
combination of theta-functions. Moreover, from our previous
investigations we recall that precisely this trace appeared as a
solution to \emph{elliptic} $(q,t)$-KZ equations. Since the result is
expressed through the theta-functions, we can use it to effectively study
the modular properties of the partition function. Of course, this
strategy works only for modular transformation on the
\emph{compactification} torus, but not for the $S$-duality. The $S$-duality
transformation corresponds to the modular transformation of the
\emph{vertical} compactified direction, where the trace is a lot harder
to compute. However, we can use the slicing invariance of the
partition function and safely change the preferred direction to the
horizontal one. This would allow us to analyze the $S$-duality
transformation as easily as the modular transformation.

Using the intertwiner picture, we can also analyze periodicity
properties of the partition function. As an example consider moving
the incoming vertical lines around the circle (i.e.\ under the
trace). The lines necessarily have to pass through the grading
operators $Q^dQ_{\perp}^{d_{\perp}}$. The action of grading operators gives a
shift in the positions of the incoming vertical lines with respect to
the outgoing vertical lines. Investigating this move in detail, we
find that it is actually accompanied by a certain change in the
complexified coupling of the gauge theory.

\subsection{Tactics of computations}
\label{sec:strat-comp}

\subsubsection{DIM algebra}
\label{sec:dim-algebra}

DIM algebra $U_{q,t}(\widehat{\widehat{\mathfrak{g}}})_{(k_1,\, k_2)}$
is a quantum toroidal algebra (hence, two hats) with two central elements fixed
to levels $k_1$ and $k_2$, and two deformation parameters $q$ and $t$;
we usually consider $\mathfrak{g}=\mathfrak{gl}_n$.  The physical
model, associated with it in just the same way as the WZW model is
associated with the Kac-Moody algebra $U(\widehat{G})_k$ is introduced
in \cite{MZ,MMZ,Mironov:2016cyq,Mironov:2016yue,Awata:2016riz} and named {\it network model}.  We refer
the reader to these papers for terminology and basic
logic. \footnote{For other related references, see \cite{DI}-\cite{DIMl} for various aspects of the DIM algebras and \cite{O1,O2,O3} for a $K$-theory approach.}  The
basic ingredients of construction of the KZ equation for the network
models generalizing \cite{FR} and used in \cite{Awata:2017cnz} are:

$\bullet$ {\bf ``Horizontal'' representations} which is given in terms
of operators acting in the infinite dimensional vector space ${\cal
  F}_{(k_1,M)}(u)$

$\bullet$ {\bf ``Vertical'' representations} given by combinatorial
formulas in some basis in another infinite dimensional space
$\mathcal{F}_{(0,k_2)}(u)$

$\bullet$ {\bf Intertwiners.} The intertwiners $\Psi^{\lambda}$, $\Psi^{*}_{\lambda}$
carry the index $\lambda$ (e.g.\ a set of $N$ Young diagrams
or a plain partition) which labels the element of the vertical
representation space:
\begin{align}
&\Psi^{\lambda}_{(0,k_2)}(z): \ \  {\cal F}_{(k_1,M)}(u)\otimes \mathcal{F}_{(0,k_2)}(z)
\ \longrightarrow \ {\cal F}_{(k_1,M+k_2)}(uz), \notag \\
&\Psi^{*}_{\lambda}{}_{(0,k_2)}(z): \ \  {\cal F}_{(k_1,M)}(u)
\ \longrightarrow \ {\cal F}_{(k_1,M-k_2)}(u/z) \otimes \mathcal{F}_{(0,k_2)}(z).
\end{align}
They depend on the choice of the ``vertical'' coproduct $\Delta^{\mathrm{vert}}$.
In the horizontal direction, they can be easily multiplied: $\Psi(z_1)\Psi(z_2)$
and $\Psi(z_2)\Psi(z_1)$ are just the compositions of operators.

\subsubsection{$(q,t)$-KZ equations}
\label{sec:kz-equations}

In this paper, we concentrate on the case when only algebraic solutions
are present. In this setup, we can supplement the items from the list
above with some more concrete properties.

Namely, the short list of ingredients for the derivation of the
$(q,t)$-KZ equation for the non-Abelian DIM algebra is:
\begin{enumerate}
\item \textbf{Shift identity.} We need to build up the shift operator,
  whose action on the intertwiner can be rewritten as the product of
  two ${\cal T}$-operators:
  \begin{equation}
    \label{eq:3}
    p^{z \partial_z} \Psi^{\lambda}(z) = ({\cal T}_{-}^{\lambda}(z))^{-1} \Psi^{\lambda}(z) {\cal T}_{+}^{\lambda}(z).
  \end{equation}
  This requirement by itself can be trivially satisfied, since we can
  simply write
  \begin{gather}
    ({\cal T}_{-}^{\lambda}(z))^{-1} = \Psi_{-}^{\lambda}(p z)
    (\Psi_{-}^{\lambda}(z))^{-1},\\
    {\cal T}_{+}^{\lambda}(z) = (\Psi_{+}^{\lambda}(z))^{-1}
    \Psi_{+}^{\lambda}(p z) ,
  \end{gather}
  where $\Psi_{\pm}$ denote the creation and annihilation parts of the
  intertwiner. However, we also need ${\cal T}_{\pm}$ to satisfy nontrivial
  ${\cal RTT}$ identities.

\item \textbf{Commutation of the ${\cal T}$-operators.} We need the
  commutation property as follows
  \begin{equation}
    \label{eq:4}
    {\cal T}^{\mu}(z)\Psi^{\lambda}(w) = {\cal R}_{\lambda \mu} \left( \frac{z}{w}
    \right) \Psi^{\lambda}(w) {\cal T}^{\mu}(z).
  \end{equation}
  To get these identities, we need first to find the non-Abelian
  ${\cal R}$-matrix.
\item \textbf{Diagonal ${\cal R}$-matrix.} One way to obtain the ${\cal R}$-matrix is
  to commute a pair of intertwiners:
  \begin{equation}
    \label{eq:5}
    \Psi^{\lambda}(z)\Psi^{\mu}(w)= {\cal R}_{\lambda \mu} \left(
      \frac{z}{w} \right) \Psi^{\mu}(w)\Psi^{\lambda}(z).
  \end{equation}
  The commutation can be done using the free boson formalism.
\item \textbf{Vacuum property of ${\cal T}$-operators.} After we get the
  shift and commutation identities, we can act with the shift operator
  on a string of intertwiners and get an insertion of a pair of the
  ${\cal T}$-operators. We can then move them to the ends of the string using
  the commutation identities. The last step in the derivation of the
  equation is to ensure that the ${\cal T}$-operators annihilate the vacuum:
  \begin{equation}
    \label{eq:6}
    {\cal T}_{+}|\varnothing\rangle = |\varnothing\rangle,\quad \langle
    \varnothing |{\cal T}_{-} = \langle
    \varnothing |
  \end{equation}
\end{enumerate}

Concretely for the non-Abelian DIM algebra, the intertwiner is built as a
normal ordered product of elements $E_i(z)$ taken at certain discrete
points specified by the Young diagram on the vertical leg. We can
derive the shift identity for each operator $E_i(z)$ in the product
separately and then account for the normal ordering constants. We have
\begin{equation}
  E_i(\mathfrak{q}^2z) = ({\cal T}^i_{-}(z))^{-1} E_i(z) {\cal T}^i_{+}(\mathfrak{q}^2 z)
\end{equation}
where
\begin{gather}
  {\cal T}^i_{-}(z) = E_i(z) F_i(\mathfrak{q}^{-1}z),\label{eq:7a}\\
  {\cal T}^i_{+}(z) = E_i(z) F_i(\mathfrak{q}z).\label{eq:8a}
\end{gather}
The ${\cal T}$-operators for the intertwiners are normal ordered products of
the basic ${\cal T}$-operators (\ref{eq:7a}),~(\ref{eq:8a}) over the boxes of
the Young diagram. It is important that the ${\cal T}$-operators in the
non-Abelian case are still diagonal in the vertical Young diagram, so
that no extra sums over diagrams appear.


\section{$(q,t)$-KZ equation for
  $U_{q,t}(\widehat{\widehat{\mathfrak{gl}}}_1)$ with general central
  charge}
\label{sec:q-t-kz}

The $q$-KZ equation for the conventional quantum affine algebra \cite{FR} contains an
extra parameter, which is missing in the DIM case we have considered
so far. This parameter is the central charge of the ``horizontal''
(highest weight) representation running in the conformal block. It
enters the shift operator and also the shifts of the
${\cal R}$-matrices. The conformal blocks of DIM are combinations of
intertwiners acting in horizontal and vertical Fock spaces. These
representations have definite central charges of the form either
$(0,1)$ or $(1,N)$. The shift operator determined by the first central
charge of the horizontal representation is therefore fixed and reads
$\left( \frac{q}{t} \right)^{z_k \partial_{z_k}}$.

In this section, we will try to extend the central charge parameter to
arbitrary values and find the corresponding solutions to the Abelian
$(q,t)$-KZ. Unfortunately, the intertwiners for general
representations are not known, so the solution cannot be found as
easily as that for the Fock spaces. However, the structure of the KZ equation
is very rigid and seems to give the only way of introducing the
central charge parameter into it. We will try to follow this route and
investigate the resulting solutions for the conformal blocks.

Let us consider the combination of intertwiners similar to the Fock
space case, but with an arbitrary representation living on the
horizontal line. The shift operator thus becomes
$p^{z_k \partial_{z_k}}$ with an arbitrary parameter $p$. The vertical
representations remain to be $(0,1)$ Fock spaces. This implies that the
${\cal R}$-matrices featuring in the KZ equation are still the same (e.g.\
they are diagonal), and the new parameter $p$ can enter only as a shift
of their arguments.

Having these two arguments, we can conjecture the $(q,t)$-KZ equation
with the parameter $p$ encoding the central charges of the horizontal representation in the simplest case of two vertical incoming
lines:
\begin{gather}
  \label{eq:1}
  p^{z_1 \partial_{z_1}} \mathcal{G}^{\lambda_1 \lambda_2}(z_1, z_2) =
  \widetilde{\mathcal{R}}_{\lambda_1 \lambda_2} \left( \frac{z_1}{z_2}
  \right)\mathcal{G}^{\lambda_1 \lambda_2}(z_1, z_2),\\
  p^{z_2 \partial_{z_2}} \mathcal{G}^{\lambda_1 \lambda_2}(z_1, z_2) =
  \frac{1}{\widetilde{\mathcal{R}}_{\lambda_1 \lambda_2} \left(
      \frac{z_1}{pz_2} \right)}\mathcal{G}^{\lambda_1 \lambda_2}(z_1,
  z_2)
\end{gather}
The solution up to a function independent of the Young diagrams is
given by
\begin{equation}
  \label{eq:2}
  \mathcal{G}^{\lambda_1 \lambda_2}(z_1, z_2) = f \left( \frac{z_1}{z_2} \right)\prod_{k \geq 0}
  \frac{G_{\lambda_1 \lambda_2} \left( \frac{q}{t} p^k \frac{z_1}{z_2} \right)
  }{G_{\lambda_1 \lambda_2} \left(  p^k \frac{z_1}{z_2} \right)}
\end{equation}
where $G_{\lambda\mu}(z)$ can be found, e.g., in \cite[eq.(18)]{Awata:2016bdm}.
Obviously for $p=\frac{q}{t}$ the usual solution
$\frac{1}{G_{\lambda_1 \lambda_2}\left(\frac{z_1}{z_2}\right)}$ for
the Fock space is recovered. The solution for general $p$ contains an
infinite product, which reminds us of the solution to the
\emph{elliptic} $(q,t)$-KZ equation. This similarity looks mysterious
and indeed might turn out to be only superficial.


\section{Quantum toroidal algebra
  $U_{\gq,\gd}(\widehat{\widehat{\mathfrak{gl}}}_n)$ and Fock
  representation}


For the future convenience, we introduce here the basic definitions and notation for the quantum toroidal algebras.

The quantum toroidal algebra $U_{\gq,\gd}(\widehat{\widehat{\mathfrak{gl}}}_n)$ has two deformation
parameters\footnote{We use the Gothic letters for deformation parameters
to keep $(q,t) = (e^{\epsilon_1}, e^{-\epsilon_2})$ for the equivariant parameters for torus action,
or those for the Macdonald function. In the following, we identify $\gq = (q/t)^{-1/2}$ and
$\gd = (qt)^{1/2}$ or $q_1 = q, q_3 = t^{-1}$.}
$\gq, \gd$, which are associated with the Cartan matrix $A$ of type $A_{n-1}^{(1)}$,
 $a_{ij}= 2 \deltabar_{i,j} - (\deltabar_{i-1,j} + \deltabar_{i+1,j})$ and
 a skew-symmetric matrix $M$ with $m_{ij} = \deltabar_{i-1,j} - \deltabar_{i+1,j}$,
 where $\deltabar$ is the Kronecker delta modulo $n$.
 In this paper, we consider the generic case of $(n>2)$.
 Explicitly the matrices $A$ and $M$ are given by
 \begin{equation}
 \label{AM}
 A = \left(\begin{array}{cccccc}
 2 & -1 & 0 & \cdots & 0 & -1 \\
 -1 & 2 & -1 & 0 & \cdots &  0\\
 0 & -1 & 2 & \ddots & \ddots& \vdots \\
 \vdots & \ddots & \ddots& \ddots & \ddots & 0\\
 0 & \ddots & \ddots& \ddots & 2 & -1 \\
 -1 & 0 & \hdots  &  0 & -1 &2
 \end{array}\right),
 \qquad
 M = \left(\begin{array}{cccccc}
 0 & -1 & 0 & \cdots & 0 & 1 \\
 1 & 0 & -1 & 0 & \cdots &  0\\
 0 & 1 & 0 & \ddots & \ddots& \vdots \\
 \vdots & \ddots & \ddots& \ddots & \ddots & 0 \\
 0 & \ddots & \ddots& \ddots & 0  & -1 \\
 -1 & 0 & \hdots  &  0 & 1 &0
 \end{array}\right).
\end{equation}
It seems difficult to introduce an analogue of the skew-symmetric matrix $M$
for other affine Lie algebras $\widehat{\mathfrak g}$, and it is not known if
the toroidal algebra for general affine algebra allows a
two parameter deformation. As we will see below,
if the second deformation parameter is trivial ($\gd=1$),
the structure function $g_{ij}(z,w)$ coincides with that for the quantum affinization
of $U_\gq(\mathfrak{g})$ for a Lie algebra with a symmetrizable Cartan matrix\footnote{
In general, we can define the quantum affinization based on the data of a quiver
\cite{Nak1, Nak2}, by introducing Chevalley generators associated with vertices and
the corresponding Drinfeld currents with the structure function $g_{ij}(z,w)$.}.
Thus $U_{\gq,\gd=1}(\widehat{\widehat{\mathfrak{gl}}}_n)$ can be regarded
as the quantum affinization of the affine algebra $\widehat{\mathfrak{gl}}_n$.

We introduce the structure function \cite{FJMM,Feigin:2013fga},
\begin{equation}
g_{ij}(z,w) := (z - \gd^{-m_{ij}} \gq^{a_{ij}} w) =
\begin{cases}
z - q_1 w, \quad (i \equiv j -1) \\
z- q_2 w, \quad (i \equiv j)\\
z - q_3 w, \quad (i \equiv j +1) \\
z-w, \quad (\hbox{otherwise})
\end{cases}
\end{equation}
where we have defined
\begin{equation}
q_1 = \gd \gq^{-1}, \quad q_2 = \gq^2, \quad q_3 = \gd^{-1} \gq^{-1},
\end{equation}
with $q_1q_2q_3 =1$. We also use
\begin{equation}
\widetilde{g_{ij}}(z,w) := \gd^{m_{ij}} g_{ij} (z,w) = (\gd^{m_{ij}} z - \gq^{a_{ij}} w).
\end{equation}

The generators of $U_{\gq,\gd}(\widehat{\widehat{\mathfrak{gl}}}_n)$ are
\begin{equation}
E_{i,k},\qquad F_{i,k}, \qquad H_{i, r}, \qquad K_i^{\pm 1}, \qquad   \gq^{\pm c/2},
\end{equation}
where $ i \in \mathbb{Z}/ n \mathbb{Z}$ (index set of simple roots or vertices of
the cyclic quiver), $k \in \mathbb{Z}, r \in \mathbb{Z} \setminus \{ 0 \}$
and $c$ is a central element.
It is convenient to employ the generating currents;
\begin{eqnarray}
E_i(z) &=& \sum_{k \in \mathbb{Z}} E_{i,k} z^{-k},  \qquad
F_i(z) = \sum_{k \in \mathbb{Z}} F_{i,k} z^{-k}, \\
K_i^{\pm}(z) &=& K_i^{\pm 1} \exp \left(
\pm ( \gq - \gq^{-1} ) \sum_{r=1}^\infty H_{i, \pm r} z^{\mp r}
\right),
\end{eqnarray}
which satisfy \cite{Feigin:2013fga};
\begin{eqnarray}
K_i^{\pm}(z) K_j^{\pm}(w) &=& K_j^{\pm}(w) K_i^{\pm}(z)  \label{KK}\\
\frac{g_{ij}(\gq^{-c}z,w)}{g_{ij}(\gq^{c}z,w)} K_i^{-}(z) K_j^{+}(w)
&=&  \frac{g_{ji}(w,\gq^{-c}z)}{g_{ji}(w,\gq^{c}z)} K_j^{+}(w) K_i^{-}(z)  \label{Kpm}\\
\widetilde{g}_{ij} (z,w) K_i^{\pm} (\gq^{(1\mp 1)c/2} z) E_j(w)
&+& g_{ji}(w,z) E_j(w) K_i^{\pm} (\gq^{(1\mp 1)c/2} z) =0 \label{KE} \\
\widetilde{g}_{ji} (w,z) K_i^{\pm} (\gq^{(1\pm 1)c/2} z) F_j(w)
&+& g_{ij}(z,w) F_j(w) K_i^{\pm} (\gq^{(1\pm 1)c/2} z) =0 \label{KF} \\
\left[ E_i(z), F_j(w) \right] &=& \frac{\deltabar_{i,j}}{\gq - \gq^{-1}}
\left( \delta (\gq^c \frac{w}{z}) K_i^{+}(z)  -  \delta (\gq^c \frac{z}{w}) K_i^{-}(w) \right) \label{EF} \\
\widetilde{g}_{ij} (z,w) E_i(z) E_j(w) &+& g_{ji}(w,z) E_j(w) E_i(z) =0 \label{EE} \\
\widetilde{g}_{ji} (w,z) F_i(z) F_j(w) &+& g_{ij}(z,w) F_j(w) F_i(z) =0  \label{FF}
\end{eqnarray}
with appropriate Serre relations\footnote{In this paper, we do not use the Serre relations.}.
The delta function in \eqref{EF} is defined by $\delta(z) = \displaystyle{\sum_{n \in \mathbb{Z}}} z^n$ and satisfies
$\delta(z) = \delta(z^{-1})$. Note that there is the following change of the scaling of
the Heisenberg part $H_{i,r}$ between \cite{FJMM} and \cite{Feigin:2013fga};
\begin{equation}
K_i^{\pm}(z) \longrightarrow K_i^{\pm} ( \gq^{-c/2}z) \qquad
H_{i, \pm r} \longrightarrow \gq^{\pm r c/2} H_{i, \pm r}.
\label{shift}
\end{equation}.
The coproduct is defined by\footnote{Remember the redefinition \eqref{shift}, which
implies $\Delta(K_i^\pm(z)) \to \Delta(K_i^\pm(C_1^{-1/2} C_2^{-1/2} z))$.}
\begin{eqnarray}
\Delta(E_i(z)) &=& E_i(z) \otimes 1 + K_i^{-} (C_1z) \otimes E_i(C_1z), \\
\Delta(F_i(z)) &=&  F_i(C_2 z) \otimes K_i^{+}(C_2z)  +  1 \otimes F_i(z), \\
\Delta(K_i^{+} (z)) &=& K_i^{+}(z) \otimes K_i^{+}(C_1^{-1} z), \\
\Delta(K_i^{-} (z)) &=&  K_i^{-}(C_2^{-1} z) \otimes K_i^{-}(z), \\
\Delta(\gq^c) &=& \gq^c \otimes \gq^c,
\end{eqnarray}
where $C_1 = \gq^c \otimes 1$ and $C_2 = 1 \otimes \gq^c$.

As is well known, there are two equivalent constructions of the (untwisted) affine Lie
algebra $\widehat{\mathfrak{g}}$ corresponding to a simple finite dimensional Lie algebra $\mathfrak{g}$.
One is based on the Chevalley generators with the Serre relations determined by the Cartan matrix $A$.
In this approach, the affine Lie algebra is obtained by replacing the Cartan matrix of finite type ($ \det A >0$) with
the corresponding one of affine type ($ \det A =0$). The other way, which is more familiar among physicists, employs
the loop algebra (the current algebra) $\mathfrak{g} \otimes \mathbb{C} [z, z^{-1}]$ of the finite dimensional Lie algebra $\mathfrak{g}$.
The one-dimensional central extension of the loop algebra with the additional grading operator
gives the affine Lie algebra $\widehat{\mathfrak{g}}$. Analogously there are two methods to obtain the quantum enveloping
algebra (or the quantum affine algebra) $U_q(\widehat{\mathfrak{g}})$. That is, Drinfeld and Jimbo
originally defined $U_q(\widehat{\mathfrak{g}})$ in terms of the Chevalley generators
with the $q$ deformed Serre relations \cite{D1,J1,J2}. Later Drinfeld observed \cite{D2} that the same algebra is
obtained by introducing the generating functions of the generators (the Drinfeld currents).
This is called Drinfeld realization, or the quantum affinization $U_q(\widehat{\mathfrak{g}})$ of $U_q(\mathfrak{g})$.
Note that, in this realization of the quantum affine algebra, we use the Cartan matrix of finite type.
In a sense, the quantum toroidal algebra $U_q(\widehat{\widehat{\mathfrak{g}}})$ is obtained by
combining the above two ways of \lq\lq affinization\rq\rq.
Namely, if we use the Cartan matrix of affine type in the Drinfeld realization, we obtain the relations
from \eqref{KK} to \eqref{FF}. In fact, if we compare these relations with those of the Drinfeld's realization of the quantum affine algebra
$U_{\gq}(\widehat{\mathfrak{g}})$ for a Lie algebra $\mathfrak{g}$ with symmetrizable (generalized) Cartan matrix
 (see for example \cite{Nak2} section 1.2), we see the difference is only a change of $g_{ij}(z,w)$
by the second deformation parameter $\gd$ (and a redefinition of the Cartan part \eqref{shift}).


It is convenient to introduce the following rational function
\begin{equation}
\psi(z) := \frac{\gq - \gq^{-1} z}{1-z}
= \gq \frac{1 - q_2^{-1} z}{1 -z} = \gq^{-1} \frac{1 - q_2 z^{-1}}{ 1 - z^{-1}} \label{psi}
\end{equation}
to rewrite the commutation relations \eqref{KE} and \eqref{KF} in the form of OPE relations of $E_j(w), F_j(w)$ with
$K_i^{\pm}(z)$ or $K_i^{\pm}(\gq z)$. We see the relation
\begin{equation}
\psi(q_2 z)^{-1} = \psi(z^{-1}).\label{inversion}
\end{equation}
In fact, as we shall see below, the matrix elements of the vertical representation in the basis
labelled by partitions are described by the function $\psi(z)$.
We can check that
\begin{eqnarray}
K_i^{+} (z) E_j(w) &=&
\begin{cases}\psi \left( q_1 q_2 \frac{w}{z} \right)^{-1} E_j(w) K_i^{+} (z),  \quad (i \equiv j -1) \\
 \psi \left(\frac{w}{z} \right)  \psi \left( q_2 \frac{w}{z} \right) E_j(w) K_i^{+} (z),  \quad (i \equiv j ) \\
 \psi \left( q_1^{-1} \frac{w}{z} \right)^{-1} E_j(w) K_i^{+} (z),  \quad (i \equiv j +1)
 \end{cases} \label{OPE} \\
K_i^{+} (z) E_j(w) &=& E_j(w) K_i^{+} (z), \quad (\hbox{otherwise}),
\end{eqnarray}
imply the exchange relation \eqref{KE}. Similarly,
the OPE relations of the same form with $K_i^{+}(z)$ being replaced by $K_i^{-}(\gq z)$ or $E_i(z)$
imply \eqref{KE} and \eqref{EE}.
We can also see that \eqref{KF} and \eqref{FF} follow from
similar relations for $F_j(w)$ with $K_i^{+} (\gq z), K_i^{-}(z)$ and $F_i(z)$, where
$\psi$ is replaced by $\psi^{-1}$ with the same argument.

The quantum toroidal algebra $U_{\gq,\gd}(\widehat{\widehat{\mathfrak{gl}}}_n)$ has
two central elements, $\gq^{c}$ and $\kappa := \prod_i K_i$.
We will call a representation of $U_{\gq,\gd}(\widehat{\widehat{\mathfrak{gl}}}_n)$ that of level $(k,\ell)$,
when $(\gq^c, \kappa) = (\gq^k, \gq^{-\ell})$.
In analogy with the case of $\mathfrak{gl}_1$, we will call the level $(0,1)$ representation vertical and the
level $(1,N)$ representation horizontal.
It is known that $U_{\gq,\gd}(\widehat{\widehat{\mathfrak{gl}}}_n)$ has two subalgebras
which are isomorphic to the quantum affine algebra $U_{\gq} (\widehat{\mathfrak{sl}}_n )$.
The horizontal subalgebra $U_{\gq}^\mathrm{hor} (\widehat{\mathfrak{sl}}_n )$
is generated by \lq\lq zero modes\rq\rq\ $E_{i,0}, F_{i,0}$ and $K_i^\pm~( i \in \mathbb{Z}/n \mathbb{Z})$
and their relations take the original form by Drinfeld-Jimbo.
On the other hand, the vertical subalgebra $U_{\gq}^\mathrm{ver} (\widehat{\mathfrak{sl}}_n )$
is most conveniently described by
restricting the Drinfeld currents to the \lq\lq finite algebra\rq\rq\ part
$E_i(z), F_i(z), K_i^{\pm}(z), \gq^{\pm c/2}$ with $i \neq 0$.
As in the case of DIM algebra,  there is an algebra automorphism that exchanges
the horizontal and vertical algebra. It also exchanges two central elements and
hence the level $(k,\ell)$ representation is mapped to the level $(-\ell, k)$ representation.
In the case of DIM,  there exists the Heisenberg subalgebra labeled by rational number $k/\ell$
and related by $SL(2, \mathbb{Z})$ transformation. It is interesting to see that the
$\mathbb{Z}_2$ symmetry of vertical and horizontal subalgebra of $U_{\gq,\gd}(\widehat{\widehat{\mathfrak{gl}}}_n)$
can be enhanced to $SL(2, \mathbb{Z})$.

When $c=0$ (corresponding to the vertical representation), the factor in \eqref{Kpm} becomes
trivial, and $K_i^{\pm} (z)$ are completely commuting. Hence, there exist simultaneous eigenstates
of $H_{i,{\pm r}}$ in this case.
In the geometric construction of representation of the quantum toroidal algebra,
we can obtain the vertical representation, the fixed points of torus action give
simultaneous eigenstates of $H_{i,{\pm r}}$. Hence, in refs \cite{VV2}-\cite{Neg}
the defining relations with $c=0$ have been provided from the very beginning.


\subsection{Vertical representation and color selection rule}

In the vertical representation or level $(0,1)$ representation,
the Heisenberg part is completely commuting
and they are diagonalizable by simultaneous eigenstates.
This representation is what we will obtain by the geometric
construction based on the (Nakajima) quiver variety
\cite{Nak1,Nak2}, \cite{VV2}-\cite{Neg}.
The relevant geometry is the instanton moduli space of
the ALE space of $A_n$ type, which is a resolution of
the orbifold $\mathbb{C}^2/ \mathbb{Z}_{n+1}$.
Originally Nakajima constructed a representation of the
affine Kac-Moody algebra from the (equivariant) cohomology of
the moduli space. A representation of the quantum affinization
is expected to be obtained, if we replace the cohomology with the corresponding
$K$ theory. When we consider the instantons on
the ALE space of $A_n$ type, the affine algebra is $A_{n-1}^{(1)}$,\
and we have a representation of the quantum toroidal algebra
by the $K$ theory of the quiver variety (the instanton moduli space).

In the vertical representation, we can label simultaneous eigenstates of $K_i^{\pm}(z)$ by a partition
(Young diagram) $\lambda$. We take a basis $\{ \vert \lambda ) \}$ of the Fock space\footnote{
We reserve the standard bra-ket notation for states in the horizontal representation to be introduced in the
next subsection.}, which simultaneously diagonalizes $K^{\pm}_\ell (z)$. Since the eigenvalues are non-degenerate,
the freedom is only in the normalization of each eigenvector $\vert \lambda )$.
In the following computation, we assume that $\{ \vert \lambda ) \}$ is an orthonormal basis.
Then $\vert \lambda )$ is canonically identified with the dual basis $( \lambda \vert $
with $( \lambda \vert \mu ) = \delta_{\lambda \mu}$.
The vertical representation with the spectral parameter $v$ is introduced in \cite{FJMM}.
For a partition $\lambda = (\lambda_1 \geq \lambda_2 \geq \cdots)$,
let $\lambda \pm 1_j =  (\lambda_1 \geq \cdots \geq \lambda_j \pm 1 \geq \cdots)$.
We assign the color $i-j +k$ modulo $n$ to the box $(i,j) \in \lambda$,
so that the empty partition and the box on the diagonal have the color $k$.
With the notation $x_s := q_1^{\lambda_s-1} q_3^{s-1} = q_2 q_1^{\lambda_s} q_3^{s}$,
the non-vanishing matrix elements of the vertical representation can be written as follows;
\begin{eqnarray}
( \lambda + 1_j \vert E_i(z) \vert \lambda )
&=&
\prod_{\substack{s=1, \\ \lambda_s + i \equiv s+k}}^{j-1} \psi(q_1^{-1}{x_s}/{x_j})
\prod_{\substack{s=1, \\ \lambda_s + i +1 \equiv s+k}}^{j-1} \psi({x_j}/{x_s})
~\delta(q_1 x_j v/z)~\deltabar_{i,j- \lambda_j +k-1} , \label{Evertical}
\\
( \lambda \vert F_i(z) \vert \lambda+1_j )
&=&
\prod_{\substack{s=j+1, \\ \lambda_s + i \equiv s+k}}^{\infty} \psi(q_1^{-1} {x_s}/{x_j})
\prod_{\substack{s=j+1, \\ \lambda_s + i +1  \equiv s+k}}^{\infty}  \psi({x_j}/{x_s})
~\delta(q_1 x_j v/z)~\deltabar_{i,j- \lambda_j +k-1}, \label{Fvertical}
\\
( \lambda \vert K^{+}_i(z) \vert \lambda )
&=&
\prod_{\substack{s=1, \\ \lambda_s + i \equiv s+k}}^{\infty} \psi(x_s v/z)
\prod_{\substack{s=1, \\ \lambda_s + i +1 \equiv s+k}}^{\infty} \psi(q_3^{-1} x_s v/z)^{-1},
\label{Kvertical+}
\\
( \lambda \vert K^{-}_i(z) \vert \lambda )
&=&
\prod_{\substack{s=1, \\ \lambda_s + i \equiv s+k}}^{\infty} \psi(q_2 z/x_s v)^{-1}
\prod_{\substack{s=1, \\ \lambda_s + i +1 \equiv s+k}}^{\infty} \psi(q_1^{-1} z/x_s v),
\label{Kvertical-}
\end{eqnarray}
where $k$ is the charge (coloring) of the empty partition, and $\psi(z)$ is defined by \eqref{psi}.
The parameter $v$ is called spectral parameter of the vertical representation.
The factor $\deltabar_{i,j- \lambda_j +k-1}$
stands for the color selection rule stating that the color of the box that is added to or removed from
the Young diagram is $i$.  Note that the matrix elements of $K^{-}_i(z)$ are related by \eqref{inversion}.
The matrix elements of the generators $E_{i,k}$ and $F_{i,k}$ are easily identified
by expanding $\delta(q_1 x_j v/z)$.
We note that, except for the range of the product over $1 \leq s \leq j-1$ or $j+1 \leq s < \infty$,
$E_i(z)$ and $F_i(z)$ have the common factors in their matrix elements.
In fact, these factors emerge as a consequence of substituting $v/z = (q_1 x_j)^{-1}$
imposed by the delta-function to the corresponding factors in \eqref{Kvertical+}.
The difference of the ranges for $E_i(z)$ and $F_i(z)$ is due to the semi-infinite
product construction of the Fock module in \cite{FJMM}.
There are the restrictions on the product in the right hand side:
\begin{equation}
\lambda_s + i \equiv s+k, \qquad  \lambda_s + i +1 \equiv s+k.
\label{selection}
\end{equation}
The meaning of these restrictions becomes clear, if one recalls that
we have assigned the color $i-j +k$ modulo $n$ to the box $(i,j) \in \lambda$.
Then the first condition of \eqref{selection} means that the last box $(s, \lambda_s)$ in the $s$-th row
has the color $i$. Note that the box $(s, \lambda_s)$ may be removed from the diagram
if $\lambda_{s} \neq \lambda_{s+1}$. Similarly, the second condition means
 the box $(s, \lambda_s+1)$, which may be added to the diagram if $\lambda_{s-1} \neq \lambda_s$,
 has the color $i$.

For a Young diagram $\lambda$, let us define the set of addable (or concave) and removable (or convex) corner
of $\lambda$. The addable corner $A(\lambda)$ is the set of boxes $(x,y) \notin \lambda$
such that we can add $(x,y)$ to $\lambda$ without violating the Young diagram condition.
Similarly, the removable corner $R(\lambda)$ is the set of boxes $(x,y) \in \lambda$
such that we can remove $(x,y)$ from $\lambda$ without violating  the Young diagram condition.
That is, if we follow the boundary of $\lambda$ from $x \to \infty$ to $y \to \infty$
the direction changes from up to right (from right to up) at addable (removable) corner.
Since the direction is up for $x \to \infty$ and right for $y \to \infty$, we see
for any Young diagram $\lambda$
\begin{equation}
\#A(\lambda) - \# R(\lambda) =1. \label{AR}
\end{equation}
When the vacuum charge is $k$, we introduce the set of addable and removable
corners of color $\ell$ by
\begin{eqnarray}
A_\ell^{(k)}(\lambda) &:=& \{ (i,j) \in A(\lambda) \vert k + i -j \equiv \ell \}, \\
R_\ell^{(k)} (\lambda)&:=& \{ (i,j) \in R(\lambda) \vert k + i -j \equiv \ell \}.
\end{eqnarray}
With these notations, we can rewrite \eqref{Kvertical+} as follows\footnote{\eqref{Kvertical-} can be rewritten similarly.};
\begin{equation}
( \lambda \vert K^{+}_\ell (z) \vert \lambda ) =
\prod_{(s, \lambda_s) \in R_\ell^{(k)} (\lambda)} \psi(x_s v/z)
\prod_{(s, \lambda_s +1) \in A_\ell^{(k)} (\lambda)} \psi(q_3^{-1} x_s v/z)^{-1}.
\end{equation}
In particular, we have
\begin{equation}
K_\ell \vert \lambda )
= \gq^{\#  R_\ell^{(k)} (\lambda)- \# A_\ell^{(k)}(\lambda)} \vert \lambda ).
\end{equation}
Hence, from \eqref{AR} the value of the center is $\kappa = \prod_\ell K_\ell = \gq^{-1}$,
and we see that the vertical representation has, in fact, level $(0, 1)$.


\subsection{Vertex operators and horizontal representation}

The Heisenberg subalgebra part of $U_{\gq,\gd}(\widehat{\widehat{\mathfrak{gl}}}_n)$ is
\begin{equation}
\left[ H_{i,r}, H_{j,s} \right]
= \delta_{r+s, 0} \frac{[r][cr]}{r} C_{ij}^{[r]},
\label{TorHeis}
\end{equation}
where $[n] = (\gq^n - \gq^{-n}) / (\gq - \gq^{-1})$.
We introduce the $(\gq, \gd)$ deformed Cartan matrix by
\begin{equation}
C_{ij} (\gq, \gd) := [a_{ij}] \gd^{-m_{ij}}, \label{deformedC}
\end{equation}
and define
\begin{equation}
 C_{ij}^{[r]} = C_{ij} (\gq^r, \gd^r) = \frac{[r a_{ij}]}{[r]} \gd^{-r m_{ij}}.
\end{equation}
Note that the commutation relation \eqref{TorHeis} is invariant under the redefinition \eqref{shift}.
To introduce a vertex operator representation with $c=1$,
we will employ the following vertex operators:
\begin{equation}
V_{i}^{(\pm)}(z) := \exp \left( \mp \sum_{r=1}^\infty \frac{H_{i, \pm r}}{[r]} z^{\mp r} \right).
\end{equation}
The fundamental OPE by normal ordering is
\begin{equation}
V_{i}^{(+)} (z) V_{j}^{(-)}(w) = s_{ij}(z,w) : V_{i}^{(+)} (z) V_{j}^{(-)}(w):,  \qquad |z| > |w|,
\label{VOPE}
\end{equation}
where
\begin{equation}
s_{ij}(z,w) =
\frac{\left( 1 - \frac{\gq w}{z} \right)^{\deltabar_{i,j}} \left( 1 - \frac{w}{\gq z} \right)^{\deltabar_{i,j}} }
{\left( 1 - \frac{\gd w}{z} \right)^{\deltabar_{i,j-1}}\left( 1 - \frac{w}{ \gd z} \right)^{\deltabar_{i,j+1}}}.
\end{equation}
Let us introduce notations for the oscillator part of the vertical operator representation:
\begin{eqnarray}
\eta_i(z) &=& V_{i}^{(-)} ( \gq^{-\frac{1}{2}} z ) V_{i}^{(+)} (\gq^{\frac{1}{2}} z), \\
\xi_i(z) &=& V_{i}^{(-)} ( \gq^{\frac{1}{2}} z )^{-1} V_{i}^{(+)} (\gq^{-\frac{1}{2}} z)^{-1}, \\
\varphi_i^{\pm}(z) &=& V_{i}^{(\pm)}(\gq^{\pm1} z) V_{i}^{(\pm)}
 (\gq^{\mp 1} z)^{-1}.
\end{eqnarray}
The inverse of the vertex operator $V_{\pm}^{(i)}(z)^{-1}$ is defined by flipping the sign of
the exponential and satisfies, for example,
\begin{equation}
V_{i}^{(+)} (z) V_{j}^{(-)}(w)^{-1} = s_{ij}(z,w)^{-1} : V_{i}^{(+)} (z) V_{j}^{(-)}(w)^{-1}:,  \qquad |z| > |w|.
\end{equation}
Then the vertex operator (horizontal) representation with level $(1,N)$
is given by\footnote{The shift of the argument of $K_i^{\pm}$ is due to \eqref{shift}.}
\begin{eqnarray}
E_i(z) &\to& \eta_i(z)~e^{\alphabar_i} z^{H_{i,0}+1} (z^{-N} \gq^{N} u)^{\deltabar_{i,k}},  \label{hE}\\
F_i(z) &\to& \xi_i(z)~e^{- {\alphabar_i}} z^{- H_{i,0}+1}  (z^N  \gq^{-N} u^{-1})^{\deltabar_{i,k}},  \label{hF} \\
K_i^{\pm}(\gq^{\frac{1}{2}} z) &\to& \varphi_i^{\pm}(z)~\gq^{\pm \partial_{\alphabar_i} \mp \deltabar_{i,k} N}. \label{hK}
\end{eqnarray}
The vertex operator representation was originally given in \cite{Saito, STU}, where the level was $(1,0)$.
Here we generalize it to level $(1,N)$. We have also introduced the spectral parameter
$u$ for the horizontal representation. Note that the modification by the level $N$ and the spectral parameter $u$
appears, only when the color $i$ of the currents is the same as the vacuum\footnote{The choice of
vacuum state breaks the cyclic symmetry of the affine $A_n$ Dynkin diagram.}.
The zero mode parts $e^{\pm \alphabar_i}, z^{\pm \partial_{\alphabar_i}}$ and $H_{i,0}$ satisfy
\begin{eqnarray}
e^{\alphabar_i} e^{\alphabar_j}  &=& (-1)^{a_{ij}}  e^{\alphabar_j} e^{\alphabar_i}, \label{cocycle} \\
z^{\partial_{\alphabar_i}} e^{\alphabar_j} &=& z^{a_{ij}}e^{\alphabar_j} z^{\partial_{\alphabar_i}}, \label{momentum}\\
z^{H_{i,0}} e^{\alphabar_j} &=& z^{a_{ij}} \gd^{-\frac{1}{2} m_{ij}} e^{\alphabar_j} z^{H_{i,0}} \label{Hzero}
\end{eqnarray}
and the same relation for $\gq$ replacing $z$. Note that \eqref{cocycle} defines a $\mathbb{Z}_2$ twist of
the group algebra of the root lattice $\overline{Q} = \sum_i \mathbb{Z} \cdot \alphabar_i$.
Since $K_i^{\pm 1} = K_i^{\pm}(0) \to \gq^{\pm \partial_{\alphabar_i} \mp \deltabar_{i,k} N}$,
we see that $\kappa = \prod_i K_i = \gq^{-N}$ by $\sum_i  \partial_{\alphabar_i} =0$.
The $N$ dependence of $E_i(z)$ and $F_i(z)$ is fixed by the commutation relations.
An additional factor of $\gq$ is introduced for later convenience.


Let us check the commutation relation \eqref{EF}. First of all, when $i \equiv j$, we have
\begin{eqnarray}
E_i(z) F_i(w) &=&
\frac{: \eta_i(z) \xi_i(w) :
\left( \frac{z}{w} \right)^{\partial_{\alphabar_i} -1- \deltabar_{i,k} N}}
{\left( 1 - \frac{\gq w}{z} \right) \left( 1 - \frac{w}{\gq z} \right)}, \qquad |z| > |w|, \\
F_i(w) E_i(z) &=&
\frac{: \xi_i(w) \eta_i(z) :
\left( \frac{z}{w} \right)^{\partial_{\alphabar_i} +1- \deltabar_{i,k} N}}
{\left( 1 - \frac{\gq z}{w} \right) \left( 1 - \frac{z}{\gq w} \right)}, \qquad |w| > |z|.
\end{eqnarray}
Using the formula
\begin{equation}
\frac{x}{(1 - \gq x)(1 - \gq^{-1} x)}  -  \frac{x^{-1}}{(1 - \gq x^{-1})(1 - \gq^{-1} x^{-1})}
= \frac{ \delta(\gq x) - \delta(\gq^{-1} x) }{\gq - \gq^{-1}}
\end{equation}
and the property $\delta(xy^{-1}) f(x) = \delta(xy^{-1})f(y)$, we obtain
\begin{equation}
\left[ E_i(z), F_i(w) \right]
= \delta\left(\frac{\gq w}{z}\right) \frac{K_i^{+} (z)}{\gq - \gq^{-1}}
- \delta\left( \frac{w}{\gq z} \right) \frac{K_i^{-} (w)}{\gq - \gq^{-1}}.
\end{equation}
When $i \equiv j \pm 1$, by taking the commutation relations of zero modes
\eqref{cocycle} and \eqref{Hzero} into account,  we can check
\begin{equation}
\left[ E_i(z), F_{i \pm 1} (w) \right] =0.
\end{equation}

Before concluding the section, we give another example that shows
the role of the commutation of the zero modes.
From the OPE relation \eqref{VOPE}, we see
\begin{equation}
\varphi^{+}_\ell (\gq^{-\frac{1}{2}} z) \eta_k(w)
=
\frac{ \left( 1 -  q_3^{-1} \frac{w}{z} \right)^{\deltabar_{\ell, k-1}}
\left( 1 -  q_2^{-1} \frac{w}{z} \right)^{\deltabar_{\ell, k}}
\left( 1 -  q_1^{-1} \frac{w}{z} \right)^{\deltabar_{\ell, k+1}}}
{\left( 1 -  q_1 \frac{w}{z} \right)^{\deltabar_{\ell, k-1}}
\left( 1 -  q_2\frac{w}{z} \right)^{\deltabar_{\ell, k}}
\left( 1 -  q_3\frac{w}{z} \right)^{\deltabar_{\ell, k+1}}}
:\varphi^{+}_\ell (\gq^{-\frac{1}{2}} z) \eta_k(w): , \label{VOPE2}
\end{equation}
and
\begin{equation}
 \eta_k(w) \varphi^{-}_\ell (\gq^{\frac{1}{2}} z)
=
\frac{ \left( 1 -  q_1^{-1} \frac{z}{w} \right)^{\deltabar_{\ell, k-1}}
\left( 1 -  q_2^{-1} \frac{z}{w} \right)^{\deltabar_{\ell, k}}
\left( 1 -  q_3^{-1} \frac{z}{w} \right)^{\deltabar_{\ell, k+1}}}
{\left( 1 -  q_3 \frac{z}{w} \right)^{\deltabar_{\ell, k-1}}
\left( 1 -  q_2\frac{z}{w} \right)^{\deltabar_{\ell, k}}
\left( 1 -  q_1\frac{z}{w} \right)^{\deltabar_{\ell, k+1}}}
: \eta_k(w) \varphi^{-}_\ell (\gq^{\frac{1}{2}} z): . \label{VOPE3}
\end{equation}
By combining with the commutation relation of the zero modes:
$\gq^{\pm\partial_{\alphabar_k}} e^{\alphabar_\ell} =
\gq^{\pm a_{k\ell}}e^{\alphabar_\ell} \gq^{\pm\partial_{\alphabar_k}}$, see \eqref{momentum},
we recover the relation \eqref{OPE} between $K_\ell^{+}(z), K_\ell^{-}(\gq z)$
and $E_k(w)$.


\section{Construction of the intertwining operator}


Let $\mathcal{F}_v^{(0,1)}$ and $\mathcal{F}_u^{(1,N)}$ be the Fock spaces
for the vertical and horizontal representations with spectral parameters $v$ and $u$.
We assume that the color of the vacuum (the highest weight state) in the vertical representation
is $0$ for simplicity of expressions.
Note that the color of the vacuum can be made $k$ by the shift of
the color indices: $E_i(z), F_i(z), K_i^{\pm}(z) \to E_{i+k} (z), F_{i+k} (z), K_{i+k}^{\pm}(z)$.
Following \cite{AFS}, we define the intertwiners of the quantum toroidal algebra
$U_{\gq,\gd}(\widehat{\widehat{\mathfrak{gl}}}_n)$ as follows:
\begin{equation}
\Phi(N,u \vert v) : \mathcal{F}_v^{(0,1)} \otimes \mathcal{F}_u^{(1,N)} \to \mathcal{F}_w^{(1,N+1)},
\qquad a\Phi = \Phi \Delta (a),
\end{equation}
and its dual
\begin{equation}
\Phi^{*}(N+1, w \vert v)  : \mathcal{F}_w^{(1,N+1)} \to  \mathcal{F}_u^{(1,N)} \otimes \mathcal{F}_v^{(0,1)},
\qquad \Delta(a) \Phi^{*} = \Phi ^{*} a.
\end{equation}
Later we will see that the intertwiners $\Phi(N,u \vert v)$ and $\Phi^{*}(N+1, w \vert v)$ exist
only when the spectral parameters satisfy a conservation law $w = -uv$.
Note that, in our notation, the level and the spectral parameter of the horizontal representation refer
to the source Fock space.  In the following, we often suppress them for simplicity.
They are indicated explicitly, whenever it is helpful.
In terms of the basis $\vert \lambda )$ of the vertical representation $\mathcal{F}_v^{(0,1)}$,
we introduce components of the intertwiner by
\begin{equation}
\Phi_\lambda = \Phi (\vert \lambda ) \otimes \bullet),
\end{equation}
where $\{ \vert \lambda ) \}$ is the basis of $\mathcal{F}_v^{(0,1)}$ introduced in section 3.1.
The component $\Phi_\lambda$ is a map between horizontal representations, and
our task is to express it in terms of the vertex operators.
Since $C_1=1$ and $C_2 = \gq$ for the vertical and the horizontal representations,
the definition of the coproduct implies the following intertwining relations for $\Phi_\lambda$:
\begin{eqnarray}
E_i(z) \Phi_\lambda(v) &=& \sum_{j=1}^{\ell(\lambda)+1} ( \lambda + 1_j \vert E_i(z) \vert \lambda )
\Phi_{\lambda + 1_j}(v) + ( \lambda \vert K_i^{-}(z) \vert \lambda ) \Phi_{\lambda}(v) E_i(z),
\label{Eintertwin}
\\
F_i(z) \Phi_\lambda(v) &=& \sum_{j=1}^{\ell(\lambda)} ( \lambda - 1_j \vert F_i(\gq z) \vert \lambda )
\Phi_{\lambda - 1_j}(v) K_i^{+} (\gq z) + \Phi_\lambda (v) F_i(z),
\\
K^{+}_i(z) \Phi_\lambda(v) &=& ( \lambda \vert  K_i^{+}(z) \vert \lambda )
\Phi_{\lambda} (v) K_i^{+}(z),
\\
K^{-}_i(\gq z) \Phi_\lambda(v) &=&  ( \lambda \vert  K_i^{-}(z) \vert \lambda )
\Phi_{\lambda} (v) K_i^{-}(\gq z).
\end{eqnarray}
At the left hand side, the currents $E_i(z), F_i(z), K_i^{+}(z)$ and $K_i^{-}(\gq z)$
are taken in the level $(1,N+1)$ representation, while, at the right hand side, the representations are at level $(1,N)$.
The argument $v$ is the spectral parameter of the vertical representation so that $(\lambda,v)$
represents the data of the state on the vertical side.
We have used that $1 = \sum_{\lambda} \vert \lambda ) ( \lambda \vert$ in order to derive
the intertwining relations assuming $( \lambda \vert \mu ) = \delta_{\lambda, \mu}$.
If we employ a different normalization,  the intertwining relation will involve the normalization factor.

The component of the dual intertwiner is defined by\footnote{Here we normalize the dual intertwiner
in a way distinct from \cite{AFS}.}
\begin{equation}
\Phi^{*} = \sum_{\lambda} \Phi^{*}_\lambda (\bullet) \otimes \vert \lambda ).
\end{equation}
Since $C_1 = \gq, C_2 =1$ for $\Phi^{*}_\lambda$, we find the following intertwining relations:
\begin{eqnarray}
 \Phi^{*}_\lambda(v) E_i(z)&=& E_i(z)\Phi^{*}_{\lambda}(v)
 +  K_i^{-}(\gq z) \sum_{j=1}^{\ell(\lambda)} \Phi^{*}_{\lambda - 1_j}(v)
 ( \lambda  \vert E_i(\gq z) \vert \lambda  - 1_j ),
\\
\Phi^{*}_\lambda(v) F_i(z) &=&
 ( \lambda  \vert K_i^{+}(z) \vert \lambda ) F_i(z) \Phi^{*}_\lambda (v)
+ \sum_{j=1}^{\ell(\lambda)+1} \Phi^{*}_{\lambda + 1_j} (v)
( \lambda \vert F_i(z) \vert \lambda + 1_j ),
\\
\Phi^{*}_\lambda(v) K^{+}_i(\gq z) &=&
( \lambda \vert  K_i^{+}(z) \vert \lambda )
 K_i^{+}(\gq z)  \Phi^{*}_{\lambda} (v),
\\
\Phi^{*}_\lambda(v) K^{-}_i(z) &=&
( \lambda \vert  K_i^{-}(z) \vert \lambda )
K_i^{-}(z) \Phi^{*}_{\lambda} (v) .
\end{eqnarray}


\subsection{Structure of the intertwining operator}

It turns out that the components of the intertwiner have the same structure
as in the DIM case \cite{AFS}. Namely, with the normalization factor $C_\lambda(q_1, q_3)$
which is related to the normalization of the basis $\vert \lambda )$
of the vertical representation, we have
\begin{equation}
  \Phi_\lambda (v)
  = \frac{t_\lambda(q_1, q_3)}{C_\lambda(q_1, q_3)}
  : \prod_{1\leq i \leq \ell(\lambda)}^{\leftarrow} \left(
  \prod_{1\leq j \leq \lambda_i}^{\leftarrow}
  E_{\bar c(i,j)}(q_1^{j-1} q_3^{i-1} v)
  \right)
  \cdot \Phi_{\varnothing} (v) :,  \label{intertwiner}
\end{equation}
where the vacuum component is (formally) given by an infinite product:
\begin{equation}
\Phi_{\varnothing}(v) = : \prod_{i,j = 1}^{\infty}  \eta_{\bar c(i,j)} (q_1^{j-1} q_3^{i-1} v)^{-1}:.
\end{equation}
$\bar c(i,j) \equiv i -j $ is the content of the box $(i,j)$ modulo $n$, which defines the coloring of boxes.
Since the zero modes are non-commutative, we have to fix the ordering of $E_i (z)$ in \eqref{intertwiner}.
This is the reason why we used the notation $\displaystyle \prod^{\leftarrow}$,
which means we take the product in the \lq\lq reversed\rq\rq\ order, namely
$\displaystyle \prod_{1\leq i \leq n}^{\leftarrow} a_i = a_n a_{n-1} \cdots a_1$.
In the case of \eqref{intertwiner}, it is more complicated, since we have double indices.
We first order $E_{\bar c(i,j)}$ in each row with respect to the second index $j$,
then we order the blocks of each row from the first (rightmost) to the last  (leftmost).
See \eqref{ordering} below more about the ordering of the product in \eqref{intertwiner}.
We impose $C_\varnothing(q_1, q_3)= t_\varnothing(q_1, q_3)=1$ as the normalization condition.
Then, later we will see the intertwining relation fixes the normalization factor as
\begin{equation}
 C_\lambda(q_1, q_3) = {\displaystyle{\prod_{\substack{\square \in \lambda \\ h_\lambda(\square) \equiv 0}}}
 (1-  q_1^{a_\lambda(\square)} q_3^{-\ell_\lambda(\square)-1})}. \label{normalization}
\end{equation}
We define the arm-length,  the leg-length and the hook length of $\square = (i,j)  \in \lambda$ by
\begin{equation}
a_\lambda(i,j) := \lambda_i -j, \qquad \ell_\lambda(i,j) = \lambda^\prime_j -i,
\end{equation}
and
\begin{equation}
h_{\lambda}(\square) = a_\lambda (\square) + \ell_\lambda (\square) +1,
\end{equation}
where $\lambda^\prime$ is the transpose of the Young diagram.
Note that if we do not have the restriction that the hook length $h_\lambda(\square)$ is a multiple of $n$,
the normalization factor appears in the norm of the Macdonald function $P_\lambda(x)$.
As we will show in section 4.3, the intertwining relation with $E_\ell(z)$ gives
the following recursion relation for the prefactor $t_{\lambda}$ of $\Phi_\lambda(v)$:
\begin{equation}
\frac{t_{\lambda + 1_j}}{t_\lambda}
= \left( - \gq q_1^{\lambda_j+1} \right)^{\deltabar_{\ell,j}}
  \prod_{\substack{s=1, \\ s-\lambda_s \equiv \ell} }^{\ell(\lambda)} \gq^{-1}
  \prod_{\substack{s=1, \\ s-\lambda_s \equiv \ell+1, \\ s \neq j}}^{\ell(\lambda)+1} \gq.
 \label{t-recursion}
\end{equation}
With the initial condition $t_{\varnothing} =1$, we obtain
\begin{equation}
 t_{\lambda}(q_1, q_3)
  =  \prod_{\substack{\square \in \lambda \\ h_\lambda(\square) \equiv 0} } \gq
  \prod_{\substack{(i,j)\in \lambda, \\ j \equiv 0}} \left( - \gq q_1^{j} \right). \label{tfactor}
\end{equation}


In formula \eqref{intertwiner}, we employ the vertex operator $E_{\bar c(i, j)}(q_1^{j-1}q_3^{i-1}v)$
in the representation with level $(1,N+1)$ and the spectral parameter $-uv$.
Recall that the component of the intertwiner $\Phi_\lambda (v)$ can be regarded as a map
between two Fock spaces $\mathcal{F}_{u}^{(1,N)} \to \mathcal{F}_{w}^{(1,N+1)}$.
We will see that the relation $w=-uv$ is required for the existence of the intertwiner.
The level and the spectral parameter of the horizontal representation affect only
the zero mode part of $E_i (z)$  (see \eqref{hE}) and
it turns out that it is natural to use the vertex operators referring to the target Fock space.

For the convenience of forthcoming computations, let us separate the zero mode part of
the intertwiner as follows\footnote{Since there is no ordering problem in the oscillator part,
we use the usual notation $\prod$ in the normal product as compared with \eqref{intertwiner}.}
\begin{equation}
  \Phi_\lambda (v)
= \frac{\widetilde{t}_\lambda(u,v; q_1, q_3)}{C_\lambda(q_1, q_3)}
  : \prod_{(i,j) \in \lambda} \eta_{\bar c(i,j)} (q_1^{j-1} q_3^{i-1} v)
  \cdot \Phi_{\varnothing} (v) :, \label{intertwiner2}
\end{equation}
where
\begin{align}
  \widetilde{t}_{\lambda}(u,v; q_1, q_3)
  &= t_\lambda(q_1, q_3) u^{|\lambda|_0} (-v)^{- N |\lambda|_0}
  f_{\lambda}(q_1, q_3)^{-N-1}  z_{\lambda}(v), \label{separation}
\end{align}
and $|\lambda|_0$ denotes the number of boxes with color $0$ in $\lambda$.
The monomial factor $\tilde{t}_\lambda(u,v; q_1, q_3)$ now depends on the horizontal spectral parameter $u$
and takes values in the group algebra of the root lattice.
The group algebra part of $\Phi_\lambda(v)$ is
\begin{equation}
  z_{\lambda}(v)
  = \prod_{1\leq i \leq \ell(\lambda)}^{\leftarrow} \left(
  \prod_{1\leq j \leq \lambda_i}^{\leftarrow}  e_{i,j}(v)
  \right),
 \qquad
  e_{i,j}(v) = e^{\alphabar_{i-j} } ~(q_1^{j-1} q_3^{i-1} v)^{H_{i-j,0 }+1}.
  \label{ordering}
\end{equation}
The factor in \eqref{separation}
\begin{equation}
f_{\lambda}(q_1, q_3)
= \prod_{\substack{ (i,j) \in \lambda \\ \bar c(i,j) \equiv 0}}
 (-1) q_1^{j-\frac{1}{2}} q_3^{i- \frac{1}{2}}, \label{framing}
\end{equation}
is the generalized framing factor arising from the commutation of zero modes.
If we do not impose the restriction $\bar c(i,j) \equiv 0$,
$f_{\lambda}(q, t^{-1})$ is nothing but the framing factor of the refined topological vertex
\cite{Taki:2007dh, Awata:2008ed}.
The dependence of the intertwiner $\Phi_\lambda(v)$ on the level $(1,N)$
can be arranged simply in the powers of $f_\lambda(q_1, q_3)$ and $-v$.
As we emphasized before,  since $e_{i,j}(v)$ are non-commutative, we have to fix the ordering of $e_{i,j}(v)$ in the product.
Our choice of the ordering in \eqref{ordering} is for convenience of computing of the intertwining relation with $E_\ell(z), F_\ell(z)$.
For example, it means that $z_{(3,2)} = e_{2,2} e_{2,1} e_{1,3} e_{1,2} e_{1,1}$.
The spectral parameter $u$ of the horizontal Fock space counts the number of boxes with the same color as the vacuum
and only appears in the second factor of \eqref{separation}. From now on, we write only the $v$-dependence explicitly.
The condition on the vacuum component $F_0(z) \Phi_\varnothing(v) =\Phi_\varnothing(v)  F_0(z)$
imposes the relation $w = -uv$ among the spectral parameters of the horizontal and the vertical Fock spaces.


Similarly, the dual intertwiner is given by replacing $E_i(z)$ by $F_i(z)$:
\begin{align}
  \Phi_\lambda^{*} (v)
  &= \frac{t^*_\lambda(q_1, q_3)}{C^\prime_\lambda(q_1, q_3)}
  : \prod_{1\leq i \leq \ell(\lambda)}^{\leftarrow} \left(
  \prod_{1\leq j \leq \lambda_i}^{\leftarrow}
  F_{\bar c(i,j)}(q_1^{j-1} q_3^{i-1} v)
  \right)
  \cdot \Phi_{\varnothing}^{*} (v) :  \label{dualint} \\
  &= \frac{\widetilde{t}^*_\lambda(u,v; q_1, q_3)}{C^\prime_\lambda(q_1, q_3)}
  : \prod_{(i,j) \in \lambda} \xi_{\bar c(i,j)} (q_1^{j-1} q_3^{i-1} v)
  \cdot \Phi_{\varnothing}^{*} (v) :,
\end{align}
with
\begin{equation}
\Phi_{\varnothing}^{*}(v) = : \prod_{i,j = 1}^{\infty}  \xi_{\bar c(i,j)} (q_1^{j-1} q_3^{i-1} v)^{-1}:.
\end{equation}
The normalization of the dual intertwiners is
\begin{equation}
 C^\prime_\lambda(q_1, q_3) = {\displaystyle{\prod_{\substack{\square \in \lambda \\ h_\lambda(\square) \equiv 0}}}
 (1-  q_1^{a_\lambda(\square)+1} q_3^{-\ell_\lambda(\square)})}. \label{dualnormalization}
\end{equation}
As in the case of $\Phi_\lambda(v)$, the vertex operator $F_{\bar c(i, j)}(q_1^{j-1}q_3^{i-1}v)$ in
\eqref{dualint} refers to the {\it target} Fock space of $\Phi_\lambda^{*} (v)$.
That is, it has the level $(1,N)$ and the horizontal spectral parameter $u$ (see \eqref{hF}).
Let us decompose the monomial factor as before,
\begin{align}
 \widetilde{t}^*_{\lambda}(u,v; q_1, q_3)
  = t^*_\lambda(q_1, q_3) (-v)^{N |\lambda|_0} u^{- |\lambda|_0}
  f_{\lambda}(q_1, q_3)^{N}  z^*_{\lambda}(v). \label{dualseparation}
\end{align}
with\footnote{The rule of ordering is the same as in the case of $z_\lambda$.}
\begin{equation}
  z^*_{\lambda}(v)
  = \prod_{1\leq i \leq \ell(\lambda)}^{\leftarrow} \left(
  \prod_{1\leq j \leq \lambda_i}^{\leftarrow}  f_{i,j}(v)
  \right),
  \qquad
  f_{i,j}(v) = e^{-\alphabar_{i-j} } ~(q_1^{j-1} q_3^{i-1} v)^{-H_{i-j,0 }+1},
 \label{dualordering}
\end{equation}
and the same generalized framing factor \eqref{framing}.
Then we have the recursion relation
\begin{equation}
  \frac{t^{*}_{\lambda+1_j}}{t^{*}_{\lambda}}
  =  (-\gq^{-1}) \left( -\gq q_1^{\lambda_j+1} \right)^{\deltabar_{\ell,j}}
  \prod_{\substack{s=1, \\ s-\lambda_s \equiv \ell} }^{\ell(\lambda)} \gq
  \prod_{\substack{s=1, \\ s-\lambda_s \equiv \ell+1, \\ s \neq j}}^{\ell(\lambda)+1} \gq^{-1}.
\end{equation}
By solving the recursion relation with the initial condition $t^{*}_{\varnothing} =1$, we obtain
\begin{equation}
  {t^{*}_{\lambda}} (q_1, q_3)
  = (-\gq)^{-|\lambda|}
  \prod_{\substack{\square \in \lambda \\ h_\lambda(\square) \equiv 0} } \gq^{-1}
  \prod_{\substack{(i,j)\in \lambda, \\ j \equiv 0}} \left( - \gq q_1^{j} \right).
\end{equation}


\subsection{Vacuum component of the intertwiner}

Let us first check that the vacuum component $\Phi_{\varnothing }(u)$ satisfies
the following intertwining relations:
\begin{eqnarray}
E_\ell(z) \Phi_\varnothing(v) &=& \delta(v/z) \deltabar_{\ell, 0}
\Phi_{(1)}(v) +  \psi( {z}/{v} )^{\deltabar_{\ell,0}} \Phi_{\varnothing}(v) E_\ell(z),
\label{Evac} \\
F_\ell(z) \Phi_\varnothing(v) &=& \Phi_\varnothing (v) F_\ell(z),
\label{Fvac} \\
K^{+}_\ell(z) \Phi_\varnothing(v) &=& \psi( q_2 {v}/{z} )^{-\deltabar_{\ell,0}} \Phi_{\varnothing} (v) K_i^{+}(z),
\label{K+vac}\\
K^{-}_\ell(\gq z) \Phi_\varnothing(v) &=& \psi( {z}/{v} )^{\deltabar_{\ell,0}} \Phi_{\varnothing} (v) K_\ell^{-}(\gq z).
\label{K-vac}
\end{eqnarray}
The color selection rule tells us that $ ( (1) \vert E_\ell(z) \vert \varnothing ) =0$
and $( \varnothing \vert  K_\ell^{\pm}(z) \vert \varnothing ) =1$, unless the color
$\ell$ is the same as that of the vacuum state $ \vert \varnothing )$, which we chose $0$.
Hence, if $\ell \neq 0$, all the currents $E_\ell(z), F_\ell(z) $ and $K_\ell^{\pm}(z)$
commute with the vacuum component $\Phi_{\varnothing }(v)$. This is consistent
with the fact that the dependence on the level $(1,N)$ and the spectral parameter
of the horizontal representation appear only in $E_0(z), F_0(z) $ and $K_0^{\pm}(z)$.

Since $C_\varnothing = t_\varnothing =1$, the non-trivial commutation relation
comes only from the vertex operator part.
A crucial point for the check of the intertwining relations is the following fact. Let
\begin{equation}
\widetilde{\eta}_{i-j}^{(\mathrm{tri})} (w) := \eta_{i-(j-1)} (q_1^{j-2} w)
\eta_{i-j} (q_1^{j-1} w) \eta_{i-(j+1)} (q_1^{j} w).
\end{equation}
Then the shift of the power of $q_1$ combined with \eqref{VOPE2} implies that non-trivial OPE factors
with $\varphi_\ell^{+}(\gq^{-\frac{1}{2}}z)$ and $\varphi_\ell^{-}(\gq^{\frac{1}{2}} z)$ cancel:
\begin{equation}
\varphi_\ell^{+}(\gq^{-\frac{1}{2}}z) \widetilde{\eta}_{i-j}^{(\mathrm{tri})} (w)
= \widetilde{\eta}_{i-j}^{(\mathrm{tri})} (w) \varphi_\ell^{+}(\gq^{-\frac{1}{2}}z),
\qquad
\varphi_\ell^{-}(\gq^{\frac{1}{2}} z) \widetilde{\eta}_{i-j}^{(\mathrm{tri})} (w)
= \widetilde{\eta}_{i-j}^{(\mathrm{tri})} (w) \varphi_\ell^{-}(\gq^{\frac{1}{2}} z).
\end{equation}
Because of this \lq\lq triplet\rq\rq\ cancellation, for each row a non-trivial OPE factor of $\varphi_\ell^{+}(\gq^{-\frac{1}{2}} z)$
with $\Phi_{\varnothing }(v)$ arises only when the first box $(i,1)$ has color $\ell -1$ or $\ell$.
When $i \equiv \ell$, we have\footnote{
Note that we are looking at OPE with the inverse of $\eta$.}
$\gq^{-1} \psi(q_1^{-1} q_3^{i-1} v/z)$. And when $i -1 \equiv \ell$, we have $\gq\cdot\psi(q_2 q_3^{i-1} v/z)^{-1}$.
Hence, if we take the product over rows, these factors cancel in general. But a non-trivial factor
$\gq\cdot\psi(q_2\frac{v}{z})^{-1}$ survives when $1 \equiv \ell +1$.
Recall that, according to our choice of the color of the vacuum, the box $(1,1)$ has color $0$.
From our definition of level $(1,N)$ representation, when the color of $K^{+}_\ell(z)$ is the same as the vacuum,
there is a change of the power of $\gq$, since the level of the horizontal representation changes
from $(1,N)$ to $(1,N+1)$. Taking this factor of $\gq$ into account, we can confirm \eqref{K+vac}.
We can also check \eqref{K-vac}.

By the same reasoning, we see that $E_\ell(z)$ and $F_\ell(z) $ commute
with the vacuum component $\Phi_{\varnothing }(v)$, unless $\ell = 0$.
When $\ell = 0$, we have
\begin{eqnarray}
F_0(z) \Phi_\varnothing(v) &=&
\left(\frac{z}{\gq}\right)^{N+1} w^{-1} \left( 1 - \frac{\gq v}{z} \right) e^{-\alphabar_0} z^{-H_{0,0} +1}
: \xi_0(z) \Phi_\varnothing(v) :, \nonumber \\
\Phi_\varnothing(v) F_0(z) &=&
\left(\frac{z}{\gq}\right)^{N} u^{-1} \left( 1 - \frac{z}{\gq v} \right)e^{-\alphabar_0} z^{-H_{0,0} +1}
 :  \Phi_\varnothing(v) \xi_0(z):.
\end{eqnarray}
Hence, the condition $F_0(z) \Phi_\varnothing(v) = \Phi_\varnothing(v) F_0(z)$ implies
\begin{equation}
w = -uv. \label{conservation}
\end{equation}
On the other hand, the substitution of \eqref{conservation} to
\begin{eqnarray}
E_0(z) \Phi_\varnothing(v) &=&
\left(\frac{z}{\gq}\right)^{-N-1} w\left( 1 - \frac{v}{z} \right)^{-1}e^{\alphabar_0} z^{H_{0,0} +1}
 : \eta_0(z) \Phi_\varnothing(v) :, \nonumber \\
\Phi_\varnothing(v) E_0(z) &=&
\left(\frac{z}{\gq}\right)^{-N} u\left( 1 - \frac{z}{q_2 v} \right)^{-1}e^{\alphabar_0} z^{H_{0,0} +1}
:  \Phi_\varnothing(v) \eta_0(z):,
\end{eqnarray}
gives
\begin{equation}
E_0(z) \Phi_\varnothing(v) - \psi(\frac{z}{v})\Phi_\varnothing(v) E_0(z)
= - \gq^{N+1} u v^{-N} \delta(\frac{v}{z}) e^{\alphabar_0} z^{H_{0,0} +1} : \eta_0(z) \Phi_\varnothing(v): .
\end{equation}
This means that \eqref{Evac} holds with
\begin{equation}
C_{(1)}(q_1, q_3) = 1, \qquad \widetilde{t}_{(1)} (u,v: q_1, q_3) = - \gq^{N+1} v^{-N} u e^{\alphabar_0} v^{H_{0,0} +1}.
\end{equation}


It may be useful to mention that the intertwining relation for $\Phi_{\varnothing }(v)$ can be
also reproduced by introducing the dual vertex operator
\begin{equation}
\widetilde{V}_i^{(\pm)} (z) := \exp \left( \mp \sum_{r=1}^\infty \Lambda_{i, \pm r} z^{\mp r} \right),
\end{equation}
with the commutation relation
\begin{equation}
\left[ \Lambda_{i,r}, H_{j,s} \right]
= \delta_{r+s,0} \deltabar_{i,j} \frac{[r]}{r}.
\end{equation}
More explicitly, $\Lambda_{i,r}$ is a linear combination of $H_{j,r}$
\begin{equation}
\Lambda_{i,r} = \sum_{j=0}^{n-1} b_{ij}^{[r]} H_{j,r},
\end{equation}
where
\begin{equation}
b_{ij}^{[r]}= b_{ij}(\gq^r, \gd^r),
\end{equation}
and $b_{ij}(\gq, \gd)$ are the components of the inverse of
the deformed Cartan matrix \eqref{deformedC};
\begin{equation}
(q_1^{\frac{n}{2}} - q_1^{-\frac{n}{2}})  (q_3^{\frac{n}{2}} - q_3^{-\frac{n}{2}})
b_{ij} (\gq, \gd)=
\begin{cases}
[i-j] \gd^{n+j-i} + [n+j-i] \gd^{j-i} \quad(i \geq j),\\
[j-i] \gd^{j-i-n} + [n+i-j] \gd^{j-i}  \quad (i \leq j) .
\end{cases}
\end{equation}
The fundamental OPE relation is
\begin{eqnarray}
V_i^{(+)} (z) \widetilde{V}_j^{(-)} (w) &=& \left( 1 - \frac{w}{z} \right)^{-\deltabar_{i,j}}
: V_i^{(+)} (z) \widetilde{V}_j^{(-)} (w) :, \\
 \widetilde{V}_i^{(+)} (z) V_j^{(-)} (w) &=& \left( 1 - \frac{w}{z} \right)^{\deltabar_{i,j}}
:  \widetilde{V}_i^{(+)} (z) V_j^{(-)} (w):.
\end{eqnarray}
Then another formula for $\Phi_{\varnothing }(v)$ is
\begin{equation}
\Phi_{\varnothing }(v) = : \widetilde{V}_k^{(-)}(\gq^{\frac{1}{2}} v)~\widetilde{V}_k^{(+)}(\gq^{\frac{3}{2}} v)^{-1} :,
\end{equation}
where $k$ is the color of the vacuum. We can check the intertwining relation for the vacuum component
with general $\deltabar_{\ell,  k}$.
Similarly, the vacuum component of the dual intertwiners can be expressed as
\begin{equation}
  \Phi_{\varnothing}^{*}(v)
  =  \widetilde{V}_k^{(-)}(\gq^{\frac{3}{2}} v)^{-1} ~\widetilde{V}_k^{(+)}(\gq^{\frac{1}{2}} v).
\end{equation}


\subsection{Zero mode part and intertwining relations}

We can use the same idea to compute the OPE relation of $\Phi_\lambda (v)$ with $K_\ell^{+}(z)$ and $K_\ell^{-}(\gq z)$.
When we compute the OPE relation of $\prod_{(i,j) \in \lambda} E_{\bar c(i,j)} (q_1^{j-1} q_3^{i-1} v)$ with $K_\ell^{+}(z)$, using \eqref{OPE} for the $(1,N+1)_{-uv}$ representation, for each row (fixed index $i$) a non-trivial OPE factor arises from the first box $(i,1)$ and the last box $(i,  \lambda_i)$ when they satisfy the color selection rule.
The factor from the first box $(i,1)$ exactly cancels the contribution from $\Phi_{\varnothing }(v)$ discussed above.
Thus the remaining factor comes from the last box $(i,  \lambda_i)$ with the color selection rule that $(i,\lambda_i)$ has the color $\ell$ or $\ell+1$.
From \eqref{OPE}, we obtain
\begin{eqnarray}
  \psi\left( \frac{x_i v}{z} \right) \quad \mathrm{for} \quad \lambda_i +\ell \equiv i, \qquad
  \psi\left( q_1 q_2 \frac{x_i v}{z} \right)^{-1} \quad \mathrm{for} \quad \lambda_i +\ell +1 \equiv i,
\end{eqnarray}
and, hence,
\begin{align}
  &\varphi_\ell^{+}(\gq^{-\frac{1}{2} } z) \gq^{\partial_{\alphabar_{\ell}} - \deltabar_{\ell,0}(N+1)}
  \prod_{1\leq i \leq \ell(\lambda)}^{\leftarrow} \left(
  \prod_{1\leq j \leq \lambda_i}^{\leftarrow}
  E_{\bar c(i,j)}(q_1^{j-1} q_3^{i-1} v) \right) \\
  &=\psi(q_2 v/z)^{\deltabar_{\ell,0}}
  \prod_{\substack{s=1, \\ s-\lambda_s \equiv \ell}}^{\ell(\lambda)} \psi(x_s v/z)
  \prod_{\substack{s=1, \\ s-\lambda_s \equiv \ell+1}}^{\ell(\lambda)+1} \psi(q_3^{-1} x_s v/z)^{-1}  \nonumber \\
  &~~\prod_{1\leq i \leq \ell(\lambda)}^{\leftarrow} \left(
  \prod_{1\leq j \leq \lambda_i}^{\leftarrow}
  E_{\bar c(i,j)}(q_1^{j-1} q_3^{i-1} v) \right)
  \varphi_\ell^{+}(\gq^{-\frac{1}{2} } z) \gq^{\partial_{\alphabar_{\ell}} - \deltabar_{\ell,0}(N+1)},
\end{align}
and
\begin{equation}
  \varphi_\ell^{+}(\gq^{-\frac{1}{2} } z) \gq^{\partial_{\alphabar_{\ell}} - \deltabar_{\ell,0}(N+1)} \Phi_{\lambda} (v)
  = \prod_{\substack{s=1, \\ s-\lambda_s \equiv \ell}}^{\ell(\lambda)} \psi(x_s v/z)
   \prod_{\substack{s=1, \\ s-\lambda_s \equiv \ell+1}}^{\ell(\lambda)+1} \psi(q_3^{-1} x_s v/z)^{-1}
   ~\Phi_{\lambda} (v)  \varphi_\ell^{+}(\gq^{-\frac{1}{2} } z) \gq^{\partial_{\alphabar_{\ell}} - \deltabar_{\ell,0}N},
\end{equation}
where we also used relation \eqref{K+vac}.
This implies
\begin{equation}
  K_\ell^{+}(z) \Phi_{\lambda} (v)
  = \prod_{\substack{s=1, \\ s-\lambda_s \equiv \ell}}^{\ell(\lambda)} \psi(x_s v/z)
   \prod_{\substack{s=1, \\ s-\lambda_s \equiv \ell+1}}^{\ell(\lambda)+1} \psi(q_3^{-1} x_s v/z)^{-1}
   ~\Phi_{\lambda} (v)  K_\ell^{+}(z).
\end{equation}
Similar computation is valid for $K_\ell^{-} (\gq z)$, since \eqref{OPE} also holds for $K_\ell^{-} (\gq z)$.


Let us move to the intertwining relation with $E_\ell(z)$.
Since \eqref{OPE} still holds even after replacing $K_\ell^{+}(z)$ by $E_\ell(z)$, we have
\begin{align}
E_\ell(z) \Phi_\lambda(v) - ( \lambda \vert K_\ell^{-}(z) \vert \lambda ) \Phi_{\lambda}(v) E_\ell(z)
  = \left[( \lambda \vert K_\ell^{+}(z) \vert \lambda )-( \lambda \vert K_\ell^{-}(z) \vert \lambda ) \right] \Phi_{\lambda}(v) E_\ell(z),
\end{align}
where we also used relation \eqref{Evac}.
Note that the first term at right hand side of \eqref{Evac} vanishes due to the coefficient $\psi(q_2 v/z) \delta(v/z)$.
To obtain the delta functions in the intertwining relation,
we make use of the following formal series identity for a rational function $\gamma(z)$ regular at $z=0, \infty$
and with simple poles at most (Lemma 3.3 of \cite{FFJMM});
\begin{align}
  \gamma^+(z) - \gamma^-(z)
  = \sum_{t} \gamma^{(t)} \delta(z/z^{(t)}), \label{formal}
\end{align}
where $\gamma^\pm(z)$ denote the Taylor expansions of $\gamma(z)$ in $z^{\mp 1}$ at $z=\infty$ and $z=0$.
The sum at the right hand side runs over all poles $z^{(t)}$ of $\gamma(z)$
with $\gamma^{(t)} = {\mathrm{Res}}_{z=z^{(t)}} \gamma(z)\frac{dz}{z}$ being the residues.
One can prove the identity by the partial fraction decomposition of rational functions.
The formula \eqref{formal} implies
\begin{align}
  ( \lambda \vert K_\ell^{+}(z) \vert \lambda )-( \lambda \vert K_\ell^{-}(z) \vert \lambda )
  &= \sum_{\substack{j=1, \\ j-\lambda_j \equiv \ell}}^{\ell(\lambda)}
  \delta\left( \frac{x_j v}{z} \right) \gq (1-q_2^{-1})
  \prod_{\substack{s=1, \\ s-\lambda_s \equiv \ell, \\ x\neq j} }^{\ell(\lambda)} \psi\left( \frac{x_s}{x_j} \right)
  \prod_{\substack{s=1, \\ s-\lambda_s \equiv \ell+1}}^{\ell(\lambda)+1} \psi\left( \frac{x_s}{q_3 x_j} \right)^{-1} \\
  &+ \sum_{\substack{j=1, \\ j-\lambda_j \equiv \ell+1}}^{\ell(\lambda)+1}
  \delta\left( \frac{q_1 x_j v}{z} \right) \gq^{-1} (1-q_2)
  \prod_{\substack{s=1, \\ s-\lambda_s \equiv \ell} }^{\ell(\lambda)} \psi\left( \frac{x_s}{q_1 x_j} \right)
  \prod_{\substack{s=1, \\ s-\lambda_s \equiv \ell+1, \\ s\neq j}}^{\ell(\lambda)+1} \psi\left( \frac{q_2 x_s}{x_j} \right)^{-1}.
\end{align}
To get $\Phi_{\lambda+1_j}(v)$ out of the difference $E_\ell(z) \Phi_\lambda(v) - ( \lambda \vert K_\ell^{-}(z) \vert \lambda )
\Phi_{\lambda}(v) E_\ell(z)$, we have to compute the normal ordered product
and take contributions of zero modes and spectral parameters into account.
We need the following OPE relation for $|z| > |w|$:
\begin{eqnarray}
\eta_i(z) \eta_j(w) &=& s_{ij} (\gq^{\frac{1}{2}} z, \gq^{-\frac{1}{2}} w)  :\eta_i(z) \eta_j(w): \nonumber \\
&=& \begin{cases}
\left( 1 - q_3 \frac{w}{z} \right)^{-1} :\eta_i(z) \eta_j(w): \quad j \equiv i-1 \\
\left( 1 - \frac{w}{z} \right) \left( 1 - q_1 q_3\frac{w}{z} \right) :\eta_i(z) \eta_j(w):  \quad j \equiv i \\
\left( 1 - q_1 \frac{w}{z} \right)^{-1}  :\eta_i(z) \eta_j(w): \quad j \equiv i+1 \\
\end{cases}
\end{eqnarray}
The \lq\lq triplet\rq\rq\ cancellation also holds in this case.
Thus, a non-trivial OPE of $\Phi_\lambda(v) \eta_\ell(z)$ appears
when the selection rule $i - \lambda_i \equiv \ell$ or $i - \lambda_i \equiv \ell+1$ is satisfied in each row.
The contribution of the $i$-th row is
\begin{align}
  \left( 1-\frac{z}{x_i v} \right) \quad \mathrm{for} \quad \lambda_i +\ell \equiv i, \quad
  \left( 1-\frac{q_3 z}{x_i v} \right)^{-1} \mathrm{for} \quad \lambda_i +\ell +1 \equiv i.
\end{align}
Hence, the product over the rows gives
\begin{align}
  &E_\ell(z) \Phi_\lambda(v) - ( \lambda \vert K_\ell^{-}(z) \vert \lambda ) \Phi_{\lambda}(v) E_\ell(z) \\
  &= \sum_{\substack{j=1, \\ j-\lambda_j \equiv \ell+1}}^{\ell(\lambda)+1}
  \delta\left( \frac{q_1 x_j v}{z} \right) (-\gq)
  \prod_{\substack{s=1, \\ s-\lambda_s \equiv \ell} }^{\ell(\lambda)}
  \psi\left( \frac{x_s}{q_1 x_j} \right) \left( 1-\frac{q_1 x_j}{x_s} \right) \nonumber \\
  &~~~\prod_{\substack{s=1, \\ s-\lambda_s \equiv \ell+1, \\ s\neq j}}^{\ell(\lambda)+1} \psi
  \left( \frac{q_2 x_s}{x_j} \right)^{-1} \left( 1-\frac{q_2^{-1}x_j}{x_s} \right)^{-1}
  :\Phi_{\lambda}(v) E_\ell(z): \\
  &= \sum_{\substack{j=1, \\ j-\lambda_j \equiv \ell+1}}^{\ell(\lambda)+1}
  \delta\left( \frac{q_1 x_j v}{z} \right) (-\gq)
  \prod_{\substack{s=1, \\ s-\lambda_s \equiv \ell} }^{\ell(\lambda)} \gq^{-1} \left( 1-\frac{x_j}{q_3 x_s} \right)
  \prod_{\substack{s=1, \\ s-\lambda_s \equiv \ell+1, \\ s\neq j}}^{\ell(\lambda)+1} \gq \left( 1-\frac{x_j}{x_s} \right)^{-1}
  :\Phi_{\lambda}(v) E_\ell(z):.
\end{align}
Note that the delta-function $\delta(q_1 x_j v/z)$ appears when $j - \lambda_j \equiv \ell+1$, that is, when we may add a box with color $\ell$ in the $j$-th row.
Then we move $e_{j,\lambda_j+1}$ to get\footnote{See Appendix B for the definition of $z_{\lambda}^{(j\pm)}(v)$.} $z_{\lambda+1_j}(v) = z_{\lambda}^{(j-)}(v) e_{j, \lambda_j+1}(v) z_{\lambda}^{(j+)}(v)$ by using Lemma 4 in Appendix B, a necessary technical result is worked out in Appendix B.
Taking the level dependence of the zero modes part into account,
we finally obtain
\begin{align}
  &E_\ell(z) \Phi_\lambda(v) - ( \lambda \vert K_\ell^{-}(z) \vert \lambda ) \Phi_{\lambda}(v) E_\ell(z) \CR
  =& \sum_{\substack{j=1, \\ j-\lambda_j \equiv \ell+1}}^{\ell(\lambda)+1}
  \left( - \gq q_1^{\lambda_j+1} \right)^{\deltabar_{\ell,j}}
  \prod_{\substack{s=1, \\ s-\lambda_s \equiv \ell} }^{j-1} \left(1-\frac{q_3 x_s}{x_j} \right)
   \prod_{\substack{s=1, \\ s-\lambda_s \equiv \ell+1}}^{j-1} \left(1-\frac{x_s}{x_j} \right)^{-1}  \CR
  &~~\times \prod_{\substack{s=j+1, \\ s-\lambda_s \equiv \ell} }^{\ell(\lambda)} \gq^{-1} \left(1-\frac{x_j}{q_3 x_s} \right)
  \prod_{\substack{s=j+1, \\ s-\lambda_s \equiv \ell+1}}^{\ell(\lambda)+1} \gq \left(1-\frac{x_j}{x_s} \right)^{-1}
  ~\delta\left(\frac{q_1 x_j v}{z}\right) \frac{t_{\lambda}}{t_{\lambda+1_j}} \frac{C_{\lambda+1_j} }{C_{\lambda}} \Phi_{\lambda+1_j}(v).
\end{align}


Now we employ the following combinatorial identity for the normalization factor
$C_\lambda$\footnote{To obtain trivial cancellations with this factor,
we have chosen the product order of $z_{\lambda}$.}
\begin{align}
  \frac{C_{\lambda + 1_j}}{C_\lambda} =
\frac{\displaystyle{\prod_{s=1}^{j-1}}\left( 1 - q_2 \frac{x_s}{x_j} \right)^{\deltabar_{s- \lambda_s, \ell+1}}}
{\displaystyle{\prod_{s=1}^{j-1}} \left( 1 - q_1^{-1} \frac{x_s}{x_j} \right)^{\deltabar_{s- \lambda_s, \ell}}}
\frac{\displaystyle{\prod_{s=j+1}^{\ell(\lambda)+1}} \left( 1 -  \frac{x_j}{x_s} \right)^{\deltabar_{s- \lambda_s, \ell+1}}}
{\displaystyle{\prod_{s=j+1}^{\ell(\lambda)}} \left( 1 - q_3^{-1} \frac{x_j}{x_s} \right)^{\deltabar_{s- \lambda_s, \ell}}}.
\label{lemma1}
\end{align}
See a related computation in the DIM case, Lemma 6.4 in \cite{AFS}.
In Appendix A, we prove \eqref{lemma1} which also appeared in section 7.2.1 of \cite{Nag2}.
Taking this into account, we arrive at
\begin{align}
  &E_\ell(z) \Phi_\lambda(v) - ( \lambda \vert K_\ell^{-}(z) \vert \lambda ) \Phi_{\lambda}(v) E_\ell(z) \CR
  &= \sum_{\substack{j=1, \\ j-\lambda_j \equiv \ell+1}}^{\ell(\lambda)+1}
  \left( - \gq q_1^{\lambda_j+1} \right)^{\deltabar_{\ell,j}}
  \prod_{\substack{s=1, \\ s-\lambda_s \equiv \ell} }^{\ell(\lambda)} \gq^{-1}
  \prod_{\substack{s=1, \\ s-\lambda_s \equiv \ell+1, \\ s \neq j}}^{\ell(\lambda)+1} \gq
  \prod_{\substack{s=1, \\ s-\lambda_s \equiv \ell} }^{j-1} \psi(\frac{x_s}{q_1 x_j})
  \prod_{\substack{s=1, \\ s-\lambda_s \equiv \ell+1}}^{j-1} \psi(\frac{x_j}{x_s})
  ~\delta \left(\frac{q_1 x_j v}{z} \right) \frac{t_{\lambda}}{t_{\lambda+1_j}} \Phi_{\lambda+1_j}(v).
\end{align}
By comparing with the intertwining relation \eqref{Eintertwin}, we obtain the recursion relation \eqref{t-recursion}
which gives the formula \eqref{tfactor}.


The OPE computation of $F_\ell(z)$ and $\Phi_\lambda(u)$ involves the commutation relation
of $E_i(z)$ and $F_j(w)$. But it can be performed similarly based on
\begin{eqnarray}
\xi_\ell(z) \eta_i(w)
&=& s_{\ell i}(z,w)^{-1} :\xi_\ell(z) \eta_i(w) :  \nonumber \\
&=&\begin{cases}
\left( 1 - \frac{\gd w}{z} \right) :\xi_\ell(z) \eta_i(w) : \quad i \equiv \ell +1 \\
\left( 1 - \frac{\gq w}{z} \right)^{-1} \left( 1 - \frac{w}{\gq z} \right)^{-1} :\xi_\ell(z) \eta_i(w) :  \quad i \equiv \ell \\
\left( 1 -  \frac{w}{\gd z} \right) :\xi_\ell(z) \eta_i(w) : \quad i \equiv \ell-1 \\
\end{cases}
\end{eqnarray}
for $|z| > |w|$.
We can deduce that
\begin{align}
  &F_\ell(z) \Phi_\lambda(v) - \Phi_\lambda(v) F_\ell(z)
  = (\gamma^+(z)-\gamma^-(z)) :F_\ell(z) \Phi_\lambda(v):, \\
  &\gamma(z)
  = \prod_{\substack{s=1, \\ s-\lambda_s \equiv \ell} }^{\ell(\lambda)} \left(  1-\frac{x_s u}{\gq z} \right)^{-1}
  \prod_{\substack{s=1, \\ s-\lambda_s \equiv \ell+1,}}^{\ell(\lambda)+1} \left( 1-\frac{x_s u}{\gq q_3 z} \right).
\end{align}
This time the delta-function $\delta\left( \frac{x_j v}{\gq z} \right)$ appears
when $j - \lambda_j \equiv \ell$, that is, when we may remove a box with color $\ell$ from the $j$-th row.
Using
\begin{equation}
: \xi_\ell (z) \eta_{j - \lambda_j} (x_j v) : \delta\left( \frac{x_j v}{\gq z} \right)
= \varphi_\ell^{+}(\gq z) \delta\left( \frac{x_j v}{\gq z} \right),
\end{equation}
we can check the intertwining relation with $F_\ell(z)$ by \eqref{lemma1}
with $\lambda_j \to \lambda_j -1$.

Finally, the dual intertwining relations can be demonstrated in the same way.


\subsection{Network matrix model and screening operator}

Network matrix model is a matrix model of the Dotsenko-Fateev type (conformal matrix model)\footnote{See  \cite{confMAMO1}-\cite{confMAMO6} for an original and generic issue of the conformal matrix models, and \cite{AGTmamo1}-\cite{AGTmamo8}, \cite{MMSh1,MMSh2,MMSh3} for AGT-related conformal matrix models.} whose
measure is determined by a trivalent planar diagram  (5 brane-web) representing a toric Calabi-Yau threefold
\cite{Mironov:2016cyq,Mironov:2016yue}. The correlation functions of the model are computed
as the (vacuum) expectation values or the traces of appropriate products of the intertwiners
glued together.
They reproduce refined topological
string amplitudes or five dimensional lift of the Nekrasov partition function for $\mathcal{N}=2$ quiver gauge theories.
Using the intertwiners constructed in this section, we can define a network matrix model
with $U_{\gq,\gd}(\widehat{\widehat{\mathfrak{gl}}}_n)$ symmetry.
There are two fundamental ways of gluing intertwiners (see Fig.3). The gluing along the horizontal line is simply
the successive action of operators on the horizontal Fock space. A particular example is
the product of $\Phi^\lambda(z)$ and the dual intertwiner $\Phi_\mu^*(w)$,
which we call $\mathcal T$ operator \cite{Awata:2016mxc}. The $\mathcal T$ operator satisfies the
$\mathcal{RTT}$ relation and plays an important role in deriving $(q,t)$-KZ equation,
since it realizes the $\gq$-shift operator (see the next section).
Note that such a product of the intertwiners along the horizontal line gives again an intertwining operator
which satisfies, for example, $a \Phi(z_1)\Phi(z_2) = \Phi(z_1)\Phi(z_2) ((1 \otimes \Delta) \circ \Delta) (a)$.
On the other hand, the gluing along the vertical line means taking the tensor product in the horizontal direction
with summation over the intermediate Young diagrams on the vertical line.
This gives the screening operator of the network matrix model \cite{Awata:2016riz};
\begin{equation}
\mathcal{S}(z) = \sum_\lambda \Phi^{*}_\lambda(z) \otimes \Phi^\lambda(z). \label{screening}
\end{equation}
In the DIM case, there should be the inverse of the square norm of the Macdonald function $|| M_\lambda||^{-2}$
as a weight in the summation over $\lambda$. However, we have changed the normalization of the dual intertwiner
$\Phi^{*}_\lambda$ and consequently there appears no weight factor in \eqref{screening}.

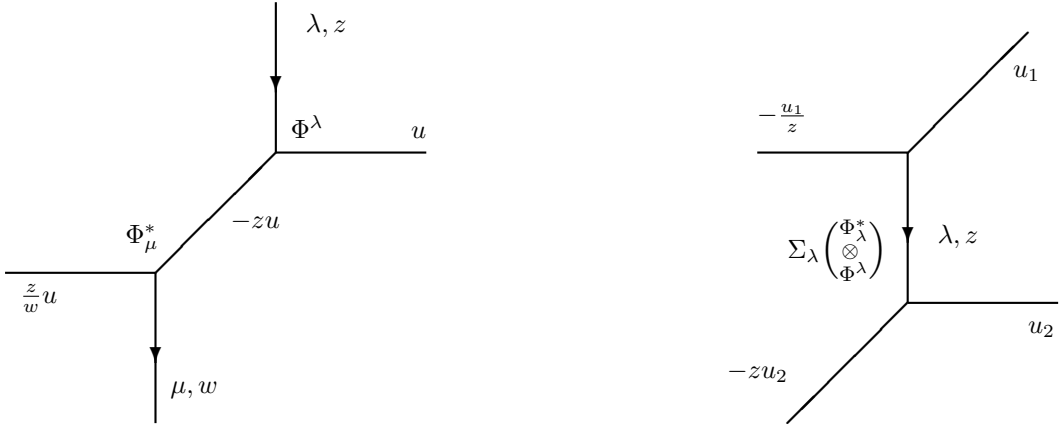
\begin{figure}[thb]
\unitlength 2mm
\begin{picture}(80,40)
\thicklines
\put(10,12){\vector(0,-1){6}}
\put(10,7){\line(0,-1){5}}
\put(0,12){\line(1,0){10}}
\put(10,12){\line(1,1){8}}
\put(18,20){\line(1,0){10}}
\put(18,30){\vector(0,-1){6}}
\put(18,25){\line(0,-1){5}}
\put(19,21){$\Phi^\lambda$}
\put(8,14){$\Phi^{*}_\mu$}
\put(20,28){$\lambda,z$}
\put(27,21){$u$}
\put(15,15){$-zu$}
\put(11,4){$\mu, w$}
\put(1,10){$\frac{z}{w} u$}
\put(0,35){$\mathcal{T}^\lambda_\mu(z,w)$}
\put(2,32){$= \Phi^{*}_\mu(w) \cdot \Phi^\lambda(z)$}
\put(60,20){\vector(0,-1){6}}
\put(60,15){\line(0,-1){5}}
\put(60,20){\line(1,1){8}}
\put(60,20){\line(-1,0){10}}
\put(60,10){\line(1,0){10}}
\put(60,10){\line(-1,-1){8}}
\put(52,13){$\Sigma_\lambda$\!\!
\large$\left(\substack{\Phi^{*}_\lambda\\\!\!\otimes\\\Phi^\lambda}\right)$}
\put(62,14){$\lambda,z$}
\put(67,25){$u_1$}
\put(68,8){$u_2$}
\put(50,22){$- \frac{u_1}{z}$}
\put(48,5){$- z u_2$}
\put(45,34){$\mathcal{S}(z) =
\sum_\lambda \Phi^{*}_\lambda(z) \otimes \Phi^\lambda(z)$}
\end{picture}
\caption{Gluing two intertwiners $\Phi$ and $\Phi^{*}$
gives the $\mathcal{T}$-operator (horizontal gluing) or
the screening operator (vertical gluing).}
\end{figure}


An important property of the screening operator is a commutativity
with $\Delta(X)$ for any element $X$ in $U_{\gq,\gd}(\widehat{\widehat{\mathfrak{gl}}}_n)$:
\begin{equation}
\left[ \Delta(X) , \mathcal{S}(z) \right] =0. \label{commutative}
\end{equation}
This relation gives constraints (Schwinger-Dyson equations) for the correlation functions
of the network matrix model. Since the coproduct $\Delta$ is a homomorphism of the algebra,
it is enough to check \eqref{commutative} for generating currents $E_\ell(w), F_\ell(w)$ and $K^\pm_\ell(w)$.
For $X=K^\pm_\ell(w)$, the commutativity is easily checked by using the definition of $\Delta$ and
the intertwining relations for $\Phi_\lambda(z)$ and $\Phi^{*}_\lambda(z)$.
For $X=E_\ell(w)$, we have
\begin{eqnarray}
\left[ \Delta(E_\ell(w)) , \mathcal{S}(z) \right]
&=& (K^{-}_\ell(\gq w) \otimes 1)
\left( \sum_\lambda \sum_{j=1}^{\ell(\lambda)+1} ( \lambda + 1_j \vert E_\ell(\gq w) \vert \lambda )
\Phi^{*}_\lambda (z) \otimes \Phi^{\lambda + 1_j} (z) \right.\CR
& &~~\left. - \sum_\lambda \sum_{j=1}^{\ell(\lambda)} ( \lambda\vert E_\ell(\gq w) \vert \lambda - 1_j )
\Phi^{*}_{\lambda - 1_j} (z) \otimes \Phi^\lambda(z)\right).
\end{eqnarray}
The right hand side vanishes inductively in the number of boxes $\vert \lambda \vert$ of the Young diagram.
A similar computation is valid for $F_\ell(w)$.


\subsection{Abelianization of the DIM intertwiner}
\label{sec:abel-nonab-intertw}

We would like to reexpress the intertwiner \eqref{intertwiner2} so that
it explicitly depends on the \emph{quotients} $\lambda^{(c)}$ and
\emph{shifts} $p_c$ of the vertical diagram $\lambda$. We will see
that the intertwiner factorizes into a product of commuting operators,
each depending on its own quotient $\lambda^{(c)}$ and shift
$p_c$. Thus the intertwiner for the \emph{non-Abelian} DIM algebra breaks
down into a product of intertwiners for the \emph{Abelian}\footnote{We
  abuse the terminology and call Abelian the DIM algebra associated with the double
  loops on the Abelian $\mathfrak{gl}_1 = \mathbb{C}$, though, of course,
  the DIM commutation relations are nontrivial. The non-Abelian DIM in this
  terminology is the deformation of the double loop algebra on
  $\mathfrak{gl}_N$.} DIM algebra.

To minimize technical steps in the derivation in this section, we limit
ourselves to the \emph{unrefined} non-Abelian DIM algebra. The unrefined
limit corresponds to setting $q_1 = q_3^{-1}$, or, equivalently, to $\gq=1$
with arbitrary $\gd$. Let us first write down the
expression \eqref{intertwiner2} for the intertwiner $\Phi_{\lambda}(v)$
using the colored characters:
\begin{multline}
  \label{eq:16}
  \Phi_{\lambda}(v) =
  \frac{\tilde{t}_{\lambda}(u,v)}{C_{\lambda}(q_1,q^{-1}_1)}\times\\
  \times :\exp \left\{ \sum_{r\geq 1} \frac{1}{r} \left[
      \sum_{c=0}^{N-1} \left( -\ch_{\lambda}^{(c)}(q_1^r) + \frac{c
          q_1^{r(c-N)}}{1 - q_1^{-rN}} - \frac{N q_1^{r(c-N)}}{(1 -
          q_1^{-rN})^2} \right)v^r H_{c,-r} - (r \leftrightarrow -r)
    \right] \right\}:~.
\end{multline}
Now we use the formula~(\ref{eq:12}) expressing the colored character
in terms of characters of the quotients:
\begin{multline}
  \label{eq:17}
  \Phi_{\lambda}(v) =
  \frac{\tilde{t}_{\lambda}(u,v)}{C_{\lambda}(q_1,q^{-1}_1)}\times\\
  \times :\exp \Bigg\{ \sum_{r\geq 1} \frac{1}{r} \Bigg[
      \sum_{c=0}^{N-1} \Bigg( -\sum_{d=0}^{N-1} L_{cd}(q_1^r) q_1^{r(N p_d
          - d)} \left(
        \ch_{\lambda^{(d)}}(q_1^{rN}) - \frac{1}{1 - q_1^{-rN}} \frac{1 -
          q_1^{-r N p_d}}{1- q_1^{rN}} \right) +\\
      + \frac{q_1^{r(N-c)}(c(1-q_1^{rN})+N)}{(1-q_1^{rN})^2}
  \Bigg) v^r
      H_{c,-r} - (r \leftrightarrow -r) \Bigg] \Bigg\}:~,
\end{multline}
where the matrix $L_{cd}(q)$ is from Eq.~\eqref{eq:15}. We introduce
modified Cartan generators $\tilde{H}_{c,r}$ which are given by the
following linear combinations of the original ones:
\begin{equation}
  \label{eq:18}
  \tilde{H}_{d,r} = \sum_{c=0}^{N-1} L_{cd}(q_1^{-r}) H_{c,r}.
\end{equation}
The modified Cartan generators~(\ref{eq:18}) satisfy very simple
commutation relations:
\begin{equation}
  \label{eq:19}
  [\tilde{H}_{d,r}, \tilde{H}_{f,s}] = \sum_{c=0}^{N-1}
  \sum_{e=0}^{N-1} L_{cd}(q_1^{-r}) L_{ef}(q_1^r) a_{ce} q_1^{-r m_{ce}} r
  \delta_{r+s,0} = (1-q_1^{rN})(1 - q_1^{-rN}) r \delta_{df} \delta_{r+s,0},
\end{equation}
where $a_{ij}$ and $m_{ij}$ are the Cartan and adjacency matrices
introduced in Eq.~(\ref{AM}), and we have used the crucial property
of $L^{-1}$:
\begin{equation}
  \label{eq:21}
  \left( (L^{-1}(q_1^{-1}))^{\mathrm{T}} L^{-1}(q_1)\right)_{ij} = \frac{a_{ij} q_1^{-m_{ij}}}{(1-q_1^N)(1-q_1^{-N})}.
\end{equation}
This property is easy to verify from the explicit
expression~(\ref{eq:14}). We therefore conclude that $\tilde{H}_{i,r}$
are $N$ \emph{independent} bosonic generators.

We notice a further simplification which occurs when we rewrite the
vacuum part of the intertwiner in terms of the new Cartan generators:
\begin{equation}
  \label{eq:22}
  \sum_{c=0}^{N-1} \frac{q_1^{r(N-c)}(c(1-q_1^{rN})+N)}{(1-q_1^{rN})^2}
  H_{c, -r} = \frac{1}{(1-q_1^N)(1-q_1^{-N})} \sum_{d=0}^{N-1} q_1^{-rd} \tilde{H}_{d, -r}.
\end{equation}
Plugging the identities~(\ref{eq:18}) and~(\ref{eq:22}) into the
intertwiner, we obtain
\begin{multline}
  \label{eq:23}
  \Phi_{\lambda}(v) =
  \frac{\tilde{t}_{\lambda}(u,v)}{C_{\lambda}(q_1,q^{-1}_1)}\times\\
  \times :\exp \left\{ \sum_{r\geq 1} \frac{1}{r} \left[
      \sum_{d=0}^{N-1} \Bigg( -
        \ch_{\lambda^{(d)}}(q_1^{rN}) + \frac{1}{(1 - q_1^{-r N}) (1 -
          q_1^{r N})}
  \Bigg) \left( q_1^{N p_d - d} v \right)^r
      \tilde{H}_{d,-r} - (r \leftrightarrow -r) \right] \right\}:.
\end{multline}
It is remarkable that, since $\tilde{H}_{r,d}$ for different $d$
commute, the intertwiner is a product of \emph{commuting} operators
each depending on its own \emph{quotient} diagram $\lambda^{(d)}$ and
the shift $p_d$, the latter entering only in the shift of the spectral
parameter. Moreover, upon closer examination each of the commuting
operators is nothing but the \emph{Abelian} DIM intertwiner!  Let us
denote the Abelian DIM intertwiner by $\Psi_{\lambda}(z)$ as
in~\cite{Awata:2016riz}:
\begin{equation}
  \label{eq:24}
  \Psi_{\lambda}^{(q_1)}(z,\tilde{H}_r) = \tilde{c}_{\lambda}(z) :\exp \left\{ \sum_{r\geq 1} \frac{1}{r} \left[
      \Bigg( -
      \ch_{\lambda}(q_1^r) + \frac{1}{(1 - q_1^{-r}) (1 -
        q_1^r)}
      \Bigg) z^r
      \tilde{H}_{-r} - (r \leftrightarrow -r) \right] \right\}:,
\end{equation}
where $\tilde{c}_{\lambda}(z)$ denotes the scalar prefactor, which we
omit in what follows, and we have explicitly written the
$q_1$-dependence of the intertwiner and also indicated that it acts in
the horizontal representation with the bosonic generators
$\tilde{H}_r$. The generators $\tilde{H}_r$ satisfy the commutation
relations
\begin{equation}
  \label{eq:26}
  [\tilde{H}_r, \tilde{H}_s] = (1-q_1^r) (1-q_1^{-r}) r \delta_{r+s,0}.
\end{equation}
Notice that here the normalization of the generators is nonstandard,
though the expression for the Abelian intertwiner~(\ref{eq:24}) is
correct. We can introduce a more convenient set of operators $a_r =
\frac{\tilde{H}_r}{1-q_1^r}$, for which we have
\begin{equation}
  \label{eq:27}
  [a_r , a_s] = r \delta_{r+s,0}.
\end{equation}
Let us also introduce modified zero modes $\tilde{\bar{\alpha}}_d$ and
$\tilde{H}_{d,0}$
\begin{gather}
  \label{eq:20}
  \bar{\alpha}_c = \sum_{d=0}^{N-1}
  (\delta_{c,d}-\delta_{(c-d)\, \mathrm{mod}\, N,1})
  \tilde{\bar{\alpha}}_d,\\
  H_{c,0} = \sum_{d=0}^{N-1}
  (\delta_{c,d}-\delta_{(c-d)\, \mathrm{mod}\, N,1})
  \tilde{H}_{d,0}
\end{gather}
which are independent (we assume that $\sum_{d=0}^{N-1}
\tilde{\bar{\alpha}}_d = \sum_{d=0}^{N-1} \tilde{H}_{d,0}=0$). Then the zero modes also factorize into a
product of independent factors:
\begin{equation}
  \label{eq:60}
  \prod_{(i,j)\in \lambda} e^{-\bar{\alpha}_{(i-j)\, \mathrm{mod}\, N}}
  (q_1^{j-i} v)^{H_{_{(i-j)\, \mathrm{mod}\, N},0}}= \prod_{d=0}^{N-1}
  e^{p_d \tilde{\bar{\alpha}}_{d}} \left( q_1^{-N|\lambda^{(d)}|-
      \frac{1}{2} N p_d(p_d +1) + d p_d + \sum_{f=0}^{N-1}\lfloor \frac{d-f}{N}
    \rfloor p_f } v^{-p_d} \right)^{\tilde{H}_{d,0}}
\end{equation}

Eventually, we get the key result
\begin{equation}
  \label{eq:25}
  \boxed{\Phi_{\lambda}^{(q_1)}(v) \sim \prod_{d=0}^{N-1}  e^{p_d \tilde{\bar{\alpha}}_{d}} \left( q_1^{-N|\lambda^{(d)}|-
      \frac{1}{2} N p_d(p_d +1) + d p_d + \sum_{f=0}^{N-1}\lfloor \frac{d-f}{N}
    \rfloor p_f } v^{-p_d} \right)^{\tilde{H}_{d,0}} \Psi_{\lambda^{(d)}}^{(q_1^N)}\left( q_1^{Np_d - d} v,
    \tilde{H}_{d,r} \right)}
\end{equation}
where we have omitted the scalar prefactors. Several remarks are in
order:
\begin{enumerate}
\item Each operator $\Psi$ acts on its own horizontal Fock space with the
  bosonic operators $\tilde{H}_{d,r}$, which are completely decoupled
  from each other. The vertical quotient diagrams $\lambda^{(d)}$ are
  also independent.

\item The normal ordering in the product in Eq.~(\ref{eq:25}) has been
  omitted, since each factor is already normal ordered, and the bosons
  $\tilde{H}_{d,r}$ commute for different $d$.

\item The Abelian intertwiners in the r.h.s.\ of Eq.~(\ref{eq:25}) have
  the equivariant parameter $q_1^N$, i.e.\ one can view the
  corresponding $\Omega$-background as a $N$-sheeted covering of the
  original one with parameter $q_1$.

\item The \emph{shifts} $p_d$ enter only as shifts of the spectral
  parameter $v$. Thus, the vertical legs on which the intertwiner acts
  do not coincide, but are shifted with respect to their center of
  mass position $v$ by $q^{N p_d - d}$. Notice also that the shifts
  are \emph{integer} powers of $q_1$.
\end{enumerate}

\begin{figure}[h]
  \centering
  \includegraphics[width=7cm]{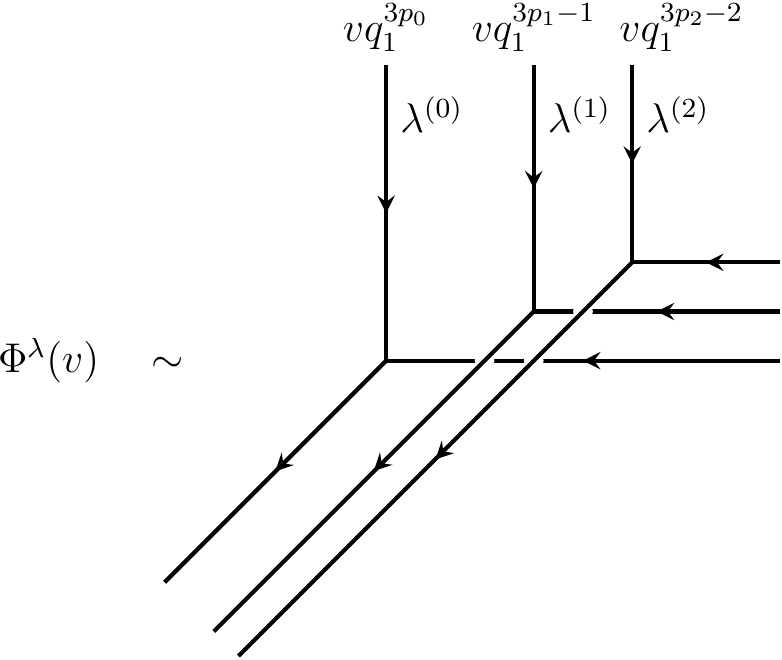}
  \caption{The intertwiner of Fock representations of
    $U_{\mathfrak{q}=1, \mathfrak{d}}
    (\widehat{\widehat{\mathfrak{gl}}}_3)$ drawn in terms of the
    intertwiners of Fock representations of $U_{\mathfrak{q}=1,
      \mathfrak{d}} (\widehat{\widehat{\mathfrak{gl}}}_1)$. The Young
    diagrams $\lambda^{(d)}$ on the vertical legs are the
    \emph{quotients} of the Young diagram $\lambda$, and the spectral
    parameters of the vertical legs depend on the \emph{shifts} $p_c$
    obtained from the quotient construction. More details on the
    quotients of Young diagrams are collected in
    Appendix~\ref{sec:from-colored-young}.}
  \label{fig:ab}
\end{figure}

Overall, since we have expressed the non-Abelian intertwiner as a
product of the Abelian ones, we can now draw a \emph{network matrix model}
picture for it, see Fig.~\ref{fig:ab}.
The non-Abelian intertwiner acts in the tensor product of $N$
horizontal Fock spaces and $N$ vertical Fock spaces. The latter have
the basis labelled by the $N$-tuple of Young diagrams
$\lambda^{(d)}$. The spectral parameters on the vertical legs are
obtained from the original one $v$ and the shifts $p_c$, while those
on the horizontal legs are encoded in the momenta (zero modes) of the
corresponding bosonic fields. The Fock spaces are intertwined pairwise
by the usual triple topological vertices so that, as a result, one gets
a tensor product of $N$ horizontal Fock spaces.

From the physical point of view, the phenomenon we observe in this
computation is that of symmetry enhancement. The Abelian intertwiner
corresponds to a triple junction of three Type IIB $(p,q)$-branes,
each of them being represented by a Fock space in the algebraic picture.
The DIM algebra plays the role of the ``worldvolume gauge symmetry''
of the brane. Since there is only one brane, the symmetry is
essentially Abelian, hence, represented by the \emph{Abelian} DIM algebra
$U_{q,t}(\widehat{\widehat{\mathfrak{gl}}}_1)$.

One can consider a triple junction of a \emph{stack} of $N$
$(p,q)$-branes. If the branes in the stack are far apart then on each
of them there is still an Abelian algebra acting. However, when we
move the branes closer together, the symmetry will be enhanced. The
natural candidate for the enhanced symmetry algebra is
$U_{\gq,\gd}(\widehat{\widehat{\mathfrak{gl}}}_N)$. \emph{A priori} it
is nontrivial that the triple junction of stacks of branes factorizes
into a product of non-interacting triple junctions. Our computation
shows that at least in the unrefined limit this is, indeed, the case:
the branes pass through each other and form the junctions just in the
way they used to when they were far apart. Perhaps
one can interpret this effect as conservation of certain protected
quantities.


\section{Level one KZ equation and Nekrasov function for ALE space}

Let us define the $\mathcal{T}$-operator \cite{Awata:2016mxc, Awata:2016bdm}
as a bilinear composition of the intertwiners
\begin{equation}
\mathcal{T}^\lambda_\mu (N, u \vert z,w) := \Phi^{*}_\mu(N +1 , -zu \vert w) \Phi^{\lambda}(N, u \vert z) :
\mathcal{F}^{(1,N)}_u \longrightarrow \mathcal{F}^{(1,N+1)}_{-zu}
\longrightarrow \mathcal{F}^{(1,N)}_{zu/w},
\end{equation}
where $(z, \lambda)$ and $(w,\mu)$ label the states in the incoming and the outgoing
vertical Fock space, respectively. In some of the computations below, it is necessary to
change the level $N$ and the spectral parameter $u$ of the horizontal Fock space.
We also introduce a function $\widetilde{G}_{\lambda\mu} (z)$ by the normal
ordering of the oscillator part\footnote{We keep the ordering of the zero mode part in
\eqref{tildeG}. The insertion of $\gq^{-1}$ is for later convenience.}
\begin{equation}
 \Phi^{*}_{\mu}(w) \Phi^{\lambda}(z) = \widetilde{G}_{\lambda\mu}(\gq^{-1} z/w) :  \Phi^{*}_{\mu}(w) \Phi^{\lambda}(z) :. \label{tildeG}
\end{equation}
As we will see below, in the construction of algebraic solutions to the $(q,t)$-KZ equation,
$\widetilde{G}_{\lambda\mu}(\gq^{-1} z/w)$ plays a role similar to the two point function
(the propagator) in the computation of correlation functions based on the Wick theorem for the free fields.
From the structure of the intertwiners
\begin{equation}
\Phi^\lambda (z) \sim : \prod_{(i,j) \in \lambda} \eta_{\bar c(i,j)} (q_1^{j-1} q_3^{i-1} z) \cdot \Phi_{\varnothing} (z) :,
\quad
\Phi_\lambda^{*} (w) \sim : \prod_{(i,j) \in \lambda} \xi_{\bar c(i,j)} (q_1^{j-1} q_3^{i-1} w)
\cdot \Phi_{\varnothing}^{*} (w) :,
\end{equation}
with the free field realization
\begin{equation}
\eta_i(z) = V_{i}^{(-)} ( \gq^{-\frac{1}{2}} z ) V_{i}^{(+)} (\gq^{\frac{1}{2}} z), \quad
\xi_i(z) = V_{i}^{(-)} ( \gq^{\frac{1}{2}} z )^{-1} V_{i}^{(+)} (\gq^{-\frac{1}{2}} z)^{-1},
\end{equation}
all the OPE relations (the two point functions) of the intertwiners are expressed in terms of
the single function $\widetilde{G}_{\lambda\mu} (z)$;
\begin{eqnarray}
 \Phi^{\mu}(w) \Phi^{\lambda}(z) &=& \widetilde{G}_{\lambda\mu}(\gq^{-2} z/w)^{-1} : \Phi^{\mu}(w) \Phi^{\lambda}(z): ,\\
 \Phi^{\mu}(w) \Phi^{*}_{\lambda}(z) &=& \widetilde{G}_{\lambda\mu}(\gq^{-1} z/w) :  \Phi^{\mu}(w) \Phi^{*}_{\lambda}(z) :, \\
 \Phi^{*}_{\mu}(w) \Phi^{*}_{\lambda}(z) &=& \widetilde{G}_{\lambda\mu}( z/w)^{-1} : \Phi^{*}_{\mu}(w) \Phi^{*}_{\lambda}(z) : .
\end{eqnarray}
Recall that the vacuum components are given by
\begin{equation}
\Phi_{\varnothing }(z) = \widetilde{V}_k^{(-)}(\gq^{\frac{1}{2}} z)~\widetilde{V}_k^{(+)}(\gq^{\frac{3}{2}} z)^{-1},
\qquad
  \Phi_{\varnothing}^{*}(w)
  =  \widetilde{V}_k^{(-)}(\gq^{\frac{3}{2}} w)^{-1} ~\widetilde{V}_k^{(+)}(\gq^{\frac{1}{2}} w)
\end{equation}
with the commutation relation
\begin{equation}
\left[ \Lambda_{i,r}, \Lambda_{j,s} \right]
= - \delta_{r+s,0} \frac{ b_{ij}^{[r]}}{r}.
\end{equation}
Since the diagonal component of inverse of the deformed Cartan matrix is
\begin{equation}
b_{ii} (\gq, \gd)=
\frac{[n]}{(q_1^{\frac{n}{2}} - q_1^{-\frac{n}{2}})  (q_3^{\frac{n}{2}} - q_3^{-\frac{n}{2}})},
\end{equation}
we find
\begin{equation}
\widetilde{G}_{\varnothing \varnothing}(z) =
\exp \left( \sum_{r=1}^\infty \frac{1}{r}
\frac{[nr]}{[r]}
\frac{\gq^r z^r}{(q_1^{\frac{nr}{2}} - q_1^{-\frac{nr}{2}})  (q_3^{\frac{nr}{2}} - q_3^{-\frac{nr}{2}})} \right).
\end{equation}
When $n \to 1$, $\widetilde{G}_{\varnothing \varnothing}(z)$ simplifies to
\begin{equation}
\widetilde{G}_{\varnothing \varnothing}(z) \to
\exp \left( \sum_{r=1}^\infty \frac{1}{r}
\frac{z^r}{(1- q_1^{r})  (1- q_3^{r})} \right).
\end{equation}
In topological string theory, $\widetilde{G}_{\varnothing \varnothing}(z)$ gives
the amplitude of the conifold and is the origin of the \lq\lq anomalous\rq\rq\ factor
in the $\mathcal{RTT}$ relation \cite{Awata:2016mxc}.
In the following, we renormalize the function $\widetilde{G}_{\lambda\mu}$ by
\begin{equation}
G_{\lambda\mu}(z) = \widetilde{G}_{\lambda\mu}(z) / \widetilde{G}_{\varnothing\varnothing}(z),
\end{equation}
so that $G_{\varnothing\varnothing}(z)=1$.


\subsection{Shift operator and ${\cal R}$-matrix}

From the combinations of the $\gq$-shift in the vertex operators, we can see that
when the ratio of the incoming and the outgoing spectral parameters is
$\gq^{\pm 1}$, the diagonal components of the $\mathcal{T}$-operator have
no positive or negative modes. Namely, if we define
\begin{eqnarray}
T^{+}_\lambda (N,u \vert z) := \mathcal{T}^\lambda_\lambda (N, u \vert \gq z, z) &~:~&
{\cal F}^{(1,N)}_u \longrightarrow {\cal F}^{(1,N)}_{\gq u}, \\
T^{-}_\lambda (N, u \vert z) := \mathcal{T}^\lambda_\lambda (N, u \vert z, \gq z) &~:~&
{\cal F}^{(1,N)}_u \longrightarrow {\cal F}^{(1,N)}_{\gq^{-1} u},
\end{eqnarray}
these operators satisfy
\begin{equation}
T^{+}_\lambda (N,u \vert z) \vert \varnothing \rangle =  c_\lambda^{+} \vert \varnothing \rangle,
\qquad
\langle \varnothing \vert T^{-}_\lambda (N,u \vert z) =  c_\lambda^{-} \langle \varnothing \vert, \label{prefactor}
\end{equation}
with
\begin{equation}
c_\lambda^{\pm} = \frac{t_\lambda t_\lambda^* \gq^{\mp(N+1) |\lambda|_0}}
{C_\lambda C'_\lambda f_\lambda} \widetilde{G}_{\lambda\lambda}(\gq^{-1\pm1}).
\end{equation}
Actually $T^{\pm}_\lambda (N,u \vert z)$ is independent of the horizontal spectral parameter $u$,
and the dependence on the level $N$ is simply
\begin{equation}
T^{\pm}_\lambda (N+1,u \vert z) = \gq^{\mp |\lambda|_0} T^{\pm}_\lambda (N,u \vert z).
\end{equation}
A crucial point in deriving the KZ equation is that
the $\gq^{-2}$-shift of the intertwining operators is realized as the action of
$T^{\pm}_\lambda (z)$ and their inverses as follows:
\begin{eqnarray}
&&\Phi^\lambda (N, \gq u \vert \gq^{-2} z)  \nonumber \\
&=& \gq^{-2|\lambda|_0}
\frac{\widetilde{G}_{\lambda\lambda}(1)}{\widetilde{G}_{\lambda\lambda}(\gq^{-2})}
T^{-}_\lambda (N+1,  -uz \vert \gq^{-2} z) \Phi^\lambda (N,u \vert z)
T^{+}_\lambda (N,  u \vert \gq^{-1} z)^{-1},   \label{shift1}\\
&& \Phi^{*}_\lambda (N+1, \gq^{-1} u \vert \gq^{-2} z) \nonumber \\
&=&\gq^{2|\lambda|_0}
\frac{\widetilde{G}_{\lambda\lambda}(\gq^{-2})}{\widetilde{G}_{\lambda\lambda}(1)}
T^{-}_\lambda (N, -\frac{\gq u}{z} \vert \gq^{-1} z)^{-1} \Phi^{*}_\lambda (N+1,  u \vert z)
T^{+}_\lambda (N+1, \gq^{-1} u \vert \gq^{-2} z),
 \label{shift2}
\end{eqnarray}
where we have used
\begin{equation}
\Phi^\lambda (N, \gq u \vert z) = \gq^{|\lambda|_0} \Phi^\lambda(N, u \vert z),
\qquad
\Phi_\lambda^* (N+1, \gq u \vert z) = \gq^{-|\lambda|_0} \Phi_\lambda^* (N+1, u \vert z).
\end{equation}
Due to relation \eqref{Ginv} to be discussed below, the prefactor can be simplified to
\begin{equation}
\left(\frac{\widetilde{G}_{\varnothing\varnothing}(1)}
{\widetilde{G}_{\varnothing\varnothing}(\gq^{-2})}\right)^{\pm 1}
= \exp \left( \pm \sum_{r=1}^\infty \frac{1}{r} \frac{ (q_2^{\frac{nr}{2}} - q_2^{-\frac{nr}{2}})}
{(q_1^{\frac{nr}{2}} - q_1^{-\frac{nr}{2}})  (q_3^{\frac{nr}{2}} - q_3^{-\frac{nr}{2}})} \right).
\end{equation}
It is instructive to count the power of $\gq$ on both sides of these relations.
We can see the factor $\gq^{\pm(1+ 2N) |\lambda|_0}$ (the positive sign for \eqref{shift1} and the negative sign for \eqref{shift2}).
Since $\gq = (t/q)^{1/2}$, the shift parameter is the same as in the $\mathfrak{gl}_1$ case \cite{Awata:2017cnz}.
It should be noticed that the $\gq^{-2}$-shift of the vertical spectral parameter is accompanied by
a shift of the horizontal parameter $\gq^{\pm 1} u$,  which is consistent with the fact that
$T^{\pm}_\lambda$ and $(T^{\mp}_\lambda)^{-1}$ shift the horizontal spectral parameter by $\gq^{\pm 1}$,
while keeping the level $N$.


In order to derive the KZ equation based on  \eqref{shift1} and \eqref{shift2}, we have to use
the commutation relations between the intertwiners and $T^{\pm}_\lambda (N,u \vert z)$ which
follow from those among the intertwiners.
Let us begin with the commutation relations of $\Phi^\lambda(z)$ and $\Phi^{*}_\mu(w)$.
We want to require them to commute up to the anomalous factor $\widetilde{G}_{\varnothing \varnothing}(z)$
so that we have a simple algebra of two copies of the Zamolodchikov algebra satisfied separately by
$\Phi^\lambda(z)$ and $\Phi^{*}_\mu(w)$. In the computation of the commutation relation,
it is important that the exchange of $\Phi^\lambda(z)$ and $\Phi^{*}_\mu(w)$ changes the level and the spectral parameter of the horizontal representation.
It also involves the exchange of the zero mode factors $z^\lambda$ and $z_\mu^{*}$,
which are the group algebra parts \eqref{ordering}, \eqref{dualordering} of the (dual) intertwiner.
We find that the above requirement is satisfied if and only if\footnote{
As we will see in the next subsection, $G_{\lambda\mu}(z)$ agrees with the Nekrasov factor $N_{\lambda\mu}(z)$
on $ALE_n \times S^1$ \eqref{ALEG}.
Thus the formula \eqref{Ginv} is a generalization of the usual symmetry of the Nekrasov factor incorporating
the contribution of zero modes.}
\begin{equation}
  G_{\lambda\mu} (\gq^{-1} u/v) z^*_{\mu}(v) z_{\lambda}(u)
  = \left( \frac{u}{v} \right)^{|\lambda|_0 + |\mu|_0} \frac{f_\lambda(q_1,q_3)}{f_\mu(q_1,q_3)}
  G_{\mu\lambda} (\gq^{-1} (u/v)^{-1}) z_{\lambda}(u) z^*_{\mu}(v),
\label{Ginv}
\end{equation}
where $f_\lambda(q_1,q_3)$ is the generalized framing factor \eqref{framing},
and $|\lambda|_0$ denotes the number of boxes with color $0$ in $\lambda$.
In Appendix D, we prove \eqref{Ginv}.

Using the relation \eqref{Ginv}, we can write down the commutation relations
of the intertwiners as follows\footnote{See also the computations in Appendix D.};
\begin{eqnarray}
\Phi_\mu^{*} (w) \Phi^\lambda(z) &=& \Upsilon^{(0)} (z/w)  \Phi^\lambda(z) \Phi_\mu^{*} (w), \\
\Phi^\mu(w)  \Phi^\lambda(z) &=&  \Upsilon^{(+)} (z/w) \mathcal{R}_{\lambda\mu}(z/w)
\Phi^\lambda(w) \Phi^\mu(z), \\
\Phi_\mu^{*} (w) \Phi_\lambda^{*}(z) &=&  \Upsilon^{(-)} (z/w)  \mathcal{R}_{\lambda\mu}^{-1}(z/w)
 \Phi_\lambda^{*}(w) \Phi_\mu^{*} (z),
\end{eqnarray}
where we have introduced the $\mathcal R$-matrix defined by
\begin{equation}
\mathcal{R}_{\lambda\mu} (z)
  = \gq^{-H(\lambda,\mu)} \frac{G_{\lambda\mu}(z)}{G_{\lambda\mu}(\gq^{-2} z)},
\label{Rdef}
\end{equation}
where
\begin{equation}
  H(\lambda,\mu) = \#\{ s\in\lambda|h_{\mu,\lambda}(s)\equiv0\} + \#\{ t\in\mu|h_{\lambda,\mu}(t)\equiv0\}.
\end{equation}
When there are no constraints on the relative hook length, $H(\lambda,\mu)= |\lambda| + |\mu|$, and
\eqref{Rdef} reduces to the definition for the $\mathfrak{gl}_1$ case \cite{Awata:2016mxc, Awata:2016bdm}.
$\Upsilon^{(0, \pm)}(z)$ stands for the anomalous factor
\begin{equation}
\Upsilon^{(+)}(z) = \frac{\widetilde{G}_{\varnothing\varnothing}(\gq^{-2} z^{-1})}
{\widetilde{G}_{\varnothing\varnothing}(\gq^{-2} z)},
\quad
\Upsilon^{(0)}(z) = \frac{\widetilde{G}_{\varnothing\varnothing}(\gq^{-1} z)}
{\widetilde{G}_{\varnothing\varnothing}(\gq^{-1} z^{-1})},
\quad
\Upsilon^{(-)}(z) = \frac{\widetilde{G}_{\varnothing\varnothing}(z^{-1})}
{\widetilde{G}_{\varnothing\varnothing}(z)},
\end{equation}
which satisfies $\Upsilon^{(0, \pm)}(z) \Upsilon^{(0, \pm)}(z^{-1}) =1$.
Note that the relation \eqref{Ginv} implies
\begin{equation}
\mathcal{R}_{\lambda\mu} (z)  \mathcal{R}_{\mu\lambda} (z^{-1}) =1. \label{unitary}
\end{equation}
The definition of the $\mathcal{R}$-matrix \eqref{Rdef} is justified by the fact that we can derive
the following ${\cal RTT}$ relation by a computation similar to the $\mathfrak{gl}_1$ case \cite{Awata:2016mxc, Awata:2016bdm},
\begin{equation}
\mathcal{R}_{\lambda\mu} (z_1/z_2) \mathcal{T}^\lambda_\nu (z_1,w_1) \mathcal{T}^\mu_\rho (z_2,w_2)
= \mathcal{T}^\mu_\rho (z_2,w_2) \mathcal{T}^\lambda_\nu (z_1,w_1)
\mathcal{R}_{\nu\rho} (w_1/w_2),
\end{equation}
up to the anomalous factor from the vacuum contribution.


\subsection{Relation to $K$-theoretic Nekrasov function for ALE space}

Difference of the Nekrasov functions for the flat space and the ALE space
is in the selection rule for the boxes of the Young diagram. The selection rule
is a consequence of taking the invariant part of the character
under the orbifold action of $\mathbb{Z}_{n+1}$ on $\mathbb{C}^2$.
To define the selection rule, we introduce the relative hook length
\begin{equation}
h_{\lambda, \mu}(s) = a_\mu (s) + \ell_\lambda (s) +1.
\end{equation}
Then the building block (the bifundamental matter contribution) of
the five-dimensional Nekrasov function for instanton counting on
$ALE_n \times S^1$ is given by \cite{Fucito:2004ry, Fujii:2005dk}
\begin{equation}
N_{\lambda \mu} (u; q_1, q_3)
= \prod_{\substack{s \in \lambda \\
h_{\mu, \lambda}(s) \equiv 0}} ( 1- u q_1^{a_\lambda(s)} q_3^{- \ell_\mu(s) -1})
 \prod_{\substack{t \in \mu \\ h_{\lambda, \mu}(t) \equiv 0}} ( 1- u q_1^{-a_\mu(t)-1} q_3^{\ell_\lambda(t)}).
 \label{ALEG}
\end{equation}
The following specialization of $N_{\lambda \mu} (u; q_1, q_3)$
 is related to the normalization factor of the intertwiners:
\begin{equation}
N_{\lambda\lambda}(1; q_1, q_3) = C_\lambda(q_1, q_3) C^\prime_\lambda(q_1^{-1}, q_3^{-1}).
\end{equation}
In \cite{TU, Uglov:1997ia}, the $\mathfrak{gl}_n$ version of the Jack polynomials (Uglov polynomials)
is obtained by taking the roots of unity limit of the Macdonald polynomials.
The Uglov polynomials play an important role in the four-dimensional (Yangian) version of
AGT correspondence for the Nekrasov partition function on the ALE space
\cite{Belavin:2011pp}-\cite{Itoyama:2014pca}. Since our current problem should be related to
a five-dimensional uplift of this story, we expect the normalization factor of the intertwiner is
closely related to the norm of an uplift of the Uglov polynomials as a generalization of the uplift of the Jack polynomials
to the Macdonald polynomials. Though such an uplift is not available at the moment,
we can guess the normalization factor from that of the Uglov polynomials given in \cite{Uglov:1997ia}.
\begin{equation}
( \lambda \vert \lambda )_{q_1, q_3}= \frac{C^\prime_\lambda (q_1, q_3)}{C_\lambda(q_1, q_3)}.
\end{equation}
See also Lemma 2 in Appendix A. At the CFT side, the uplift might be related to $q$-deformed $\mathcal{W}$-coset models.
We can expect the same level-rank duality as in the undeformed case, since the character is invariant under the $q$-deformation.
It is interesting to see how the level-rank duality is realized in the setting of quantum toroidal algebras.

Now we argue that the renormalized two-point function $G_{\lambda\mu}(z)$ of
the intertwiners is nothing but the bifundamental matter contribution \eqref{ALEG} on the ALE space.
By a direct computation of the OPE factors between $\Phi_\varnothing^{*}(v)$ and $\Phi^{\lambda}(u)$,
we obtain
\begin{eqnarray}
G_{ \lambda \varnothing}(\gq^{-1} \frac{z}{w})  &=& \prod_{\substack{(i,j) \in \lambda \\  i-j \equiv 0}}
( 1 - \gq^{-1} q_1^{j-1} q_3^{i-1} \frac{z}{w}) \nonumber \\
&=& \prod_{\substack{s \in \lambda \\
a_\lambda(s) + \ell_\varnothing(s) +1 \equiv 0}}
( 1 - \gq^{-1} q_1^{a_\lambda(s)} q_3^{-\ell_\varnothing(s)-1} \frac{z}{w})   
= N_{\lambda \varnothing} (\gq^{-1} \frac{z}{w}; q_1, q_3),
\end{eqnarray}
where, in the second equality, we convert the summation over the co-arm length
$j -1$ in each row ($1 \leq j \leq \lambda_i$) to that over the arm length $\lambda_i - j$
and use $\ell_\varnothing(i,j) =-i$.

In general, from the normal ordering
\begin{equation}
  \xi_\ell(q_1^{\mu_j} q_3^{j-1}w) \Phi^\lambda(z)
  = \frac{\displaystyle{\prod_{\substack{i=1, \\ i-\lambda_i \equiv \ell+1}}^{\ell(\lambda)+1}}
  (1-\gq^{-1} q_1^{\lambda_i-\mu_j-1} q_3^{i-j-1} z/w) }
{\displaystyle{\prod_{\substack{i=1, \\ i-\lambda_i \equiv \ell}}^{\ell(\lambda)}}
 (1-\gq^{-1} q_1^{\lambda_i-\mu_j-1} q_3^{i-j} z/w)}
   : \xi_\ell(q_1^{\mu_j} q_3^{j-1}w) \Phi^\lambda(z) :
\end{equation}
for $\ell \equiv j-\mu_j-1$, we obtain a recursion relation
for $G_{\lambda\mu}(u)$ with respect to the second diagram $\mu$,
\begin{equation}
  G_{\lambda\mu+1_j}(\gq^{-1}z/w)
  = \frac{\displaystyle{\prod_{\substack{i=1, \\ i-\lambda_i \equiv \ell+1}}^{\ell(\lambda)+1}}
   (1-\gq^{-1} q_1^{\lambda_i-\mu_j-1} q_3^{i-j-1} z/w)}
   {\displaystyle{\prod_{\substack{i=1, \\ i-\lambda_i \equiv \ell}}^{\ell(\lambda)}}
  (1-\gq^{-1} q_1^{\lambda_i-\mu_j-1} q_3^{i-j} z/w)} G_{\lambda\mu}(\gq^{-1}z/w).
\end{equation}
In Appendix C, we prove that $N_{\lambda \mu} (u; q_1, q_3)$ satisfies exactly the same recursion relation.
Thus, we have
\begin{equation}
G_{ \lambda \mu}(u) = N_{\lambda \mu} (u; q_1, q_3).
\end{equation}


\subsection{Level one KZ equation for $U_{\gq,\gd}(\widehat{\widehat{\mathfrak{gl}}}_n)$}

From the commutation relations between the intertwiners, we obtain
the following commutation relations of the $\mathcal{T}$-operator
and the intertwiners:
\begin{eqnarray}
\mathcal{T}_\mu^\lambda(z,w) \Phi^\kappa (z') &=& \Upsilon^{(0)} (z'/w) \Upsilon^{(+)} (z'/z)
\mathcal{R}_{\kappa\lambda}(z'/z) \Phi^\kappa (z') \mathcal{T}_\mu^\lambda(z,w), \\
\mathcal{T}_\mu^\lambda(z,w) \Phi^{*}_\nu (w') &=&  \Upsilon^{(0)} (w'/z) \Upsilon^{(-)} (w'/w)
\mathcal{R}_{\nu\mu}(w'/w)^{-1} \Phi^{*}_\nu (w') \mathcal{T}_\mu^\lambda(z,w).
\end{eqnarray}
It is convenient to introduce a universal function
that is the $\gq^{-2}$-difference of the vacuum anomalous factor
\begin{equation}
\overline{\Upsilon} (z) := \frac{\widetilde{G}_{\varnothing\varnothing}(z)}
{\widetilde{G}_{\varnothing\varnothing}(\gq^{-2} z)}
\end{equation}
and define a renormalized $\mathcal R$-matrix by
\begin{equation}
\overline{\mathcal{R}}_{\lambda\mu}(z) := \overline{\Upsilon} (z) \mathcal{R}_{\lambda\mu}(z)
= \gq^{- H(\lambda,\mu)} \frac{\widetilde{G}_{\lambda\mu}(z)}
{\widetilde{G}_{\lambda\mu}(\gq^{-2} z)}.
\end{equation}
It is amusing that a similar decomposition takes place for the $\mathcal R$-matrix for
the tensor product of evaluation representations of the quantum affine algebra \cite{EFK}.
Here the vertical Fock representation of the quantum toroidal algebra plays the role
of the evaluation representation \cite{eval}.
With the renormalized $\mathcal R$-matrix, the commutation relations between
the shift operator $T_\lambda^{\pm}(z)$ and the intertwiners take the following simple form:
\begin{eqnarray}
T_\lambda^{+}(N+1, -wu \vert z) \Phi^\mu (N,  u \vert w) &=&
\overline{\mathcal{R}}_{\mu\lambda}(w/\gq z)  \Phi^\mu (N, \gq u \vert w) T_\lambda^{+}(N,  u \vert z), \label{Tcom1}\\
 \Phi^\mu (N, \gq^{-1} u \vert w) T_\lambda^{-}(N,  u \vert z)&=&
 \overline{\mathcal{R}}_{\lambda\mu}(z/w)  T_\lambda^{-}(N+1, -wu \vert z) \Phi^\mu (N,  u \vert w) , \label{Tcom2}\\
T_\lambda^{+}(N, - \frac{u}{w} \vert z) \Phi^{*}_\mu (N+1, u \vert  w) &=&
\overline{\mathcal{R}}_{\mu\lambda}(w/z)^{-1} \Phi^{*}_\mu (N+1, \gq u \vert w) T_\lambda^{+}(N+1, u \vert  z), \label{Tcom3}\\
 \Phi^{*}_\mu (N+1, \gq^{-1} u \vert w) T_\lambda^{-}(N+1, u \vert z) &=&
 \overline{\mathcal{R}}_{\lambda\mu}(\gq z/w)^{-1}  T_\lambda^{-}(N, - \frac{u}{w} \vert z) \Phi^\mu (N+1, u \vert  w).\label{Tcom4}
\end{eqnarray}
After commuting with $T^{\pm}$-operators, there is the $\gq^{\pm 1}$-shift of the horizontal parameter of the intertwiners,
while the level of $T^{\pm}$ itself changes.
Let us derive an ($(q,t)$-KZ) equation for the correlation function of the intertwiners
\begin{equation}
\mathcal{G}^{(n,m)} (v \vert \vec{z}, \vec{\lambda} ; \vec{w},  \vec{\mu})
= \langle \varnothing \vert \Phi^{*}_{\mu_1}(w_1)  \cdots \Phi^{*}_{\mu_m}(w_m)
\Phi^{\lambda_1}(z_1)  \cdots \Phi^{\lambda_n}(z_n) \vert \varnothing \rangle,
\end{equation}
where $v$ is the incoming (rightmost) spectral parameter of the horizontal representations.
Without loss of generality, we can assume that the right vacuum belongs to the level $(1,0)$ representation.
Then the left vacuum belongs to the level $(1, n-m)$ representation.
The other horizontal spectral parameters are determined by the conservation law for the existence of
the intertwining operator. The difference operator $\gq^{-2 x \partial_x}$ acting on each intertwiner
produces an insertion of $T_\lambda^{\pm}(z)$ according to \eqref{shift1} and \eqref{shift2}.
Using the commutation relations \eqref{Tcom1} -- \eqref{Tcom4}, we can then move $T_\lambda^{-}$
to the left and $T_\lambda^{+}$ to the right. Finally acting on the (dual) vacuum, $T_\lambda^{\pm}(z)$
produces the prefactors $c_\lambda^{\pm}$ given by \eqref{prefactor}
which cancel $\widetilde{G}$ function in \eqref{shift1} and \eqref{shift2}.
The remaining factor comes only from the level dependent part of $c_\lambda^{\pm}$
and from the assumption on the level of vacua $\vert \varnothing \rangle$ and $\langle \varnothing \vert$
we obtain the factor $\gq^{\pm (n-m) |\lambda_k |_0}$.
In this way we can write down the $(q,t)$-KZ equation;
\begin{align}
&\gq^{-2 z_k \partial_{z_k} + v \partial_{v} } \cdot \mathcal{G}^{(n,m)} (v \vert \vec{z}, \vec{\lambda} ; \vec{w},  \vec{\mu}) \nonumber \\
&~~= \gq^{(n-m) |\lambda_k |_0} \prod_{\ell=1}^m \overline{\mathcal{R}}_{\lambda_k \mu_\ell} (z_k/ \gq w_\ell)^{-1}
\prod_{i < k} \overline{\mathcal{R}}_{\lambda_k \lambda_i}  (z_k/\gq^2 z_i)
\prod_{k < j} \overline{\mathcal{R}}_{\lambda_j \lambda_k} (z_j/ z_k)^{-1}
~\mathcal{G}^{(n,m)}(v), \label{KZ1}
\end{align}
and
\begin{align}
&\gq^{-2 w_k \partial_{w_k}  - v \partial_{v}} \cdot \mathcal{G}^{(n,m)} (v \vert \vec{z}, \vec{\lambda} ; \vec{w},  \vec{\mu}) \nonumber \\
&~~= \gq^{(m-n) |\mu_k |_0}\prod_{i < k} \overline{\mathcal{R}}_{\mu_k \mu_i}  (w_k/w_i)
\prod_{k < j} \overline{\mathcal{R}}_{\mu_j \mu_k} (\gq^2 w_j/ w_k)^{-1}
\prod_{\ell=1}^n \overline{\mathcal{R}}_{\lambda_\ell \mu_k} (\gq z_\ell/w_k)
 ~\mathcal{G}^{(n,m)}(v). \label{KZ2}
\end{align}
The additional operator $\gq^{\pm v \partial_{v}}$  accounts for the shift of the horizontal parameter
in \eqref{shift1} and \eqref{shift2}. Note that since all the horizontal parameters are proportional
to the initial parameter $v$, all of them are shifted by $\gq^{\pm1}$.
After commuting the intertwiners with the operators $T^{\pm}_{\lambda_k}$ or $T^{\pm}_{\mu_k}$,
the shifted horizontal parameters get back to the original values.


\subsection{Nekrasov function as algebraic solutions to KZ equation}

The Nekrasov function for the $U(1)$ gauge theory on the ALE space satisfies
the $(q,t)$-KZ equation derived in the last subsection.
This solution does not require any screening operators and, hence, there are no integrations
associated with the screening operators. In this sense, the Nekrasov function for the $U(1)$ gauge theory
gives an algebraic solution to the $(q,t)$-KZ equation. For the $U(N)$ gauge theory, one could need
screening operators, and we have to glue $N$ building blocks, with each block being
an appropriate Nekrasov functions for the $U(1)$ theory.

To construct algebraic solutions to the $(q,t)$-KZ equation, let us introduce
the \lq\lq modified\rq\rq\ two point function
\begin{equation}
F_{\lambda\mu} (z) := z^{- \frac{1}{2} (H(\lambda, \mu) + |\mu|_0 - |\lambda|_0) }
\cdot \widetilde{G}_{\lambda\mu}(z),
\end{equation}
which satisfies a fundamental difference equation
\begin{equation}
F_{\lambda\mu} (\gq^{-2} z) = \gq^{|\mu|_0 - |\lambda|_0} \overline{\mathcal{R}}_{\lambda\mu}^{-1} (z) F_{\lambda\mu} (z).
\end{equation}
Then one can check
\begin{equation}
\mathcal{G}^{(n,m)} (v \vert \vec{z}, \vec{\lambda} ; \vec{w},  \vec{\mu})
= v^{\vert\vec{\lambda}\vert_0 - \vert\vec{\mu}\vert_0}~
\frac{\displaystyle{\prod_{i=1}^n  \prod_{j=1}^m}  F_{\lambda_i \mu_j} \left(\gq^{-1} \frac{z_i}{w_j} \right) }
{\displaystyle{\prod_{1 \leq i< j \leq n}} F_{\lambda_j \lambda_i} \left( \gq^{-2} \frac{z_j}{z_i} \right)
\displaystyle{\prod_{1 \leq k< \ell \leq m}} F_{\mu_\ell\mu_k} \left( \frac{w_\ell}{w_k} \right)}
\end{equation}
satisfies both the $(q,t)$-KZ equations \eqref{KZ1} and \eqref{KZ2}.


\section{Modular and periodic properties of double elliptic systems
  from $U_{q,t}(\widehat{\widehat{\mathfrak{gl}}}_1)$ network matrix
  model}
\label{sec:modul-prop-double}

Let us use solutions to the elliptic KZ equations for
$U_{q,t}(\widehat{\widehat{\mathfrak{gl}}}_1)$ obtained
in~\cite{Awata:2017cnz} to deduce the properties of the $6d$ $U(N)$ gauge
theories with adjoint hypermultiplet of mass $m$ compactified on a torus
$T^2$. These systems are described by the double elliptic integrable
systems \cite{Delli1,Delli2,Delli3} and possess remarkable modular properties~\cite{Gleb}. The
partition function, or prepotential, of the double elliptic system
depends on the bare complexified coupling constant $\tau$ of the gauge
theory and on the complex structure modulus of the compactification
torus $\hat{\tau}$. $S$-duality can be thought of as the symmetry of
the theory with respect to inversion of the coupling constant, $\tau
\mapsto - \frac{1}{\tau}$. However, in the double elliptic case, this
transformation is mixed with the transformation of the
compactification torus so that its complex structure is shifted
$\hat{\tau} \mapsto \hat{\tau} - N \frac{m(m+\varepsilon_1 + \varepsilon_2)}{\tau}$. We will derive
this transformation from the exact solution of the elliptic KZ
equations, which can be understood as the network matrix model
correlator. It also gives the basic building block of the $6d$ version
of Nekrasov functions.

From string theory considerations, one can also argue that the
partition function of the $6d$ theory should be doubly periodic in the
mass parameter, e.g.\ $m \mapsto m + 1$ or $m \mapsto m + \hat{\tau}$
should leave it invariant. Upon closer look, however, there is a
surprise here: when shifting the mass by $\hat{\tau}$, one also needs
to shift the coupling constant $\tau \mapsto \tau + N(\hat{\tau} + 2
m+\varepsilon_1 + \varepsilon_2)$. This seemingly mysterious shift can be also explained from the
network matrix model picture, which we redraw here (for $N=2$):
\begin{equation}
  \label{eq:37}
  Z (Q, Q_{\perp}, P, P_{\perp}, \vec{z}) = \qquad \parbox{9cm}{\includegraphics[width=9cm]{dell-crop}}
\end{equation}
The algebraic expression corresponding to this picture reads
\begin{equation}
  \label{eq:9a}
  Z (Q, Q_{\perp}, P, P_{\perp}, \vec{z}) = \sum_{\lambda_1, \ldots ,\lambda_N} P_{\perp}^{|\lambda|}
  \mathrm{Tr}_{\mathcal{F}_u^{(1,0)}} \left( Q^d Q_{\perp}^{d_{\perp}}
    \Psi^{*}_{\lambda_1}(P z_1)
    \cdots \Psi^{*}_{\lambda_N}(P z_N) \Psi^{\lambda_1}(z_1)
    \cdots \Psi^{\lambda_N}(z_N) \right)
\end{equation}

In Eq.~\eqref{eq:37} we have introduced the grading operators $Q^d
Q_{\perp}^{d_{\perp}}$ counting the states of a given degree just as in
the ordinary characters of affine algebra representations $V_{\mu}$:
\begin{equation}
  \label{eq:33}
  \ch_{\mu} (Q) = \Tr_{V_{\mu}} Q^d
\end{equation}
where $d$ is the grading operator counting the modes in the loop
algebra. An important difference between compactified DIM networks and
the characters $\ch_{\mu}$ is that, in the DIM case, there are \emph{two}
grading directions and \emph{two} grading operators $d$ and
$d_{\perp}$. This happens because the DIM algebra is essentially a double
loop algebra. However, we cannot simply write a generalization of
character as
\begin{equation}
  \label{eq:34}
  \text{``}\ch_{\mathcal{F}_u^{(k_1, k_2)}} (Q, Q_{\perp}) =
  \Tr_{\mathcal{F}_u^{(k_1, k_2)}} Q^d Q_{\perp}^{d_{\perp}}\text{''}
\end{equation}
at least not for the simplest representation, the Fock one
$\mathcal{F}_u^{(k_1, k_2)}$. The problem is that the general grading
operator $Q^d Q_{\perp}^{d_{\perp}}$ \emph{shifts} the spectral
parameter $u$ of the Fock space $\mathcal{F}_u^{(k_1, k_2)}$, i.e.\ it
does not map the representation space into itself. A pedantic reader
might notice that what we have just stated actually means that Fock
representations are strictly speaking \emph{not} representations of
the whole DIM algebra. Indeed, the Fock representations are counterparts of
the \emph{evaluation} representation of the affine algebra
$\widehat{\mathfrak{g}}$, which are representations of the \emph{loop
  algebra} $L \mathfrak{g}$, i.e.\ with zero central charge and
\emph{without} the action of the grading operator $d$.  Evaluation
representation $\mathrm{ev}_u$ is a representation in which all the
modes $j_n^a$ of affine currents $j^a(z)$ act in the same way, up to
scalar factors:
\begin{equation}
  \label{eq:35}
  j_n^a |\vec{v}, u \rangle  = u^n | t^a \vec{v}, u \rangle,
\end{equation}
where $t^a$ denote the generators of the finite algebra. In
particular, the action of modes of the Cartan generators can be
simultaneously diagonalized. The grading operator $d$ can be
introduced, but it transforms one evaluation representation into
another one, shifting the spectral parameter $u$:
\begin{equation}
  \label{eq:36}
  e^{z d}|\vec{v}, u \rangle = |\vec{v}, u + z \rangle
\end{equation}

Similarly, the Fock representations $\mathcal{F}_u^{(k_1, k_2)}$ of the DIM algebra
are representations in which a certain \emph{linear combination} of
central charges vanishes. Therefore, a certain linear combination of
grading operators $d$ and $d_{\perp}$ transforms one Fock
representation into another one, shifting the spectral parameter, while an
orthogonal linear combination counts the level of the states in the
representation. In other words, the Fock representation is an evaluation
representation for a quantum affine (in the
$\widehat{\mathfrak{gl}}_1$ case, simply Heisenberg) subalgebra of the DIM algebra
with a particular slope $(k_1, k_2)$. The action of all the DIM
generators can be expressed in terms of the action of just one
Heisenberg subalgebra (with ``slope'' $(-k_2, k_1)$) of DIM. The
action of the ``orthogonal'' DIM subalgebra with slope $(k_1, k_2)$, on
the other hand, can be diagonalized on the whole Fock module and gives
rise to the basis of Macdonald polynomials. Taking the trace of the
grading operators~\eqref{eq:34} is forbidden for the same reason as
the trace of $Q^d$ over the evaluation representation $\mathrm{ev}_u$:
it simply makes no sense, since the operator does not act from the
representation space into itself.

In general, the grading operator $Q^d Q_{\perp}^{d_{\perp}}$ changes
the spectral parameter $u$ of the Fock representation
$\mathcal{F}_u^{(k_1,k_2)}$ of general slope $(k_1, k_2)$ into
$Q^{-k_2} Q_{\perp}^{k_1} u$:
\begin{equation}
  \label{eq:31}
  Q^d Q_{\perp}^{d_{\perp}}: \mathcal{F}_u^{(k_1,k_2)} \to \mathcal{F}_{Q^{-k_2} Q_{\perp}^{k_1} u}^{(k_1,k_2)},
\end{equation}
or, pictorially,
\begin{equation}
  \label{eq:38}
  \parbox{4cm}{\includegraphics[width=4cm]{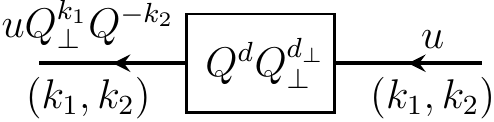}}
\end{equation}
One can notice that, at the r.h.s.\ of~\eqref{eq:31}, the slope vector
$(k_1, k_2)$ and the fugacity vector $(\ln Q, \ln Q_{\perp})$ are
paired using the skew symmetric $SL(2,\mathbb{Z})$-invariant bilinear
form.

Instead of the trace~\eqref{eq:34}, one thus needs to consider in
addition to the grading operators a nontrivial network of
the intertwiners which compensates for the shift of the spectral
parameters. Eq.~\eqref{eq:37} provides an example of such a setup. The
spectral parameters of the horizontal representations \emph{before}
and \emph{after} the wavy lines are $u Q_{\perp} P^{-2}$ and $u$
respectively. These spectral parameters have to coincide for the trace
(represented by the wavy line) to be well-defined.  We therefore tune
$Q_{\perp} = P^2$ or, for generic $N$,
\begin{equation}
  \label{eq:39}
  \boxed{Q_{\perp} = P^N}
\end{equation}
so that the whole network between the wavy lines does not shift the
horizontal spectral parameter. We will not write $Q_{\perp}$ among the
arguments of the partition function $Z(Q, Q_{\perp}, P, P_{\perp},
\vec{z})$ henceforth.

The relations between the spectral parameters of the network and the
gauge theory parameters are as follows:
\begin{gather}
  Q = e^{2\pi i \hat{\tau} },\qquad   P_{\perp} = e^{2\pi i \tau },\notag\\
  P = e^{2\pi i \left( m + \frac{\varepsilon_1 + \varepsilon_2}{2}\right) },\qquad
  z_i = e^{2\pi i a_i},\notag\label{eq:10a}\\
  q = e^{-2\pi i \varepsilon_2},\qquad t = e^{2\pi i
    \varepsilon_1},\notag
\end{gather}
where $a_i$ are the Coulomb moduli. The periodicity of the partition
function with respect to shifts of $a$, $m$, $\tau$ or $\hat{\tau}$ by
$1$ is automatic in this formalism. Notice that the picture we are
considering is \emph{almost} symmetric with respect to the exchange of
the vertical and horizontal directions (the exchange is called Miki
automorphism \cite{Miki}, spectral duality \cite{specdu1}-\cite{Sham3}, or $S$-duality of Type IIB strings
depending on the formalism). The asymmetry appears only in the number
of lines. Thus we can predict that the \emph{necklace quiver} $6d$
theory with the gauge group $U(N)^{\otimes N}$ will be spectral self-dual, i.e.\
symmetric with respect to the exchange of $\tau$ (a certain
combination of the $N$ coupling constants) and $\hat{\tau}$
accompanied by a suitable exchange of the vevs and masses of the
bifundamental hypermultiplets. In particular, the $6d$ $U(1)$ theory with
adjoint matter is spectral self-dual.

In the two subsequent sections, we will use the network matrix model
formalism to deduce the modular and periodicity properties of the
partition function~\eqref{eq:9a}. The derivation of the modular and
periodicity properties from the DIM intertwiner picture gives the
answer for the gauge theory in \emph{arbitrary $\Omega$-background}
with all corrections in $\epsilon_{1,2}$ automatically taken into
account.

\subsection{Adjoint mass shift}
\label{sec:adjoint-mass-shift}
Let us understand the transformation of the partition function when
the mass is shifted by $\hat{\tau}$. To this end, we carefully use the
commutation relations of the DIM algebra intertwiners and
automorphisms. We start with the ``double'' (vertical and horizontal)
trace of the intertwiners shown in Fig.~\ref{fig:ellqKZ}.

The procedure breaks down into two steps which can be divided further
into substeps:
\begin{enumerate}
\item \textbf{Commutation of intertwiners.} Let us move the
  intertwiners $\Psi^{\lambda_i}(z_i)$ cyclically under the
  trace. Here are the steps of this procedure:
  \begin{enumerate}
  \item \textbf{Move $\Psi^{\lambda_i}(z_i)$ to the left through the intertwiners $\Psi^{*}_{\mu_i}(w_i)$:}
    \begin{equation}
      \label{eq:28}
      Z (Q, P, P_{\perp}, \vec{z}) =\qquad   \parbox{9cm}{\includegraphics[width=9cm]{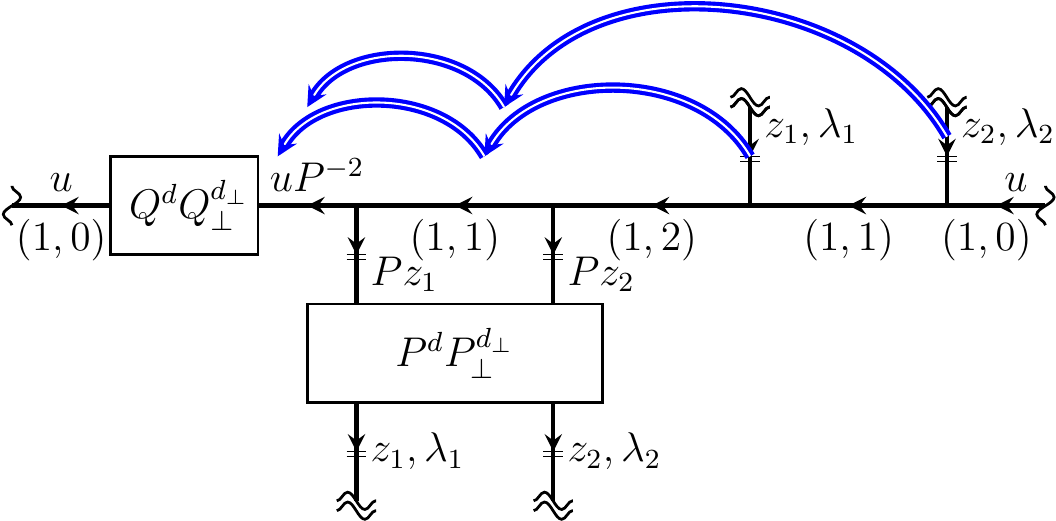}}
      \end{equation}
      Here we use the commutation relation for $\Psi$ with $\Psi^{*}$,
      which involves the scalar function $\Upsilon_{q,t}$
      \begin{equation}
        \label{eq:29}
        \parbox{11cm}{\includegraphics[width=11cm]{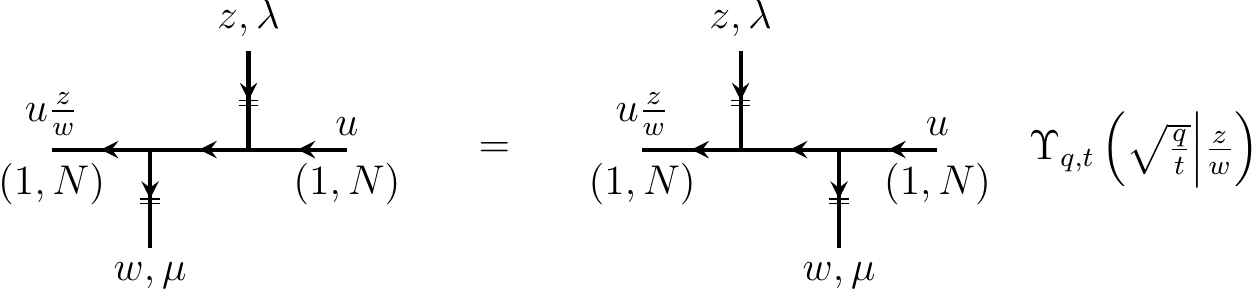}}
      \end{equation}
      with
      \begin{equation}
        \label{eq:48}
        \Upsilon_{q,t} \left( \alpha | x \right) = \exp \left[ \sum_{n
          \geq 1} \frac{\alpha^n}{n} \frac{(x^n - x^{-n})}{(1-q^n)(1-t^{-n})} \right].
      \end{equation}

    \item \textbf{Move $\Psi^{\lambda_i}(z_i)$ through the grading operators $Q^d Q_{\perp}^{d_{\perp}}$:}
      \begin{multline}
        \label{eq:30}
        Z (Q, P, P_{\perp}, \vec{z}) =\\
        = \prod_{i,j=1}^N
          \Upsilon_{q,t}\left( \sqrt{\frac{q}{t}}\Big|
            \frac{z_i}{P z_j} \right)\times \quad  \parbox{9cm}{\includegraphics[width=9cm]{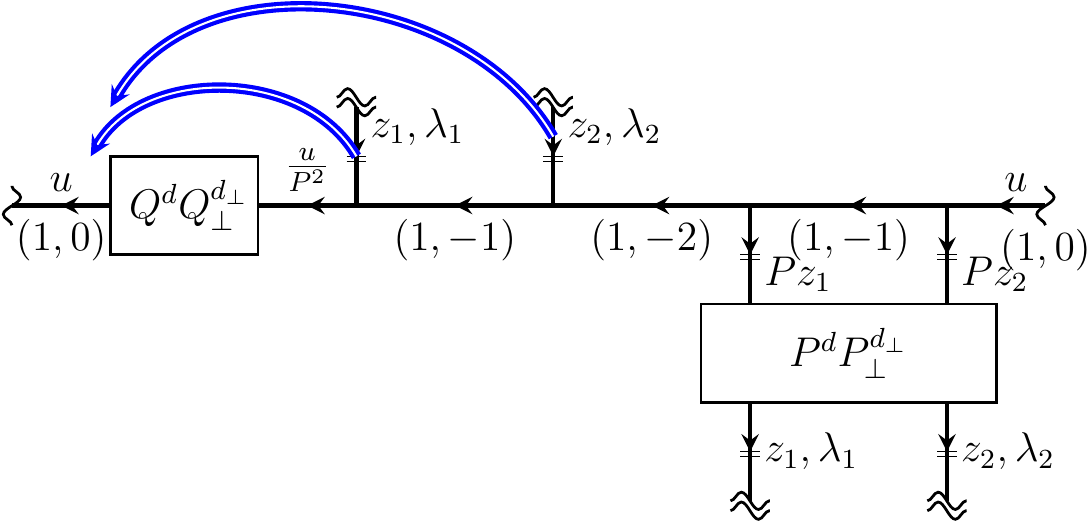}}
        \end{multline}
        The crucial point is that the grading operators satisfy the
        intertwining relations with $\Psi$ so that
        \begin{equation}
          \label{eq:11a}
          \parbox{15cm}{\includegraphics[width=15cm]{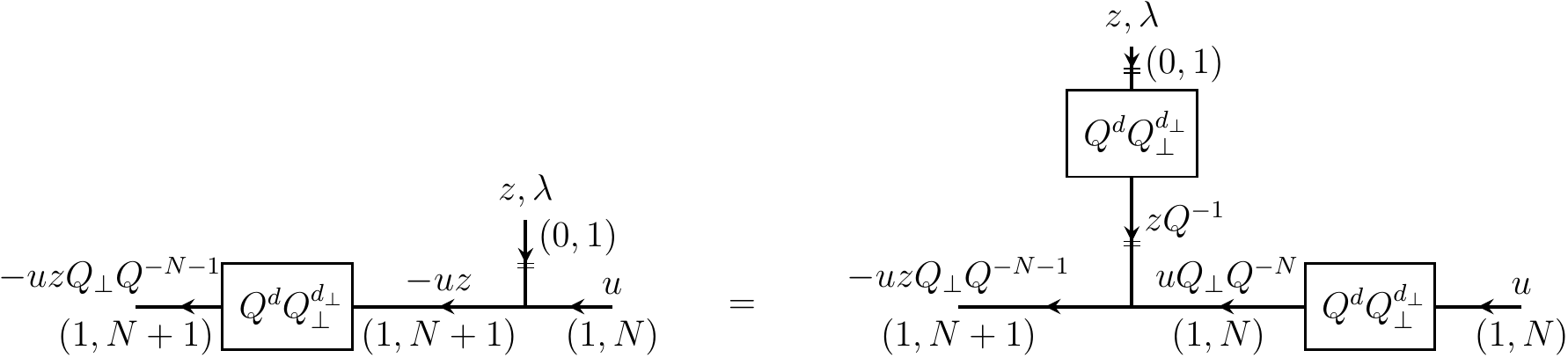}}
        \end{equation}
        where we have indicated all the slopes and spectral parameters
        of the Fock spaces corresponding to the legs explicitly.

      \item \textbf{Move $\Psi^{\lambda_i}(z_i)$ through the trace so
          that they emerge on the right:}
        \begin{multline}
          \label{eq:32}
                  Z (Q, P, P_{\perp}, \vec{z}) =\\
        = \prod_{i,j=1}^N
          \Upsilon_{q,t}\left( \sqrt{\frac{q}{t}}\Big| \frac{z_i}{P z_j} \right)\times \quad  \parbox{9cm}{\includegraphics[width=9cm]{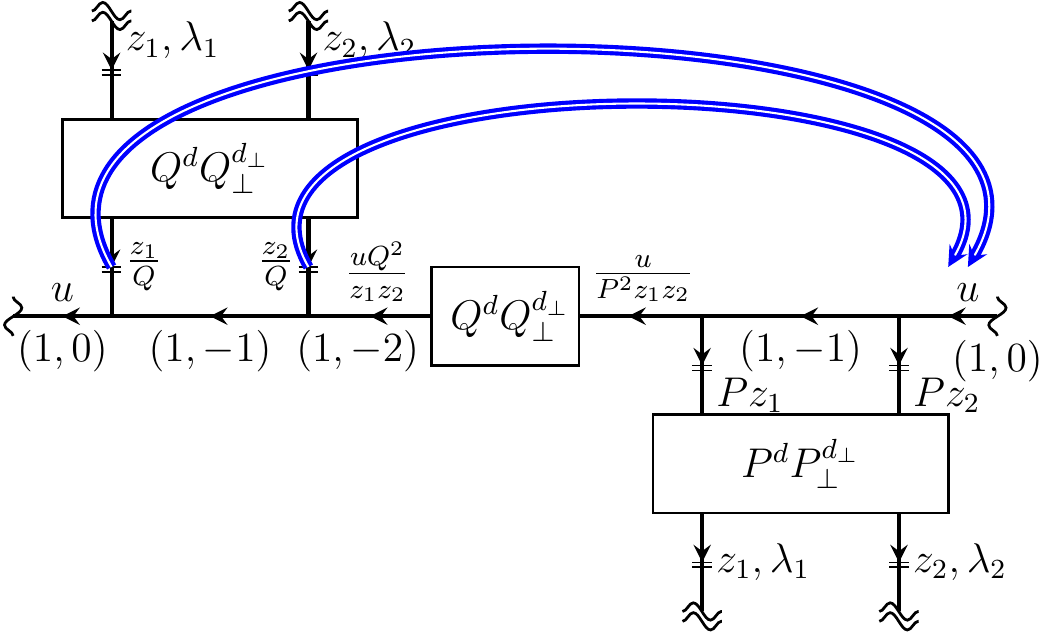}}
        \end{multline}
        Recall that the corresponding (double) wavy lines are
        identified with each other. Therefore, the grading operators $Q^d
        Q_{\perp}^{d_{\perp}}$ sitting on the upper vertical legs are
        effectively multiplied with the grading operators $P^d
        P_{\perp}^{d_{\perp}}$ sitting on the lower ones.

  \end{enumerate}
  After these steps one arrives at the following picture:
  \begin{multline}
    \label{eq:12a}
                 Z (Q, P, P_{\perp}, \vec{z}) =\\
        = \prod_{i,j=1}^N
          \Upsilon_{q,t}\left( \sqrt{\frac{q}{t}}\Big|
            \frac{z_i}{P z_j} \right)\times \quad  \parbox{9cm}{\includegraphics[width=9cm]{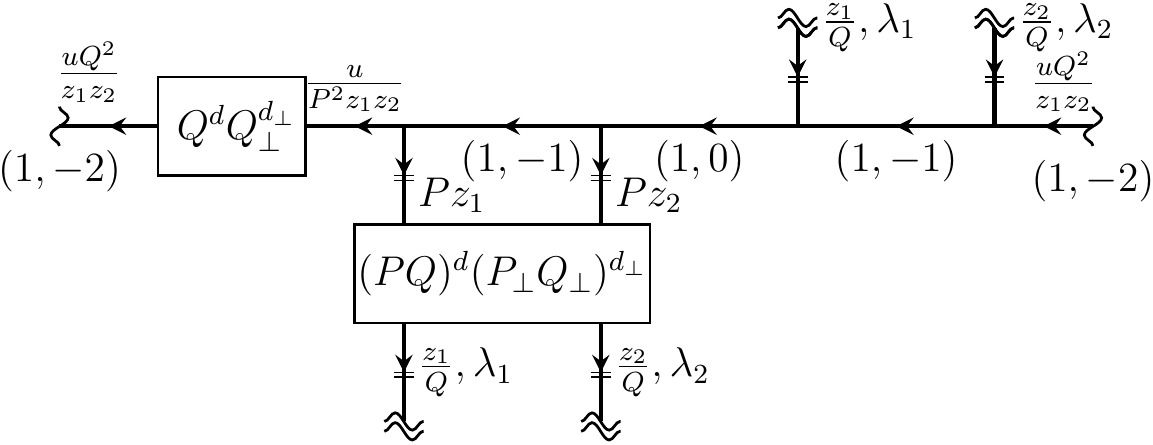}}
  \end{multline}
  The expression looks almost the same as the initial
  one~\eqref{eq:9a}. There are two key differences\footnote{The
    horizontal spectral parameter has also changed, but this is
    inessential since there is only one horizontal line and the
    overall shift eliminates this difference.}:
  \begin{enumerate}
  \item The fugacities on the vertical legs are different. They used
    to be $(P, P_{\perp})$, whereas after the cyclic movement of the
    intertwiners they became $(PQ, P_{\perp} Q_{\perp})$.

  \item The Fock space over which the trace is taken has a different
    \emph{slope,} which used to be $(1,0)$ and has become $(1, -N)$.  To
    compare the parameters of the theory with that of the initial
    setup, we need to transform the slope of the Fock space back to
    $(1,0)$ using the $T$-transformation, i.e.\ act with the
  $T$-element from the $SL(2,\mathbb{Z})$ automorphism group of DIM.
\end{enumerate}

\item \textbf{$T$-transformation.} The action of the automorphism
  $T\in SL(2, \mathbb{Z})$ of the DIM algebra on the vertical and
  horizontal representations is easy to deduce. In particular, the
  grading operators, as well as the central charges form doublets
  under $SL(2,\mathbb{Z})$. Let us consider the action of $T$ on the
  elements of the network~\eqref{eq:12a} in turn:

  \begin{enumerate}
  \item \textbf{On the legs.}  $T$-transformation naturally
  transforms the slope vector (which is the vector of central charges)
  of the horizontal Fock representation $\mathcal{F}_u^{(1,m)}$:
  \begin{gather}
    \label{eq:40}
\parbox{3.5cm}{\includegraphics[width=3.5cm]{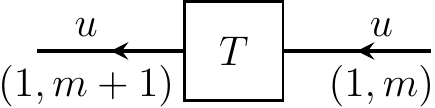}}
  \end{gather}
It acts diagonally on the vertical Fock space $\mathcal{F}_z^{(0,1)}$  in the basis of Macdonald polynomials:
  \begin{equation}
    \label{eq:41}
    \parbox{6.5cm}{\includegraphics[width=6.5cm]{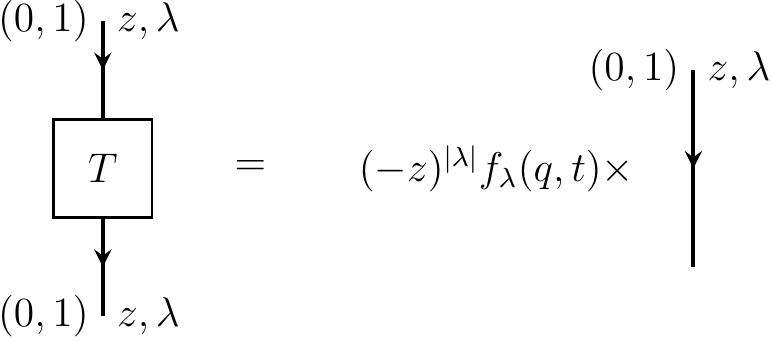}}
  \end{equation}
  where
  \begin{equation}
    \label{eq:42}
    f_{\lambda}(q,t) = (-1)^{|\lambda|} q^{\sum_{(i,j)\in \lambda}
      \left( j-\frac{1}{2} \right)} t^{\sum_{(i,j)\in \lambda} \left(
        \frac{1}{2} - i\right)}
  \end{equation}
  is the framing factor.

  These actions are, of course, consistent with the explicit expression
  for the DIM intertwiners written down in~\cite{AFS}.

  \item \textbf{On the grading operators.} The action of $T$ on the
    grading operators is explicitly given by
    \begin{equation}
      \label{eq:43}
       \parbox{15cm}{\includegraphics[width=15cm]{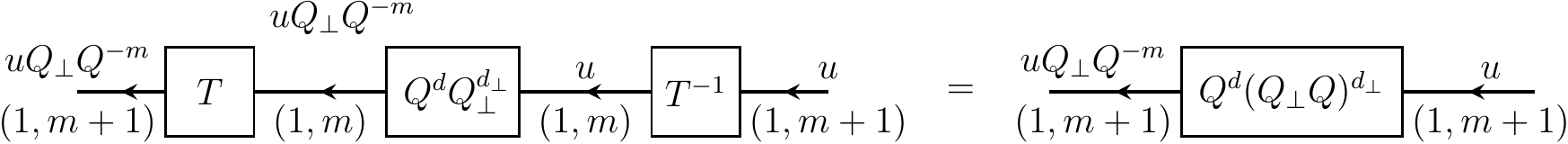}}
     \end{equation}
     Notice how the fugacities $Q$, $Q_{\perp}$ transform as a doublet
     of $SL(2,\mathbb{Z})$.

    \end{enumerate}

    Now we are ready to insert the identity operators $1 = T^{-N}T^{N}$ to
    all the intermediate legs of the network~\eqref{eq:12a} and
    commute them with the intertwiners and grading operators using
    Eqs.~\eqref{eq:40},~\eqref{eq:43}:
    \begin{multline}
      \label{eq:45}
                       Z (Q, P, P_{\perp}, \vec{z}) =\prod_{i,j=1}^N
          \Upsilon_{q,t}\left( \sqrt{\frac{q}{t}}\Big|
            \frac{z_i}{P z_j} \right)\times\\
        \times  \quad  \parbox{13cm}{\includegraphics[width=13cm]{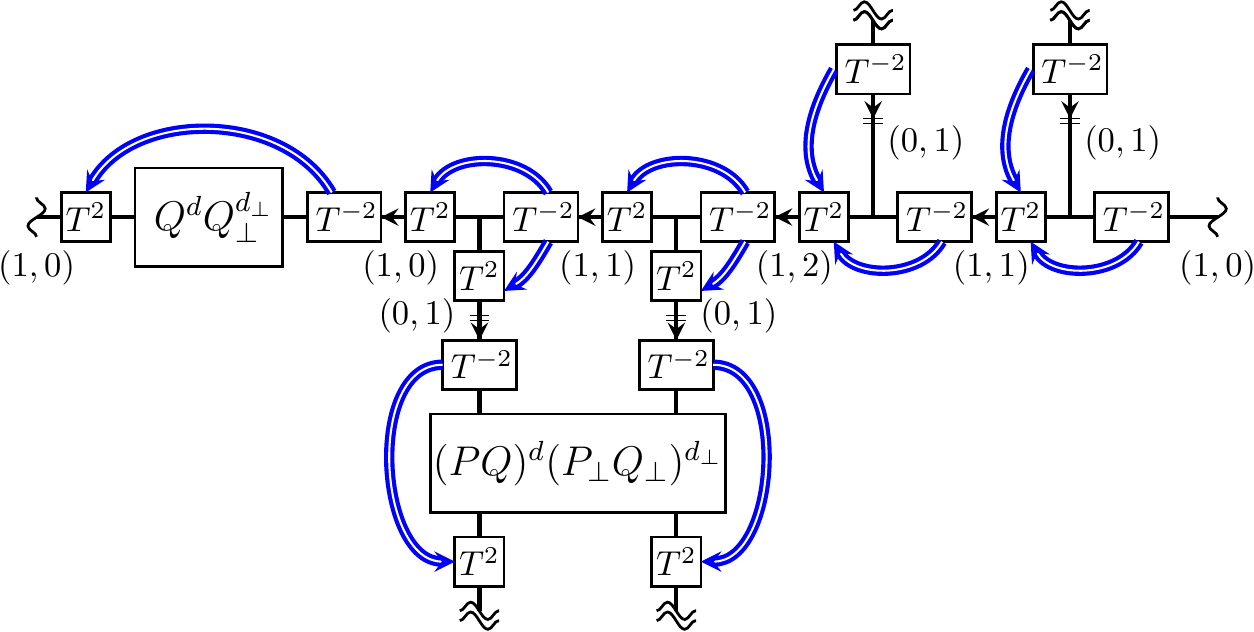}}
      \end{multline}
      The slopes indicated on the picture are those appearing \emph{in
        between} the $T^{-N}$ and $T^N$ operators.

    \end{enumerate}
     Finally, we obtain
      the equality between the initial gauge theory partition function
      $Z (Q, P, P_{\perp}, \vec{z})$ and a similar one, but with
      shifted parameters:
      \begin{multline}
        \label{eq:46}
                         Z (Q, P, P_{\perp}, \vec{z}) =\\
        = \prod_{i,j=1}^N
          \Upsilon_{q,t}\left( \sqrt{\frac{q}{t}}\Big|
            \frac{z_i}{P z_j} \right)\times
          \quad  \parbox{9cm}{\includegraphics[width=9cm]{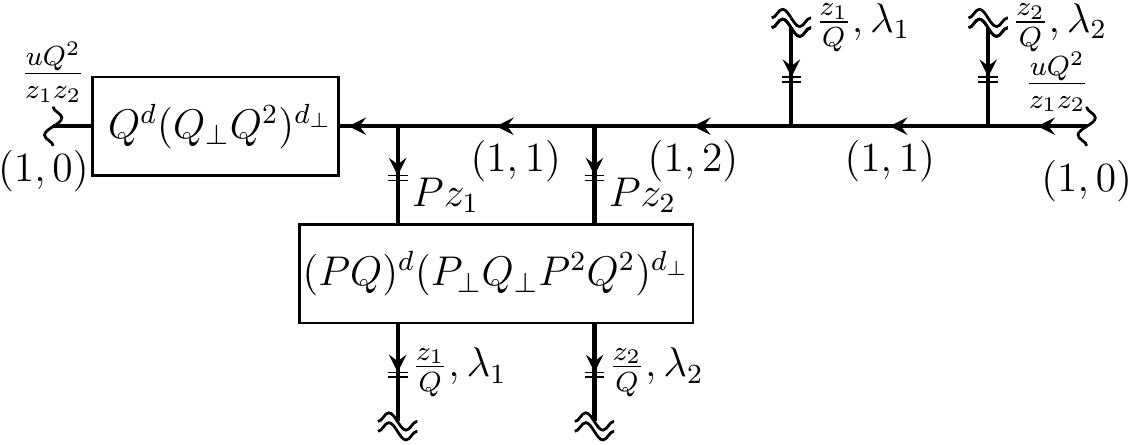}}\quad
          =\\
          =\prod_{i,j=1}^N
          \Upsilon_{q,t}\left( \sqrt{\frac{q}{t}}\Big|
            \frac{z_i}{P z_j} \right) Z (Q, Q P, P_{\perp}
          Q^N P^{2N}, \vec{z})
        \end{multline}
        Notice that we have omitted an inessential overall shift of
        $z_i$ in the argument of the partition function. The product
        of $\Upsilon_{q,t}$ functions arises from the \emph{classical}
        part of the gauge theory partition function. Indeed, the
        prefactor is independent of $P_{\perp}$, i.e.\ of the gauge
        theory coupling constant, thus we can safely send it to zero and
        still keep Eq.~\eqref{eq:46} intact. In this limit, the gauge
        theory instantons do not contribute, or, in the language of
        intertwiners, the vertical lines become uncompactified so that
        $\lambda_1 = \lambda_2 = \varnothing$. What remains is the
        strip of intertwiners compactified only along the horizontal
        direction. The prefactor can be absorbed into a simple
        redefinition of the partition function
        \begin{equation}
          \label{eq:49}
          Z_{\mathrm{periodic}}(Q, P, P_{\perp}, \vec{z}) =  \prod_{i,j}^N
          \Xi_{q,t,Q}\left( \sqrt{\frac{q}{t}} \Big| \frac{Q z_i}{P z_j} \right)  Z (Q, P, P_{\perp}, \vec{z})
        \end{equation}
        where
        \begin{equation}
          \label{eq:47}
          \Xi_{q,t,Q}\left( \alpha | x \right) = \exp \left[ - \sum_{n
            \geq 1} \frac{\alpha^n}{n} \frac{x^n + Q^n x^{-n}}{(1-q^n)(1-t^{-n})(1-Q^n)} \right]
        \end{equation}
        and we have the following difference equation\footnote{Notice
          that $\Xi_{q,t,Q}\left( (Q)^{-\frac{1}{2}} \alpha \Big|
            Q^{\frac{1}{2}} x \right) $ is symmetric in $q$, $t^{-1}$
          and $Q$.}
        \begin{equation}
          \label{eq:50}
          \frac{\Xi_{q,t,Q}\left( \alpha | Q x
            \right)}{\Xi_{q,t,Q}\left( \alpha | x \right)} =
          \Upsilon_{q,t} \left( \alpha | x \right).
        \end{equation}
        The resulting partition function $Z_{\mathrm{periodic}}$ is
        invariant with respect to the following shift of the
        parameters:
        \begin{equation}
  \boxed{Z_{\mathrm{periodic}} (Q, Q P, P_{\perp}
          Q^N P^{2N}, \vec{z}) = Z_{\mathrm{periodic}} (Q, P, P_{\perp}
          , \vec{z})}\label{eq:14a}
\end{equation}
Now the shift of the gauge theory parameters can be easily extracted
from the dictionary~\eqref{eq:10a}:
\begin{gather}
  \label{eq:44}
  m \mapsto m+ \hat{\tau},\\
  \tau \mapsto \tau + N (2 m + \varepsilon_1 + \varepsilon_2 +
  \hat{\tau}),\\
  \hat{\tau} \mapsto \hat{\tau},\\
  a_i \mapsto a_i.
\end{gather}
In absence of the $\varepsilon$-deformation, this transformation law
coincides with what was found in~\cite{Gleb} for the classical
integrable system. Here we provide an \emph{exact} expression valid
for arbitrary $\Omega$-background. In particular, setting
$\varepsilon_1 = \hbar$, $\varepsilon_2 = 0$ in Eq.~\eqref{eq:44}
gives the transformation law for the exact prepotential of the quantum
double elliptic integrable system (as usual in the Nekrasov-Shatashvili limit \cite{SWqint1,SWqint2,SWqint3,SWqint4}).

\subsection{Modular transformations}
\label{sec:modul-transf}
The properties of the partition function with respect to the modular
transformations of the compactification torus can also be deduced from
the network matrix model. We take the explicit expression for the
trace of intertwiners over the horizontal Fock space
from~\cite{Awata:2017cnz}. It is written as a product of
theta-functions, which are modular invariant up to a simple
prefactor. This prefactor gives rise to an extra shift of the
\emph{gauge coupling}~$\tau$. For $N > 1$, we will discuss only one of
the two modular transformations of the $6d$ $U(N)$ partition function,
the second being the $\tau \mapsto - \frac{1}{\tau}$. This second
transformation can also be easily analysed in our framework, but we
leave this task for the future.

\subsubsection{$U(1)$ theory}
\label{sec:u1-theory}
As a warm-up, we consider the $U(1)$ theory, where the expressions
are simpler, though the modular properties are still nontrivial. In
particular, there are no vacuum moduli in this case. We have:
\begin{multline}
  \label{eq:15a}
  Z(P,P_{\perp},Q) = \exp \left[ \sum_{n \geq 1} \frac{1}{n}
    \frac{\left( 1 - (qt)^{\frac{n}{2}} P^{-n} \right) \left( 1 -
        (qt)^{-\frac{n}{2}} P^{-n} \right) \left( \sqrt{\frac{q}{t}} P
        Q\right)^n + (1-Q^n) \left( \sqrt{\frac{q}{t}} \frac{1}{P}
      \right)^n }{(1-q^n)(1-t^{-n})(1-Q^n)} \right]\times\\
  \times \sum_{\lambda} \left( \sqrt{\frac{t}{q}} \frac{P_{\perp}}{P}
  \right)^{|\lambda|} \frac{\Theta_{\lambda \lambda}
    \left(\sqrt{\frac{q}{t}} P \Big|
      Q\right)}{\Theta_{\lambda\lambda}(1|Q)} \sim \sum_{\lambda}
  e^{2\pi i |\lambda| (\tau - m)} \frac{\Theta_{\lambda \lambda}
    \left(e^{2\pi i m} \big| e^{2\pi i
        \hat{\tau}}\right)}{\Theta_{\lambda\lambda}(1|e^{2\pi i
      \hat{\tau}})},
\end{multline}
where\footnote{We slightly change the notations as compared
  to~\cite{Awata:2017cnz}. There was also a typo
  in~\cite{Awata:2017cnz}.}
\begin{equation}
  \label{eq:16a}
  \Theta_{\lambda\mu}(x|Q)  = \prod_{(i,j)\in \lambda} \theta_Q (x
  q^{\lambda_i - j} t^{\mu_j^{\mathrm{T}}-i+1}) \prod_{(i,j)\in \mu}
  \theta_Q (x q^{j -\mu_i - 1} t^{i - \lambda_j^{\mathrm{T}}}),
\end{equation}
and $\theta_Q(x)$ is the Jacobi theta function:
\begin{equation}
  \label{eq:52}
  \theta_Q(x) = \prod_{k \geq 1} \left( 1 - Q^{k+1}  \right) \left( 1
    - Q^k x \right) \left( 1 - Q^k \frac{Q}{x}  \right).
\end{equation}
Notice that in Eq.~\eqref{eq:15a} we exclude an extra $\Xi_{q,t,Q}$
prefactor used to construct the periodic partition function
$Z_{\mathrm{periodic}}$. The modular transformation of the theta function
is given by the standard formula:
\begin{equation}
  \label{eq:51}
  \theta_{e^{-\frac{2\pi i}{\tau}}}( e^{\frac{2\pi i z}{\tau}}) =
  \left( i \tau \right)^{\frac{1}{2}} e^{\frac{\pi i}{\tau} \left( z +
    \frac{1 - \tau}{2}\right)^2}   \theta_{e^{2\pi i \tau}}( e^{2\pi i z}).
\end{equation}
Making the modular transformation
\begin{gather}
  \label{eq:18a}
  Q = e^{2\pi i \hat{\tau}}
  \mapsto \tilde{Q} = e^{-\frac{2\pi i}{\hat{\tau}}},\\
  q = e^{-2\pi i \varepsilon_2}\mapsto \tilde{q}= e^{-\frac{2\pi
      i \varepsilon_2}{\hat{\tau}}},\\
  t = e^{2\pi i \varepsilon_1}\mapsto \tilde{t}= e^{\frac{2\pi
      i\varepsilon_1}{\hat{\tau}}},\\
  P = e^{2\pi i m} \mapsto \tilde{P} = e^{\frac{2\pi i m}{\hat{\tau}}},
\end{gather}
we get a prefactor from each theta-function in the product in
Eq.~\eqref{eq:16a}. The term in the instanton expansion labelled by
the diagram $\lambda$ is multiplied with the exponential of the following expression:
\begin{multline}
  \label{eq:53}
  \frac{\pi i}{\hat{\tau}} \sum_{(i,j)\in \lambda} \Biggl[  \left( m  +
      \frac{1-\tau}{2} - \varepsilon_2 (\lambda_i - j) + \varepsilon_1
   (\lambda^{\mathrm{T}}_j - i + 1) \right)^2 + \left( m  +
      \frac{1-\tau}{2} - \varepsilon_2 (-\lambda_i + j - 1) + \varepsilon_1
      (- \lambda^{\mathrm{T}}_j + i) \right)^2 -\\
\left(
      \frac{1-\tau}{2} - \varepsilon_2 (\lambda_i - j) + \varepsilon_1
   (\lambda^{\mathrm{T}}_j - i + 1) \right)^2 + \left(
      \frac{1-\tau}{2} - \varepsilon_2 (-\lambda_i + j - 1) + \varepsilon_1
      (- \lambda^{\mathrm{T}}_j + i) \right)^2    \Biggr]~.
\end{multline}
There happen to be many cancellations between different terms in the
sum~\eqref{eq:53}. The final answer for the prefactor looks quite
simple:
\begin{equation}
  \label{eq:17a}
  \frac{\Theta_{\lambda\lambda}\left(\sqrt{\frac{\tilde{q}}{\tilde{t}}} \tilde{P} \Big|
      \tilde{Q}\right)}{\Theta_{\lambda\lambda}(1|\tilde{Q})} =
  e^{\frac{2\pi i}{\hat{\tau}} m (m+ \varepsilon_1 +
      \varepsilon_2) |\lambda|} e^{ \frac{2\pi i m}{\hat{\tau}} |\lambda| - 2
      \pi i m |\lambda|} \frac{\Theta_{\lambda\lambda}\left(\sqrt{\frac{q}{t}} P \Big|
      Q\right)}{\Theta_{\lambda\lambda}(1|Q)}~.
\end{equation}
The partition function is invariant (up to an overall scalar
factor) under the modular transformation of $\hat{\tau}$, if we perform
a shift of the complexified coupling of the gauge theory encoded in
$P_{\perp} = e^{2\pi i \tau}$:
\begin{equation}
  \label{eq:19a}
 \boxed{ Z \left( \frac{\varepsilon_1}{\hat{\tau}}, \frac{\varepsilon_2}{\hat{\tau}}, -\frac{1}{\hat{\tau}}, \tau - \frac{m(m+\varepsilon_1 + \varepsilon_2)}{\hat{\tau}},
    \frac{m}{\hat{\tau}}  \right) \sim Z \left(\varepsilon_1, \varepsilon_2 ,\hat{\tau}, \tau
    ,
    m  \right)}
\end{equation}

In the classical case, this matches the transformation law obtained
from the Seiberg-Witten theory techniques and modular anomaly
equations~\cite{Gleb}. Notice that our derivation is valid for the general
$\Omega$-background, in particular, it holds for the \emph{quantized}
double elliptic integrable system.

As we have mentioned earlier, the $U(1)$ theory is spectral self-dual,
which, in this case, means that it is invariant under the exchange of
$\tau$ and $\hat{\tau}$. This implies the second modular
transformation for $\tau$. Notice that for $m=0$ the theory becomes $6d$
$\mathcal{N} = (2,0)$ theory compactified on $T^2$ without any
punctures\footnote{In the $\Omega$-background, the theory is invariant
  with respect to the reflection $m \to -\varepsilon_1 - \varepsilon_2
  - m$, thus $m = -\varepsilon_1 - \varepsilon_2$ also leads to a symmetry
  enhancement.}. The partition function becomes a product of two
$\eta$-functions:
\begin{equation}
  \label{eq:55}
  Z(\varepsilon_1, \varepsilon_2, \hat{\tau}, \tau, 0) \sim
  \prod_{k\geq 1} \frac{1}{\left( 1 - e^{2\pi i \tau k} \right) \left( 1 - e^{2\pi i \hat{\tau} k} \right)}  ,
\end{equation}
where we omit an overall prefactor independent of $\hat{\tau}$ and
$\tau$. The spectral duality and modular invariance are evident in
this limiting case.

\subsubsection{$U(N)$ theory}
\label{sec:un-theory}
The case of $U(N)$ gauge theory can be understood along the same lines
as the $U(1)$ one. The partition function is equal to
\begin{multline}
  \label{eq:54}
  Z(P,P_{\perp},Q,\vec{z}) = \prod_{i,j}^N \exp \left[
    \sum_{n \geq 1} \frac{1}{n} \frac{\left( 1 - (qt)^{\frac{n}{2}}
        P^{-n} \right) \left( 1 - (qt)^{-\frac{n}{2}} P^{-n} \right)
      \left( \sqrt{\frac{q}{t}} P Q\right)^n + (1-Q^n) \left(
        \sqrt{\frac{q}{t}} \frac{z_i}{P z_j}
      \right)^n }{(1-q^n)(1-t^{-n})(1-Q^n)} \right]\times\\
  \times \sum_{\vec{\lambda}} \left(
    \left(\frac{t}{q}\right)^{\frac{N}{2}} \frac{P_{\perp}}{P^N}
  \right)^{|\lambda|} \prod_{i,j=1}^N\frac{\Theta_{\lambda^{(i)}
      \lambda^{(j)}} \left(\sqrt{\frac{q}{t}} P \frac{z_i}{z_j} \Big|
      Q\right)}{\Theta_{\lambda^{(i)} \lambda^{(j)}}(\frac{z_i}{z_j}\big|Q)}
  \sim \sum_{\vec{\lambda}}
  e^{2\pi i |\lambda| (\tau - N m)} \prod_{i,j}^N \frac{\Theta_{\lambda^{(i)} \lambda^{(j)}}
    \left(e^{2\pi i (m + a_i - a_j)} \big| e^{2\pi i
        \hat{\tau}}\right)}{\Theta_{\lambda^{(i)}\lambda^{(j)}}(e^{2\pi i (a_i - a_j)}|e^{2\pi i
      \hat{\tau}})}.
\end{multline}
Now we have a nontrivial dependence on $(N-1)$ vacuum moduli $a_a$,
which we assume add up to zero, $\sum_{a=1}^N a_a =0$. The modular
transformation of $a_a$ is the same as that of the mass $m$:
\begin{equation}
  \label{eq:58}
  z_a = e^{2\pi i a_a} \mapsto \tilde{z}_a = e^{2\pi i \tilde{a}_a}  =
  e^{2\pi i \frac{a_a}{\hat{\tau}}}.
\end{equation}
After the modular transformation, the theta-functions in the instanton
series give a prefactor, which is the exponential of
\begin{multline}
  \label{eq:56}
  \frac{\pi i}{\hat{\tau}} \sum_{a,b}^N \Biggl[ \sum_{(i,j)\in
    \lambda^{(a)}} \left( m + a_a - a_b + \frac{1-\tau}{2} - \varepsilon_2
    (\lambda^{(a)}_i - j) + \varepsilon_1 (\lambda^{(b)\mathrm{T}}_j
    - i + 1) \right)^2 +\\
  + \sum_{(i,j)\in \lambda^{(b)}} \left( m + a_a - a_b +
    \frac{1-\tau}{2} - \varepsilon_2 (-\lambda^{(b)}_i + j - 1) +
    \varepsilon_1
    (- \lambda^{(a)\mathrm{T}}_j + i) \right)^2 -\\
  -\sum_{(i,j)\in
    \lambda^{(a)}} \left( a_a - a_b + \frac{1-\tau}{2} - \varepsilon_2
    (\lambda^{(a)}_i - j) + \varepsilon_1 (\lambda^{(b)\mathrm{T}}_j
    - i + 1) \right)^2 -\\
  - \sum_{(i,j)\in \lambda^{(b)}} \left( a_a - a_b +
    \frac{1-\tau}{2} - \varepsilon_2 (-\lambda^{(b)}_i + j - 1) +
    \varepsilon_1
    (- \lambda^{(a)\mathrm{T}}_j + i) \right)^2 \Biggr]~.
\end{multline}
Again there are many cancellations, in particular, the dependence on
$a_a$ cancels completely. Eventually, we have a simple transformation
law:
\begin{equation}
  \label{eq:57}
  \prod_{i,j=1}^N\frac{\Theta_{\lambda^{(i)}
      \lambda^{(j)}} \left(\sqrt{\frac{\tilde{q}}{\tilde{t}}} \tilde{P} \frac{\tilde{z}_i}{\tilde{z}_j} \Big|
      \tilde{Q}\right)}{\Theta_{\lambda^{(i)}
      \lambda^{(j)}}(\frac{\tilde{z}_i}{\tilde{z}_j}\big|\tilde{Q})} =
  e^{\frac{2\pi i N}{\hat{\tau}} m (m+ \varepsilon_1 +
      \varepsilon_2) |\lambda|} e^{ \frac{2\pi i m N}{\hat{\tau}} |\lambda| - 2
      \pi i m N |\lambda|}    \prod_{i,j=1}^N\frac{\Theta_{\lambda^{(i)}
      \lambda^{(j)}} \left(\sqrt{\frac{q}{t}} P \frac{z_i}{z_j} \Big|
      Q\right)}{\Theta_{\lambda^{(i)}
      \lambda^{(j)}}(\frac{z_i}{z_j}\big|Q)}~.
\end{equation}
The partition function is therefore invariant under the following
modular transformation:
\begin{equation}
  \label{eq:59}
  \boxed{Z \left( \frac{\varepsilon_1}{\hat{\tau}},
    \frac{\varepsilon_2}{\hat{\tau}}, -\frac{1}{\hat{\tau}}, \tau - N \frac{m(m+\varepsilon_1 + \varepsilon_2)}{\hat{\tau}},
    \frac{m}{\hat{\tau}} , \frac{\vec{a}}{\hat{\tau}} \right) \sim Z \left(\varepsilon_1, \varepsilon_2 ,\hat{\tau}, \tau
    ,
    m, \vec{a}  \right)}
\end{equation}
The transformation is consistent with the classical case discussed
in~\cite{Gleb}.


\section{Discussion}
We have presented two generalizations and one application of the
$(q,t)$-KZ equation for the quantum toroidal algebra
$U_{q,t}(\widehat{\widehat{\mathfrak{gl}}}_1)$ derived
in~\cite{Awata:2017cnz}. The first generalization is the case of an
\emph{arbitrary} horizontal level in
$U_{q,t}(\widehat{\widehat{\mathfrak{gl}}}_1)$, where we postulate the
KZ equation using the analogy with the quantum affine case. The second
generalization is the KZ equation for the ``non-Abelian'' quantum
toroidal algebra $U_{\gq, \gd} (\widehat{\widehat{\mathfrak{gl}}}_n)$,
but with horizontal level one. In this setup, we find the expressions
for the intertwiners of the Fock representations and show that (at least in
the unrefined case) they factorize into products of the intertwiners for
$U_{q^n,q^n}(\widehat{\widehat{\mathfrak{gl}}}_1)$. Thus, the networks
of  the ``non-Abelian'' intertwiners can be redrawn as more complicated
networks of the Abelian ones. We call this procedure Abelianization.

In both of these cases, we still find only algebraic solutions. For
$U_{\gq,\gd} (\widehat{\widehat{\mathfrak{gl}}}_n)$ these solutions
are related to the Nekrasov functions on the ALE spaces $\mathbb{C}^2/
\mathbb{Z}_n$.

We also consider an application of network matrix models and KZ
equations to $6d$ gauge theories. We identify the compactified network
of the intertwiners corresponding to the $6d$ $U(N)$ gauge theory with
massive adjoint hypermultiplet compactified on $T^2$ and study the
properties of the partition function under shifts of the adjoint mass
and modular transformations of the compactification torus. We find all
$\varepsilon$-corrections to the known classical answer. It would be
interesting to understand the origin and interplay between two
$SL(2,\mathbb{Z})$ modular transformations and spectral duality of the
$6d$ gauge theory.

The most interesting and nontrivial generalization of $(q,t)$-KZ
equations, when \emph{both} central charges are \emph{arbitrary} still
remains to be understood. In this case, new \emph{integral} solutions
should arise, which generalize the Nekrasov functions in a nontrivial
way. We plan to study these intriguing cases elsewhere.

\subsection*{Acknowledgements}

 We very much appreciate correspondence and discussions with M.~Jimbo, S.~Minabe,  and S.~Yanagida.

\bigskip

Our work is supported in part by Grants-in-Aid for Scientific Research
(\# 17K05275) (H.A.), (\# 15H05738) (H.K.) and JSPS Bilateral Joint Projects (JSPS-RFBR collaboration)
``Topological Field Theories and String Theory: from Topological Recursion
to Quantum Toroidal Algebra'' from MEXT, Japan. It is also partly supported by the grant of the Foundation for the Advancement of Theoretical Physics ``BASIS" (A.Mor.), by  RFBR grants 16-01-00291 (A.Mir.) and 16-02-01021 (A.Mor. and Y.Z.), by joint grants 17-51-50051-YaF, 15-51-52031-NSC-a, 16-51-53034-GFEN, 16-51-45029-IND-a (A.M.'s and Y.Z.). The work of Y.Z. was supported in part by INFN and by the ERC Starting Grant 637844-HBQFTNCER.

\appendix


\section{Combinatorics of the normalization factor}

\theoremstyle{definition}
\newtheorem{thm}{Lemma}

We have normalized the components of the intertwiner $\Phi_\lambda(u)$ by the following factor:
\begin{equation}
 C_\lambda(q_1, q_3) = {\displaystyle{\prod_{\substack{\square \in \lambda \\ h_\lambda(\square) \equiv 0}}}
 (1-  q_1^{a_\lambda(\square)} q_3^{-\ell_\lambda(\square)-1})},
\end{equation}
Note that in the product there is a restriction on the length of the hook.
Our normalization of the component of the dual intertwiner $\Phi^*_\lambda(u)$
is slightly different and given by
\begin{equation}
 C^\prime_\lambda(q_1, q_3) = {\displaystyle{\prod_{\substack{\square \in \lambda \\ h_\lambda(\square) \equiv 0}}}
 (1-  q_1^{a_\lambda(\square)+1} q_3^{-\ell_\lambda(\square)})}.
\end{equation}
In this subsection, we will prove technical lemmas on these normalization factors.
\begin{thm}
When we add a box with color $\ell$ in the $k$-th row,
the change of the normalization factors $C_\lambda$ and $C'_\lambda$ is given by
\begin{equation}
\frac{C_{\lambda + 1_k}}{C_\lambda} =
\frac{\displaystyle{\prod_{j=1}^{k-1}}\left( 1 - q_2 \frac{x_j}{x_k} \right)^{\deltabar_{j- \lambda_j, \ell+1}}}
{\displaystyle{\prod_{j=1}^{k-1}}\left( 1 - q_1^{-1} \frac{x_j}{x_k} \right)^{\deltabar_{j- \lambda_j, \ell}}}
\frac{\displaystyle{\prod_{j=k+1}^{\ell(\lambda)+1}}\left( 1 -  \frac{x_k}{x_j} \right)^{\deltabar_{j- \lambda_j, \ell+1}}}
{\displaystyle{\prod_{j=k+1}^{\ell(\lambda)}}\left( 1 - q_3^{-1} \frac{x_k}{x_j} \right)^{\deltabar_{j- \lambda_j, \ell}}},
\label{Lemma1a}
\end{equation}
and
\begin{equation}
\frac{C'_{\lambda + 1_k}}{C'_\lambda} =
\frac{\displaystyle{\prod_{j=1}^{k-1}}\left( 1 - \frac{x_j}{x_k} \right)^{\deltabar_{j- \lambda_j, \ell+1}}}
{\displaystyle{\prod_{j=1}^{k-1}}\left( 1 - q_3\frac{x_j}{x_k} \right)^{\deltabar_{j- \lambda_j, \ell}}}
\frac{\displaystyle{\prod_{j=k+1}^{\ell(\lambda)+1}}\left( 1 -  q_2^{-1} \frac{x_k}{x_j} \right)^{\deltabar_{j- \lambda_j, \ell+1}}}
{\displaystyle{\prod_{j=k+1}^{\ell(\lambda)}}\left( 1 - q_1 \frac{x_k}{x_j} \right)^{\deltabar_{j- \lambda_j, \ell}}}.
\label{Lemma1b}
\end{equation}
\end{thm}

Since we have to deal with the coloring and the length of the hooks in $\lambda$,
let us introduce convenient notations for this purpose.
We define the head $(x_h, y_h)$ and the tail $(x_t, y_t)$ of a hook
by the condition $(x_h, y_h-1), (x_t, y_t) \in \lambda$ and $(x_h,  y_h), (x_t+1, y_t) \notin \lambda$.
\begin{equation}
\begin{ytableau}
\bullet & \cdots & \cdots & \bullet & *(red)\times \\
\vdots \\
*(red) \bullet
\end{ytableau}
\qquad\qquad
\hbox{The hook consists of boxes with $\bullet$.}
\nonumber
\end{equation}
Note that the tail belongs to the hook, but the head does not. The corner of the hook is $(x_h, y_t)$.
The color of the boxes is increasing along a hook from the head to the tail. If a hook satisfies
$h_\lambda(x_h, y_t) \equiv 0$, the head and the tail have the same color. In the above hook diagram,
the head is the red box with $\times$ and the tail is the red box with $\bullet$.
In terms of these notations, the normalization factor is
\begin{equation}
 C_\lambda(q_1, q_3) = {\displaystyle{\prod_{\substack{\square \in \lambda \\ h_\lambda(\square) \equiv 0}}}
 (1-  q_1^{y_h - y_t-1} q_3^{x_h - x_t -1})}.
\end{equation}
Thus, we can evaluate each factor of $C_\lambda(q_1, q_3)$ by identifying the heads and the tails of the hooks.

Now when we add a box $(k, \lambda_k +1)$ with color $\ell$ in the $k$-th row, a newly appearing
hook in $\lambda + 1_k$ has the head $(k, \lambda_k +2)$ or the tail $(k, \lambda_k +1)$.
Let us first consider the case when $(k, \lambda_k +1)$ is the tail and the hook length is a multiple of $n$.
Then the head $(j, \lambda_j +1)$ for $j \leq k-1$ has the color $\ell$. This hook gives a new factor
to $C_{\lambda + 1_k}$,  if the up-shifted hook is {\it not} a hook in the original diagram $\lambda$.
This takes place if $\lambda_{j-1} > \lambda_j$, that is, when we can add a box with color $\ell$
in the $j$-th row. Since the head and the tail of such a new hook are
$(j, \lambda_j +1)$ and $(k, \lambda_k +1)$, the new factor is
\begin{equation}
( 1 - q_1^{\lambda_j - \lambda_k -1} q_3^{j-k-1} ).
\end{equation}
Next, when $(k, \lambda_k +2)$ is the head and the hook length is a multiple of $n$,
we may have the tail $(j, \mu_j)$ for $k \leq j$ and $\lambda_{j+1} < \mu_j \leq \lambda_k$.
This time a new factor in $C_{\lambda + 1_k}$ appears, if the left-shifted hook is {\it not}
a hook in the original diagram $\lambda$. This is the case only when $\mu_j = \lambda_{j+1} +1$,
the minimum of allowed $\mu_j$. Since the tail $(j, \lambda_{j+1} +1)$ has the same color $\ell-1$
as $(k, \lambda_k +2)$, we can add a box $(j+1, \lambda_{j+1} +1)$ with color $\ell$ in the $(j+1)$-th row.
The new factor is
\begin{equation}
( 1 - q_1^{\lambda_k - \lambda_j} q_3^{k-j} ),
\end{equation}
where we have made a shift $j \to j-1$ so that $k+1 \leq j \leq \ell (\lambda) +1$.
On the other hand, thinking in the opposite way, we see that the hooks that cease to
contribute $C_\lambda(q_1, q_3)$ have either the head $(j,  \lambda_j +1)$ and the tail
$(k-1, \lambda_k +1)$ for $1 \leq j \leq k-1$, or the head $(k,  \lambda_k +1)$ and the tail
$(j, \lambda_j)$ for $k+1 \leq j \leq \ell(\lambda)$.  The selection rule is
that we can remove a box with the color $\ell$ from the $j$-th row.
The corresponding factors are
\begin{equation}
( 1 - q_1^{\lambda_j - \lambda_k -1} q_3^{j-k} ),
\end{equation}
for the former case and
\begin{equation}
( 1 - q_1^{\lambda_k - \lambda_j } q_3^{k-j-1} ).
\end{equation}
for the latter one. By taking above four factors with the color selection rule, we obtain \eqref{Lemma1a}.
Considering the difference between $C_\lambda(q_1, q_3)$ and $C^\prime_\lambda(q_1, q_3)$,
we see that \eqref{Lemma1b} follows from \eqref{Lemma1a} by a shift $q_1^{y} q_3^{x} \to q_2^{-1} q_1^{y} q_3^{x}$.

In the vertical representation, we take a basis $\{ \vert \lambda ) \}$ of the Fock space,
which simultaneously diagonalizes $K^{\pm}_\ell (z)$. Since the eigenvalues are non-degenerate,
the freedom is only the change of the norm of each eigenvector $\vert \lambda )$.
The normalization factors $C_\lambda(q_1, q_3)$ and $C^\prime_\lambda(q_1, q_3)$
are related to the relative normalization of $\vert \lambda )$.
In fact, if we define\footnote{This is motivated by the
formula for the norms of the Macdonald functions.}
\begin{equation}
( \lambda \vert \lambda )_{q_1,q_3} := \frac{C^\prime_\lambda(q_1, q_3)}
{C_\lambda(q_1, q_3)},
\end{equation}
the recursion relation for $( \lambda \vert \lambda )_{q_1,q_3}$ is
given by the matrix elements of the vertical representation as follows:
\begin{thm}
\begin{equation}
( \lambda \vert F_\ell(z) \vert \lambda + 1_k ) = \gq^{\# A_\ell^{(0)} - \# R_\ell^{(0)}}
\frac{(  \lambda + 1_k \vert \lambda + 1_k )_{q_1,q_3}}
 {(  \lambda \vert \lambda  )_{q_1,q_3}} ( \lambda + 1_k \vert E_\ell(z) \vert  \lambda ).
\end{equation}
\end{thm}
By using Lemma 1, the formula is easily checked by direct computation.
This is a generalization of Lemma 6.1 in \cite{AFS}.


\section{Zero mode factor of the intertwiner}

In this Appendix, we prove the lemmas concerning the zero mode part of the intertwiner.
First of all, we recall the zero mode algebra
\begin{align}
  k^\pm_i e_j(z) &= \gq^{\pm a_{i,j}} e_j(z) k^\pm_i, &
  k^\pm_i f_j(z) &= \gq^{\mp a_{i,j}} f_j(z) k^\pm_i, \\
  e_i(z) e_j(w) &= (w/z)^{-a_{i,j}} (-\gd)^{-m_{i,j}} e_j(w) e_i(z), &
  f_i(z) f_j(w) &= (w/z)^{-a_{i,j}} (-\gd)^{-m_{i,j}} f_j(w) f_i(z), \\
  e_i(z) f_j(w) &= (w/z)^{a_{i,j}} (-\gd)^{m_{i,j}} f_j(w) e_i(z), &
  f_i(z) e_j(w) &= (w/z)^{a_{i,j}} (-\gd)^{m_{i,j}} e_j(w) f_i(z),
\end{align}
\begin{align}
  \gq^{\pm1} e_i(\gq^{\pm1} z) f_i(z) = \gq^{\mp1} f_i(\gq^{\mp1} z) e_i(z) = k^\pm_i,
\end{align}
where $i,j \in \mathbb{Z}/ n \mathbb{Z}$, $a_{i,j} = 2\deltabar_{i,j} - \deltabar_{i-1,j} - \deltabar_{i+1,j}$ and $m_{i,j} = \deltabar_{i-1,j} - \deltabar_{i+1,j}$.
These are represented by
\begin{align}
  k^\pm_i \rightarrow \gq^{\pm \partial_{\alphabar_i}}, \quad
  e_i(z) \rightarrow e^{\alphabar_i} z^{H_{i,0}+1}, \quad
  f_i(z) \rightarrow e^{- {\alphabar_i}} z^{- H_{i,0}+1}
\end{align}
and relations \eqref{cocycle}--\eqref{Hzero}.
Then the first lemma is stated as below
\begin{thm}
\begin{align}
  z^{[k]}_{\lambda}(v)
  &= \prod_{1\leq i \leq \ell(\lambda)}^{\leftarrow} \left(
  \prod_{1\leq j \leq \lambda_i}^{\leftarrow}  e^{[k]}_{i,j}(v)
  \right), &
  e^{[k]}_{i,j}(v)
  &= e_{i-j+k}(q_1^{j-1} q_3^{i-1} v), \\
  z^{[k]*}_{\lambda}(v)
  &= \prod_{1\leq i \leq \ell(\lambda)}^{\leftarrow} \left(
  \prod_{1\leq j \leq \lambda_i}^{\leftarrow}  f^{[k]}_{i,j}(v)
  \right), &
  f^{[k]}_{i,j}(v)
  &= f_{i-j+k}(q_1^{j-1} q_3^{i-1} v) \\
\end{align}
satisfy
\begin{align}
  e_\ell(u) z^{[k]}_{\lambda}(v) &= \pi^{[k]}_{\lambda,\ell}(v/u) z^{[k]}_{\lambda}(v) e_\ell(u), &
  f_\ell(u) z^{[k]*}_{\lambda}(v) &= \pi^{[k]}_{\lambda,\ell}(v/u) z^{[k]*}_{\lambda}(v) f_\ell(u), \\
  e_\ell(u) z^{[k]*}_{\lambda}(v) &= \pi^{[k]}_{\lambda,\ell}(v/u)^{-1} z^{[k]*}_{\lambda}(v) e_\ell(u), &
  f_\ell(u) z^{[k]}_{\lambda}(v) &= \pi^{[k]}_{\lambda,\ell}(v/u)^{-1} z^{[k]}_{\lambda}(v) f_\ell(u), \\
  k^\pm_\ell z^{[k]}_{\lambda}(v) &=  (\tilde{\pi}^{[k]}_{\lambda,\ell})^{\pm1} z^{[k]}_{\lambda}(v) k^\pm_\ell, &
  k^\pm_\ell z^{[k]*}_{\lambda}(v) &=  (\tilde{\pi}^{[k]}_{\lambda,\ell})^{\mp1} z^{[k]*}_{\lambda}(v) k^\pm_\ell,
\end{align}
where
\begin{align}
  \pi^{[k]}_{\lambda,\ell}(z)
  &= \left( - \gq z \right)^{-\deltabar_{\ell,k}}
  \prod_{\substack{s=1, \\ s-\lambda_s+k \equiv \ell} }^{\ell(\lambda)} \left(- \gq^{-1} x_s z \right)^{-1}
  \prod_{\substack{s=1, \\ s-\lambda_s+k \equiv \ell+1}}^{\ell(\lambda)+1} \left(- \gq^{-1} q_3^{-1} x_s z \right), \
  x_s = q_1^{\lambda_s-1} q_3^{s-1} \\
  \tilde{\pi}^{[k]}_{\lambda,\ell}
  &= \gq^{\deltabar_{\ell,k}}
  \prod_{\substack{s=1, \\ s-\lambda_s+k \equiv \ell} }^{\ell(\lambda)} \gq
  \prod_{\substack{s=1, \\ s-\lambda_s+k \equiv \ell+1}}^{\ell(\lambda)+1} \gq^{-1},
\end{align}
and $[k]$ denotes the vacuum color $k \in \mathbb{Z}/ n \mathbb{Z}$.\footnote{We often omit the symbol $[k]$ when $k=0$.}
\end{thm}

Now we shall check the commutation relation between $e_\ell(u)$ and $z^{[k]}_\lambda(v)$ using the same idea as for the oscillator parts.
Thanks to the triplet cancellation in each row
\begin{align}
  e_\ell(u) \ e_{i,j+1}(v) e_{i,j}(v) e_{i,j-1}(v)
  = e_{i,j+1}(v) e_{i,j}(v) e_{i,j-1}(v) \ e_\ell(u), \quad \ell \equiv i-j,
\end{align}
we only have to consider the left- and right-most boxes for each row.
The factor which comes from the left-most box $(s,1)$ for $1\leq s \leq \ell(\lambda)$ is
\begin{align}
\begin{cases}
  -\gq q_3^{s} v/u
  & s+k \equiv \ell, \\
  \left( -\gq q_3^{s-1} v/u \right)^{-1}
  & s+k \equiv \ell+1.
\end{cases}
\end{align}
The non-trivial factors surviving after we take the product over the rows are $(-\gq v/u)^{-1}$ when $\ell \equiv k$ and $-\gq^{-1} q_3^{-1} x_{\ell(\lambda)+1} v/u$ when $\ell \equiv \ell(\lambda)$.
On the other hand, the factor which comes from the right-most box $(s,\lambda_s)$ for $1\leq s \leq \ell(\lambda)$ is
\begin{align}
\begin{cases}
  \left( -\gq^{-1} x_s v/u \right)^{-1}
  & s-\lambda_s+k \equiv \ell, \\
  -\gq^{-1} q_3^{-1} x_s v/u
  & s-\lambda_s+k \equiv \ell+1.
\end{cases}
\end{align}
Hence, gathering these factors, we get $\pi^{[k]}_{\lambda,\ell}(v/u)$.
The computation of other commutation relations can be performed in the same way, hence, we omit them here.

Consequently, we obtain the following results used in section 4.3.
For each $j$ we divide a partition $\lambda=(\lambda_1, \ldots, \lambda_{\ell(\lambda)})$ into two parts $\lambda^{\{ j+\}}=(\lambda_1, \ldots, \lambda_j)$, $\lambda^{\{ j-\}}=(\lambda_{j+1}, \ldots, \lambda_{\ell(\lambda)})$ so that
\begin{align}
  z_{\lambda}^{(j+)}(v)
  &:= \prod_{1\leq i \leq j}^{\leftarrow} \left(
  \prod_{1\leq k \leq \lambda_i}^{\leftarrow}  e^{[0]}_{i,k}(v)
  \right)
  = z^{[0]}_{\lambda^{\{ j+\}}}(v), \\
  z_{\lambda}^{(j-)}(v)
  &:= \prod_{j+1\leq i \leq \ell(\lambda)}^{\leftarrow} \left(
  \prod_{1\leq k \leq \lambda_i}^{\leftarrow}  e^{[0]}_{i,k}(v)
  \right)
  = z^{[j]}_{\lambda^{\{ j-\}}}(q_3^j v).
\end{align}
\begin{thm}
Under the condition $j-\lambda_j -1 \equiv \ell$, we have
\begin{align}
  e_{j,\lambda_j+1}(v) z_{\lambda}^{(j-)}(v)
  &= \pi^{[j]}_{\lambda^{\{ j-\}},\ell}(q_1^{-1}q_3^j x_j^{-1}) z_{\lambda}^{(j-)}(v) e_{j,\lambda_j+1}(v), \\
  z_{\lambda}^{(j+)}(v) e_{j,\lambda_j+1}(v)
  &= \left( \pi^{[0]}_{\lambda^{\{ j+\}},\ell}(q_1^{-1} x_j^{-1}) \right)^{-1} e_{j,\lambda_j+1}(v) z_{\lambda}^{(j+)}(v),
\end{align}
and under the condition $j-\lambda_j \equiv \ell$, we have
\begin{align}
  f_{j,\lambda_j}(\gq^{-1}v) z_{\lambda}^{(j-)}(v)
  &= \left( \pi^{[j]}_{\lambda^{\{ j-\}},\ell}(\gq q_3^j x_j^{-1}) \right)^{-1} z_{\lambda}^{(j-)}(v) f_{j,\lambda_j}(\gq^{-1}v), \\
  z_{\lambda}^{(j+)}(v) f_{j,\lambda_j}(\gq^{-1}v)
  &= \pi^{[0]}_{\lambda^{\{ j+\}},\ell}(\gq x_j^{-1}) f_{j,\lambda_j}(\gq^{-1}v) z_{\lambda}^{(j+)}(v), \\
  k^+_\ell z_{\lambda}^{(j+)}(v)
  &= \tilde{\pi}^{[0]}_{\lambda^{\{ j+\}},\ell} z_{\lambda}^{(j+)}(v) k^+_\ell, \\
  z_{\lambda}^{(j-)}(v) k^+_\ell
  &= \left( \tilde{\pi}^{[j]}_{\lambda^{\{ j-\}},\ell} \right)^{-1} k^+_\ell z_{\lambda}^{(j-)}(v).
\end{align}
We can also check the same result, if we replace $e \leftrightarrow f$, $z \leftrightarrow z^*$, $k^+ \leftrightarrow k^-$.
\end{thm}
We can write down the factors explicitly, for example,
\begin{align}
  &\pi^{[j]}_{\lambda^{\{ j-\}},\ell}(q_1^{-1}q_3^j x_j^{-1})
  = \left( -\gq q_1^{\lambda_j+1} \right)^{\deltabar_{\ell,j}}
  \prod_{\substack{s=j+1, \\ s-\lambda_s \equiv \ell} }^{\ell(\lambda)} \left(-\gq q_3 \frac{x_s}{x_j} \right)^{-1}
  \prod_{\substack{s=j+1, \\ s-\lambda_s \equiv \ell+1}}^{\ell(\lambda)+1} \left(-\gq \frac{x_s}{x_j} \right), \\
  &\pi^{[0]}_{\lambda^{\{ j+\}},\ell}(q_1^{-1} x_j^{-1})
  =  (-\gq) \left( -\frac{\gq}{q_1 x_j} \right)^{-\deltabar_{\ell,0}} \left( - \gq q_1^{\lambda_j+1} \right)^{-\deltabar_{\ell,j}}
  \prod_{\substack{s=1, \\ s-\lambda_s \equiv \ell} }^{j-1} \left(-\gq q_3 \frac{x_s}{x_j} \right)^{-1}
  \prod_{\substack{s=1, \\ s-\lambda_s \equiv \ell+1}}^{j-1} \left(-\gq \frac{x_s}{x_j} \right).
\end{align}


\section{Recursion relation for Nekrasov function}

In this Appendix, we give a proof of
\begin{equation}
G_{ \lambda \mu}(z) = N_{\lambda \mu} (z; q_1, q_3)
\end{equation}
for the Nekrasov function (bifundamental contribution) on $ALE_n \times S^1$,
\begin{equation}
N_{\lambda \mu} (z; q_1, q_3)
= \prod_{\substack{s \in \lambda \\
h_{\mu, \lambda}(s) \equiv 0}} ( 1- z q_1^{a_\lambda(s)} q_3^{- \ell_\mu(s) -1})
 \prod_{\substack{t \in \mu \\ h_{\lambda, \mu}(t) \equiv 0}} ( 1- z q_1^{-a_\mu(t)-1} q_3^{\ell_\lambda(t)})
\label{ALEappendix}
\end{equation}
by obtaining a recursion relation for $N_{\lambda \mu} (z; q_1, q_3)$.

\begin{thm}
The Nekrasov factor $N_{\lambda \mu} (z; q_1, q_3)$ satisfies the following recursion relation
\begin{align}
  &N_{\lambda \mu+1_j}(z)/N_{\lambda \mu}(z)  \CR
  &= \prod_{\substack{i=1, \\ i-\lambda_i \equiv \ell}}^{\ell(\lambda)} (1- z q_1^{\lambda_i-\mu_j-1} q_3^{i-j})^{-1}
  \prod_{\substack{i=1, \\ i-\lambda_i \equiv \ell+1}}^{\ell(\lambda)+1} (1- z q_1^{\lambda_i-\mu_j-1} q_3^{i-j-1}),
  \label{Nrecursion}
\end{align}
where $\ell = j - \mu_j -1$ is the color of the added box $(j, \mu_j +1)$.
\end{thm}

To obtain the recursion relation, we can proceed in the way similar to the proof of Lemmas 1 and 2 in Appendix A.
However, since a pair of the Young diagrams is involved, the argument necessarily becomes more sophisticated.
As the Nekrasov factor \eqref{ALEG} consists of two parts, let us consider each part separately.
\begin{enumerate}
\item $\prod_{s\in\lambda}$-part.

In this case, the condition $h_{\mu,\lambda}(s) = a_\lambda(s) + \ell_\mu(s) +1\equiv 0$ is imposed.
Only the leg length $\ell_\mu(s)$ may change, when we add the box $(j, \mu_j +1)$ to $\mu$.
Hence, suppose $s=(i,\mu_j+1)\in\lambda$ for $1\leq i\leq \ell(\lambda)$
satisfying $\mu_j+1\leq\lambda_i$. The corresponding relative hook is the left one in \eqref{hooktail},
\begin{align}
  \begin{ytableau}
  s & \cdots & x \\
  \vdots \\
  y
  \end{ytableau} \
  \begin{cases}
  x= (i, \lambda_i) \\
  y= (j, \mu_j+1)
  \end{cases}, \quad
  \begin{ytableau}
  s & \cdots & x \\
  \vdots \\
  y
  \end{ytableau} \
  \begin{cases}
  x= (i, \lambda_i) \\
  y= (j-1, \mu_j+1)
  \end{cases}.
  \label{hooktail}
\end{align}
Then the new factor $(1- z q_1^{a_\lambda(s)} q_3^{-\ell_\mu(s)-1})=(1- z q_1^{\lambda_i-\mu_j-1} q_3^{i-j-1})$
appears in $N_{\lambda\mu+1_j}$ if $h_{\mu+1_j,\lambda}(s) = a_\lambda(s) + \ell_\mu(s) +1\equiv 0$,
which is equivalent to $i-\lambda_i \equiv \ell+1$.
Conversely, for the right hook in \eqref{hooktail},
the old factor $(1- z q_1^{a_\lambda(s)} q_3^{-\ell_\mu(s)-1})=(1- z q_1^{\lambda_i-\mu_j-1} q_3^{i-j})$
ceases to contribute $N_{\lambda\mu}$ if $h_{\mu,\lambda}(s) \equiv 0$, which is equivalent to $i-\lambda_i \equiv \ell$.

\item $\prod_{t\in\mu}$-part.

This time the constraint is $h_{\lambda,\mu}(t) = a_\mu(t) + \ell_\lambda(t) +1\equiv 0$,
where only the arm length $a_\mu(t)$ may change,  and
there is the new box $(j, \mu_j +1) \in \mu$. In any case, it is enough to
consider $t=(j,\lambda_{i+1}+1)\in\mu+1_j$ for $0\leq i\leq \ell(\lambda)$
satisfying $\lambda_{i+1} \leq \mu_j$ and $\lambda_i \neq \lambda_{i+1}$\footnote{We set $\lambda_0 =0$.},
see the left hook in \eqref{hookhead},
\begin{align}
  \begin{ytableau}
  t & \cdots & x \\
  \vdots \\
  y
  \end{ytableau} \
  \begin{cases}
  x= (j, \mu_j+1) \\
  y= (i, \lambda_{i+1}+1)
  \end{cases}, \quad
  \begin{ytableau}
  t & \cdots & x \\
  \vdots \\
  y
  \end{ytableau} \
  \begin{cases}
  x= (j, \mu_j) \\
  y= (i, \lambda_{i}).
  \end{cases}
  \label{hookhead}
\end{align}
Then the new factor $(1- z q_1^{-a_\mu(t)-1} q_3^{\ell_\lambda(t)})
=(1- z q_1^{\lambda_{i+1}-\mu_j-1} q_3^{i-j})$
appears in $N_{\lambda\mu+1_j}$ if $h_{\lambda,\mu+1_j}(t) \equiv 0$,
which is equivalent to $i+1-\lambda_{i+1} \equiv \ell+1$.
Conversely, for the right hook in \eqref{hookhead}, set $t=(j,\lambda_i)\in\mu$ for $1\leq i\leq \ell(\lambda)$
satisfying $\lambda_i\leq\mu_j$ and $\lambda_i \neq \lambda_{i+1}$,
then the old factor $(1- z q_1^{-a_\mu(t)-1} q_3^{\ell_\lambda(t)})=(1- z q_1^{\lambda_i-\mu_j-1} q_3^{i-j})$
disappears if $h_{\lambda,\mu}(t) \equiv 0$, which is equivalent to $i-\lambda_i \equiv \ell$.
Note we can take these factors even when $\lambda_i = \lambda_{i+1}$ due to the cancellation.
\end{enumerate}
Thus we see that the changes for $\lambda_{i} \leq \mu_j (\lambda_{i+1} \leq \mu_j)$ come from the
$\prod_{t\in\mu}$-part and those for $\mu_j +1 \leq \lambda_{i} $ come from the $\prod_{s\in\lambda}$-part.
Combining these two contributions, we arrive at the recursion relation \eqref{Nrecursion}.


\section{Symmetry of Nekrasov function}

In this Appendix, we prove the following relation for the Nekrasov function:
\begin{thm}
\begin{align}
  N_{\lambda\mu} (\gq^{-1} u/v) z^*_{\mu}(v) z_{\lambda}(u)
  = \left( \frac{u}{v} \right)^{|\lambda|_0 + |\mu|_0} \frac{f_\lambda(q_1,q_3)}{f_\mu(q_1,q_3)}
  N_{\mu\lambda} (\gq^{-1} v/u ) z_{\lambda}(u) z^*_{\mu}(v),
\label{appsymmetry}
\end{align}
where $N_{\lambda\mu}(z)$ is the Nekrasov function on $ALE_n \times S^1$ \eqref{ALEG},
$f_\lambda$ is the generalized framing factor \eqref{framing} and
\begin{eqnarray}
  z_{\lambda}(u)
  &=& \prod_{1\leq i \leq \ell(\lambda)}^{\leftarrow} \left(
  \prod_{1\leq j \leq \lambda_i}^{\leftarrow}  e_{i,j}(u)
  \right),
 \qquad
  e_{i,j}(u) = e^{\alphabar_{i-j} } ~(q_1^{j-1} q_3^{i-1} u)^{H_{i-j,0 }+1}.
  \label{eorder} \\
  z^*_{\mu}(v)
  &=& \prod_{1\leq i \leq \ell(\mu)}^{\leftarrow} \left(
  \prod_{1\leq j \leq \mu_i}^{\leftarrow}  f_{i,j}(v)
  \right),
  \qquad
  f_{i,j}(v) = e^{-\alphabar_{i-j} } ~(q_1^{j-1} q_3^{i-1} v)^{-H_{i-j,0 }+1},
 \label{forder}
\end{eqnarray}
are the group algebra parts of the (dual) intertwiner.
\end{thm}

Since a direct computation leads us to
\begin{align}
  \frac{N_{\lambda \mu} (\gq^{-1} u/v)}{N_{\mu \lambda} (\gq^{-1} v/u)}
  =
  \prod_{\substack{s \in \lambda \\ h_{\mu, \lambda}(s) \equiv 0}} (-\gq^{-1}q_1^{a_\lambda(s)} q_3^{- \ell_\mu(s) -1} u/v)
  \prod_{\substack{t \in \mu \\ h_{\lambda, \mu}(t) \equiv 0}} (-\gq^{-1} q_1^{-a_\mu(t)-1} q_3^{\ell_\lambda(t)} u/v),
\end{align}
it suffices to show that
\begin{equation}
  z^*_{\mu}(v) z_{\lambda}(u) z^*_{\mu}(v)^{-1} z_{\lambda}(u)^{-1} = G_{\mu\lambda}^{(0)} (u/v),
  \label{zcom}
\end{equation}
where
\begin{equation}
G_{\mu\lambda}^{(0)} (z) = z^{|\lambda|_0+|\mu|_0} \frac{f_\lambda}{f_\mu}
  \prod_{\substack{s \in \lambda \\ h_{\mu, \lambda}(s) \equiv 0}} (-\gq^{-1}q_1^{a_\lambda(s)} q_3^{- \ell_\mu(s) -1} z)^{-1}
  \prod_{\substack{t \in \mu \\ h_{\lambda, \mu}(t) \equiv 0}} (-\gq^{-1} q_1^{-a_\mu(t)-1} q_3^{\ell_\lambda(t)} z)^{-1}.
  \label{G0}
\end{equation}
Since \eqref{zcom} is trivially satisfied, when $\mu = \varnothing$, we can take the same strategy which we used in Appendix C.
Thus we first derive a recursion relation for the left hand side of \eqref{zcom}.
From \eqref{forder}, we see that $z^*_{\mu + 1_j}(v)/ z^*_{\mu}(v) \sim   f_{j,\mu_j+1}(v)$.
Hence, the desired recursion relation follows from the relation
\begin{align}
  f_{j,\mu_j+1}(v) z_{\lambda}(u)
  = \left( -\gq q_1^{-\mu_j} q_3^{-j+1} u/v \right)^{\deltabar_{\ell,0}}
  \frac{\displaystyle{\prod_{\substack{i=1, \\ i-\lambda_i \equiv \ell}}^{\ell(\lambda)}}
  (-\gq^{-1} q_1^{\lambda_i-\mu_j-1} q_3^{i-j} u/v)}
  {\displaystyle{\prod_{\substack{i=1, \\ i-\lambda_i \equiv \ell+1}}^{\ell(\lambda)+1}}
  (-\gq^{-1} q_1^{\lambda_i-\mu_j-1} q_3^{i-j-1} u/v)}
  z_{\lambda}(u) f_{j,\mu_j+1}(v),
  \label{zrecursion}
\end{align}
where $\ell = j - \mu_j -1$.  We can check \eqref{zrecursion} in the way similar to the proof of Lemma $4$ in Appendix B.
Then by the same argument as in Appendix C where we translated the color selection rule in the right hand side of
\eqref{zrecursion} to the condition on the relative hook length, we see that the right hand side of \eqref{G0}
satisfies the same recursion relation. This completes the proof of Lemma 6.


The function $G_{\mu\lambda}^{(0)} (z) $ universally appears in the commutation relations among
the intertwiner $\Phi^\lambda$ and the dual intertwiner $\Phi_\mu^{*}$, since all the commutation relations
of the zero modes can be expressed in terms of $G_{\mu\lambda}^{(0)} (z)$ as follows:
\begin{eqnarray}
z^*_{\mu}(v) z_{\lambda}(u) &=& G_{\mu\lambda}^{(0)} (u/v) z_{\lambda}(u) z^*_{\mu}(v), \qquad
z_{\mu}(v) z_{\lambda}^{*} (u) = G_{\mu\lambda}^{(0)} (u/v) z_{\lambda}^{*} (u)z_{\mu}(v), \\
z_{\mu}(v) z_{\lambda}(u) &=& (G_{\mu\lambda}^{(0)} (u/v))^{-1}  z_{\lambda}(u) z_{\mu}(v), \qquad
z_{\mu}^{*} (v) z_{\lambda}^{*} (u) = (G_{\mu\lambda}^{(0)} (u/v))^{-1}  z_{\lambda}(u) z_{\mu}(v).
\end{eqnarray}
Using these relations,  we can see
\begin{align}
  N_{\lambda\mu} (\gq^{-1} u/v) z_{\mu}(v) z^*_{\lambda}(u)
  = \left( \frac{u}{v} \right)^{|\lambda|_0 + |\mu|_0} \frac{f_\lambda}{f_\mu}
  N_{\mu\lambda} (\gq^{-1} (u/v)^{-1}) z^*_{\lambda}(u) z_{\mu}(v),
\end{align}
\begin{align}
  &N_{\lambda\mu} (\gq^{-2} u/v)^{-1} z_{\mu}(v) z_{\lambda}(u)
  = \left( \left( \frac{u}{\gq v} \right)^{|\lambda|_0 + |\mu|_0} \frac{f_\lambda}{f_\mu}
  N_{\mu\lambda} ( \gq^{-1} (u/\gq v)^{-1}) \right)^{-1} z_{\mu}(v) z_{\mu}(\gq v)^{-1} z_{\lambda}(u) z_{\mu}(\gq v)  \\
  &= \left( \frac{u}{\gq v} \right)^{-|\lambda|_0 - |\mu|_0} \frac{f_\mu}{f_\lambda}
  N_{\mu\lambda} (v/u)^{-1} z_{\lambda}(u) z_{\mu}(v)
  \times \gq^{-|\lambda|_0-|\mu|_0} \gq^{H(\lambda,\mu)} \\
  &= \left( \frac{u}{v} \right)^{-|\lambda|_0 - |\mu|_0} \frac{f_\mu}{f_\lambda}
  N_{\mu\lambda} (v/u)^{-1} z_{\lambda}(u) z_{\mu}(v)
  \times \gq^{H(\lambda,\mu)},
\end{align}
\begin{align}
  &N_{\lambda\mu} (u/v)^{-1} z^*_{\mu}(v) z^*_{\lambda}(u)
  = \left( \left( \frac{\gq u}{v} \right)^{|\lambda|_0 + |\mu|_0} \frac{f_\lambda}{f_\mu}
  N_{\mu\lambda} ( \gq^{-1} (\gq u/v)^{-1}) \right)^{-1} z^*_{\lambda}(\gq u) z^*_{\mu}(v) z^*_{\lambda}(\gq u)^{-1} z^*_{\lambda}(u)  \\
  &= \left( \frac{\gq u}{v} \right)^{-|\lambda|_0 - |\mu|_0} \frac{f_\mu}{f_\lambda}
  N_{\mu\lambda} (\gq^{-2} v/u)^{-1} z^*_{\lambda}(u) z^*_{\mu}(v)
  \times \gq^{|\lambda|_0+|\mu|_0} \gq^{-H(\lambda,\mu)} \\
  &= \left( \frac{u}{v} \right)^{-|\lambda|_0 - |\mu|_0} \frac{f_\mu}{f_\lambda}
  N_{\mu\lambda} (\gq^{-2} v/u)^{-1} z^*_{\lambda}(u) z^*_{\mu}(v)
  \times \gq^{-H(\lambda,\mu)},
\end{align}
where
\begin{align}
  H(\lambda,\mu) = \#\{ s\in\lambda|h_{\mu,\lambda}(s)\equiv0\} + \#\{ t\in\mu|h_{\lambda,\mu}(t)\equiv0\}.
\end{align}


\section{From colored Young diagrams to quotients }
\label{sec:from-colored-young}
In this Appendix, we first introduce the notion of the quotient of the
Young diagram. We then use some combinatorial identities to
express the characters of colored Young diagrams in terms of their
quotients.

\subsection{Quotients of the Young diagram}
\label{sec:quot-young-diagr}
Let $Y$ be a Young diagram and $N$ a natural number. Then $Y$
determines the $N$-tuple of Young diagrams $\{ Y^{(0)}, \ldots,
Y^{(N-1)}\}$ called \emph{quotients} and the vector of integer-valued
\emph{shifts} $\{p_0, \ldots, p_{N-1}\}$, satisfying $\sum_{c=0}^{N-1}
p_c = 0$. The correspondence is described in steps:
\begin{enumerate}
\item \textbf{Transformation into Maya diagram.} By the boson-fermion
  correspondence, the Young diagram determines the Maya diagram
  specifying the fermionic state. This state consists of a Dirac sea
  of electrons with momenta $k=-\frac{1}{2},-\frac{3}{2},\ldots$ plus
  an equal number of electrons and holes, with momenta given by the
  Frobenius coordinates $d_i$, $d_i^{*}$ of $Y$:
\begin{gather}
  \text{holes at: } d_i = Y_i-i -
  \frac{1}{2},\qquad i=1,\ldots, n(Y),\\
  \text{electrons at: } d_i^{*} = i - Y^{\mathrm{T}}_i +
  \frac{1}{2},\qquad i=1,\ldots, n(Y),
\end{gather}
where $n(Y)$ is the length of the diagonal of $Y$.

\emph{Example:} For the diagram $Y=[3,2]$, the holes are at $d =
\{ -\frac{5}{2}, - \frac{1}{2}\}$, and the electrons are at $d^{*} =
\{ \frac{1}{2}, \frac{3}{2} \}$.

\item \textbf{Division of the momenta lattice.} The lattice of
  fermionic momenta is divided into $N$ subsectors labelled by the
  \emph{color} $c=0,\ldots, (N-1)$. Subsector $c$ contains the
  electrons and holes with momenta $k$ such that $k - \frac{1}{2}
  \equiv c \mod N$.
  The general formula for
  the momenta of electrons and holes in subsector $c$ is given by
  \begin{gather}
    d^{(c)} = \left\{ \frac{1}{N} \left(d_i - \frac{1}{2} - c \right)
      + \frac{1}{2} \quad \bigg| \quad d_i - \frac{1}{2} \equiv c \mod
      N, \quad i=1,\ldots , n(Y) \right\},\\
    d^{*(c)} = \left\{ \frac{1}{N} \left(d^{*}_i - \frac{1}{2} - c \right)
      + \frac{1}{2} \quad \bigg|\quad d^{*}_i - \frac{1}{2} \equiv c \mod
      N, \quad i=1,\ldots , n(Y) \right\}.
  \end{gather}

  \emph{Example:} Let $Y=[3,2]$ as in the example above and $N=3$,
  then we get three subsectors with $c=0,1,2$:
  \begin{align}
    c=0&: \qquad d^{(0)}= \left\{ -\frac{1}{2} \right\}, \quad
    d^{(0)*}=\left\{ \frac{1}{2} \right\},\notag\\
    c=1&: \qquad d^{(1)}= \varnothing , \quad
    d^{(1)*}=\left\{ \frac{1}{2} \right\}, \label{eq:2a}\\
    c=2&: \qquad d^{(2)}= \left\{ -\frac{1}{2} \right\}, \quad
    d^{(2)*}=\varnothing.\notag
  \end{align}

\item \textbf{Shifting the vacuum charge.} The collection of electrons
  and holes from each subsector determines a fermionic state. However,
  though the total number of electrons is equal to the total number of
  holes, their numbers might not match in each subsector
  separately. Equivalently, the Dirac seas in the subsectors have
  different levels, i.e.\ the states may have nonzero vacuum charge. We
  denote the \emph{negative} value\footnote{In other words, $p_c$ denotes the
    value of the momentum shift needed to eliminate the vacuum
    charge. Of course, one can use the convention, where the sign of
    $p_c$ is reversed.} of this vacuum charge by $p_c$:
  \begin{equation}
    p_c = \# \left\{ d_i^{(c)} \right\} - \# \left\{ d_i^{(c)*} \right\}.
  \end{equation}
  Notice that the sum of all $p_c$ vanishes since the charge of the
  original vacuum state is zero. Eliminating the vacuum charges in
  each sector, one can transform the corresponding fermionic states
  into \emph{quotient} Young diagrams $Y^{(c)}$.

  \emph{Example:} Let $Y=[3,2]$ and $N=3$. The subsectors are listed
  in Eq.~(\ref{eq:2a}). The shifts read:
  \begin{align}
    c=0&: \qquad p_0 = 0,\notag\\
    c=1&: \qquad p_1 = -1,\\
    c=2&: \qquad p_2 = 1.\notag
  \end{align}
  Finally, the collection of quotient Young diagrams together with
  shifts is
  \begin{equation}
    Y = [3,2] \quad \Leftrightarrow \quad
    \begin{array}{c}
      Y^{(c)} = \{
      [1],\varnothing,\varnothing \},\\
      p_c = \{ 0 , -1, 1 \}
    \end{array}
  \end{equation}
\end{enumerate}

One can easily write down the relation between the total number of
boxes in the original Young diagram and its quotients:
\begin{equation}
  |Y| = \sum_{c=0}^{N-1} \left( N |Y^{(c)}| + \frac{N}{2} p_c^2 - c p_c \right).
\end{equation}
In the next subsection, we work out a more general relation between the
character of the Young diagram and its quotients.

\subsection{Decomposing characters}
\label{sec:decomp-char}
The (uncolored) character of the Young diagram is defined as follows:
\begin{equation}
  \label{eq:7}
  \ch_Y(q) = \sum_{(i,j)\in Y} q^{j-i}.
\end{equation}
After substituting $q=e^{\hbar}$ and expanding in $\hbar$, Eq.~(\ref{eq:7})
actually gives the character of the $\mathfrak{gl}_{\infty}$
representation associated with the Young diagram $Y$. We also have
\begin{equation}
  \label{eq:8}
  \ch_Y(1) = |Y|.
\end{equation}
We would like to rewrite the character as a manifest function of the
quotient diagrams $Y^{(c)}$ and of the shifts $p_c$. This, indeed, can be
done and we obtain the following expression:
\begin{equation}
  \label{eq:9}
  \ch_Y(q) = \frac{1 - q^N}{1-q} \sum_{c=0}^{N-1} q^{N p_c - c}
  \left( \ch_{Y^{(c)}}\left( q^N \right) - \frac{1}{1-q^{-N}} \frac{1
      - q^{-Np_c}}{1 - q^N} \right).
\end{equation}

One can also introduce the \emph{colored} character
\begin{equation}
  \label{eq:10}
  \ch^{(c)}_Y (q) = \sum_{
    \begin{smallmatrix}
      (i,j)\in Y\\
      i-j \equiv c \mod N
    \end{smallmatrix}
} q^{j-i}.
\end{equation}
Naturally
\begin{equation}
  \label{eq:11}
  \ch_Y (q) = \sum_{c=0}^{N-1} \ch^{(c)}_Y (q).
\end{equation}
The colored characters can also be expressed in terms of the quotient
Young diagrams:
\begin{equation}
  \label{eq:12}
  \ch^{(c)}_Y (q)  =  \sum_{d=0}^{N-1} q^{-N \lfloor \frac{d-c}{N}
    \rfloor - c +d} q^{Np_d - d} \left( \ch_{Y^{(d)}}\left( q^N \right) -
    \frac{1}{1-q^{-N}} \frac{1- q^{-N p_d}}{1 - q^N}\right),
\end{equation}
where $\lfloor x \rfloor$ denotes the floor function of $x$. One
observes that the colored characters $\ch^{(c)}_Y(q)$ are linear
combinations of $\ch_{Y^{(c)}}(q)$ with very special coefficients
forming a matrix $L$:
\begin{equation}
  \label{eq:15}
    L_{cd}(q) = q^{-N \lfloor \frac{d-c}{N}
    \rfloor - c + d}.
\end{equation}
The matrix $L$ turns out to have a particularly nice inverse:
\begin{gather}
  \label{eq:13}
  L(q)^{-1} =  \frac{1}{1 - q^N} \left(
  \begin{array}{cccccc}
    1 &-q & 0 & \cdots &  0\\
    0 & 1 &-q & \ddots &   \vdots\\
    0 & 0 & 1 & \ddots & 0 \\
    0 &  &  & \ddots  &  -q\\
    -q& 0 & \cdots &  0  & 1\\
  \end{array}
\right),\\
\end{gather}
or, in the index notation,
\begin{equation}
  \label{eq:14}
  (L(q)^{-1})_{cd} = \frac{\delta_{cd} - q \bar{\delta}_{c,d-1}}{1-q^N},
\end{equation}
where $\bar{\delta}_{cd}$ is the Kronecker symbol modulo $N$. In
sec.~\ref{sec:abel-nonab-intertw}, we used this result for the
inverse matrix in order to transform the expression for the non-Abelian DIM intertwiner
into a product of commuting intertwiners.




\begin{thebibliography}{99}

\bibitem{CFT1} A. Belavin, A. Polyakov and A. Zamolodchikov, Nucl. Phys. {\bf B241} (1984) 333-380

\bibitem{CFT2}
A. Zamolodchikov and Al. Zamolodchikov, {\sl Conformal field theory and critical phenomena in 2d systems}, (MCCME, Moscow, Russia), ISBN 978-5-94057-520-7, 2009

\bibitem{CFT3}
L. Alvarez-Gaume, Helvetica Physica Acta {\bf 64} (1991) 359-526

\bibitem{CFT4}
P. Di Francesco, P. Mathieu and D. Senechal, {\sl Conformal Field Theory}, Springer, 1996

\bibitem{AGT1} L.~Alday, D.~Gaiotto and Y.~Tachikawa,
  Lett.\ Math.\ Phys.\ {\bf 91} (2010) 167-197, arXiv:0906.3219

\bibitem{AGT2}
  N.~Wyllard,
  JHEP {\bf 0911} (2009) 002, arXiv:0907.2189

\bibitem{AGT3}
  A.~Mironov and A.~Morozov, Nucl.\ Phys.\ {\bf B825} (2009) 1-37,
  arXiv:0908.2569

\bibitem{AGT5d1}
H.~Awata and H.~Kanno, JHEP  {\bf 0505} (2005) 039,
arXiv: hep-th/0502061

\bibitem{AGT5d2}
H.~Awata and Y.~Yamada,
  JHEP {\bf 1001} (2010) 125,
arXiv:0910.4431

\bibitem{AGT5d3}
H.~Awata and Y.~Yamada,
  Prog.\ Theor.\ Phys.\  {\bf 124} (2010) 227,
arXiv:1004.5122

\bibitem{AGT5d4}
S. Yanagida, arXiv:1005.0216

\bibitem{AGT5d5}
H.~Awata, H.~Fuji, H.~Kanno, M.~Manabe and Y.~Yamada,
  Adv.\ Theor.\ Math.\ Phys.\  {\bf 16} (2012) no.3,  725
arXiv:1008.0574

\bibitem{AGT5d6}
H.~Kanno and Y.~Tachikawa,
  JHEP {\bf 1106} (2011) 119,
arXiv:1105.0357

\bibitem{AGT5d7}
A. Mironov, A. Morozov, S. Shakirov and A. Smirnov, Nucl. Phys. {\bf B855} (2012) 128, arXiv:1105.0948

\bibitem{AGT5d8}
H.~Kanno and M.~Taki,
  JHEP {\bf 1205} (2012) 052
  arXiv:1203.1427

\bibitem{AGT5d9}
F. Nieri, S. Pasquetti and F. Passerini, arXiv:1303.2626

\bibitem{AGT5d10}
  F. Nieri, S. Pasquetti, F. Passerini and A. Torrielli, arXiv:1312.1294

\bibitem{AGT5d11}
M.-C. Tan, JHEP {\bf 12} (2013) 031, arXiv:1309.4775; arXiv:1607.08330

\bibitem{AGT5d12}
H. Itoyama, T.Oota and R. Yoshioka,arXiv:1602.01209

\bibitem{AGT5d13}
  A. Nedelin and M. Zabzine, arXiv:1511.03471

\bibitem{AGT5d14}
  R. Yoshioka, arXiv:1512.01084

\bibitem{AGT5d15}
    Y.~Ohkubo, H.~Awata and H.~Fujino,
  arXiv:1512.08016

\bibitem{AGT5d16}
S. Pasquetti, arXiv:1608.02968

\bibitem{SW1} N. Seiberg and E. Witten,
Nucl. Phys. {\bf B426} (1994) 19-52, hep-th/9407087

\bibitem{SW2} N. Seiberg and E. Witten,
Nucl. Phys. {\bf B431} (1994) 484-550, hep-th/9408099

\bibitem{GKMMM1} A. Gorsky, I. Krichever, A. Marshakov, A. Mironov and A. Morozov,
Phys.Lett. {\bf B355} (1995) 466, hep-th/9505035

\bibitem{GKMMM2}
R. Donagi and E. Witten, Nucl. Phys. {\bf B460} (1996) 299-334, hep-th/9510101

\bibitem{GKMMM3}
E. Martinec,
Phys. Lett. {\bf B367} (1996) 91-96, hep-th/9510204

\bibitem{GKMMM4}
E. Martinec and N. Warner,
Nucl. Phys. {\bf 459} (1996) 97, hep-th/9511052

\bibitem{GM} A.~Gorsky and A.~Mironov,
  In: {\sl Aratyn, H. (ed.) et al.: Integrable hierarchies and modern physical theories}, 33-176, hep-th/0011197

\bibitem{Nek1} N. Nekrasov, Adv. Theor. Math. Phys. {\bf 7} (2004) 831-864, hep-th/0206161

\bibitem{Nek2}
R. Flume and R. Pogossian, Int. J. Mod. Phys. {\bf A18} (2003) 2541, hep-th/0208176

\bibitem{Nek3}
N.Nekrasov and A.Okounkov, hep-th/0306238

\bibitem{WZW1}  J. Wess and B. Zumino, Phys. Lett. {\bf B37} (1971) 95

\bibitem{WZW2}
S. Novikov, Usp. Mat. Nauk {\bf 37} (1982) 3

\bibitem{WZW3}
E. Witten, 
Comm. Math. Phys. {\bf 92} (1984) 455

\bibitem{WZW4}
A. Polyakov and P. Wiegmann, 
Phys. Lett. {\bf B131} (1983) 121

\bibitem{WZW5}
A. Polyakov and P. Wiegmann, 
Phys. Lett. {\bf B141} (1984) 223

\bibitem{DF1} B.Feigin and D.Fuks, Funct. Anal. Appl. {\bf 16} (1982)
114-126 (Funkt. Anal. Pril. {\bf 16} (1982) 47-63)

\bibitem{DF2}
Vl. Dotsenko and V. Fateev, Nucl. Phys. {\bf B240} (1984) 312-348

\bibitem{GMMOS1} M. Wakimoto, 
Commun. Math. Phys. {\bf 104} (1986) 605-609

\bibitem{GMMOS2}
A. Gerasimov, A. Marshakov, A. Morozov, M. Olshanetsky and S. Shatashvili,
Int. J. Mod. Phys. A5 (1990) 2495-2589

\bibitem{GMMOS3}
V. Dotsenko, 
Nucl. Phys. {\bf B338} 747 (1990); 
Nucl. Phys. {\bf B358} (1991) 547

\bibitem{GMMOS4}
B. Feigin and E. Frenkel 
Phys. Lett. {\bf B246} (1990) 75-81

\bibitem{MMSh1} A. Mironov, A. Morozov and Sh. Shakirov,
JHEP {\bf 02} (2010) 030, arXiv:0911.5721

\bibitem{MMSh2} A. Mironov, A. Morozov and Sh. Shakirov,
Int. J. Mod. Phys. {\bf A25} (2010) 3173-3207, arXiv:1001.0563

\bibitem{MMSh3} A. Mironov, A. Morozov and Sh. Shakirov,
JHEP {\bf 1103} (2011) 102, arXiv:1011.3481

\bibitem{MZ} A.~Morozov and Y.~Zenkevich,
JHEP {\bf 1602} (2016) 098,  arXiv:1510.01896

\bibitem{MMZ}
A.~Mironov, A.~Morozov and Y.~Zenkevich, Phys. Lett. {\bf B756} (2016) 208-211, arXiv:1512.06701

\bibitem{Mironov:2016cyq}
  A.~Mironov, A.~Morozov and Y.~Zenkevich,
JHEP {\bf 1605}, 121 (2016), arXiv:1603.00304

\bibitem{Mironov:2016yue}
  A.~Mironov, A.~Morozov and Y.~Zenkevich,
Phys.\ Lett.\ B {\bf 762}, 196 (2016), arXiv:1603.05467

\bibitem{Awata:2016riz}
  H.~Awata, H.~Kanno, T.~Matsumoto, A.~Mironov, A.~Morozov, A.~Morozov, Y.~Ohkubo and Y.~Zenkevich,
  JHEP {\bf 1607} (2016)  103,
arXiv:1604.08366

\bibitem{Awata:2016mxc}
  H.~Awata, H.~Kanno, A.~Mironov, A.~Morozov, A.~Morozov, Y.~Ohkubo and Y.~Zenkevich,
JHEP {\bf 1610} (2016) 047, arXiv:1608.05351

\bibitem{Awata:2016bdm}
  H.~Awata, H.~Kanno, A.~Mironov, A.~Morozov, A.~Morozov, Y.~Ohkubo and Y.~Zenkevich,
Nucl.Phys. {\bf B918} (2017) 358, arXiv:1611.07304

\bibitem{Awata:2017cnz}
  H.~Awata, H.~Kanno, A.~Mironov, A.~Morozov, A.~Morozov, Y.~Ohkubo and Y.~Zenkevich,
 Phys.Rev. {\bf D96} (2017) 026021, arXiv:1703.06084

\bibitem{Matsuo1} J.-E. Bourgine, Y. Matsuo and H. Zhang, arXiv:1512.02492

\bibitem{Matsuo2}
  J.~E.~Bourgine, M.~Fukuda, Y.~Matsuo, H.~Zhang and R.~D.~Zhu,
  arXiv:1606.08020

 \bibitem{Matsuo3}
J.~E.~Bourgine, M.~Fukuda, K.~Harada, Y.~Matsuo and R.~D.~Zhu,
  JHEP {\bf 1711} (2017) 034,
  arXiv:1703.10759

\bibitem{Matsuo4}
 M.~Fukuda, K.~Harada, Y.~Matsuo and R.~D.~Zhu,
  PTEP {\bf 2017}, no. 9 (2017) 093A01,
  arXiv:1705.02941

 \bibitem{Matsuo5}
  J.~E.~Bourgine, M.~Fukuda, Y.~Matsuo and R.~D.~Zhu,
  JHEP {\bf 1712} (2017) 015,
  arXiv:1709.01954

\bibitem{KP1} T.~Kimura and V.~Pestun,
  arXiv:1512.08533

\bibitem{KP2} T.~Kimura and V.~Pestun,
arXiv:1608.04651

\bibitem{KP3} T.~Kimura and V.~Pestun, arXiv:1705.04410

\bibitem{O1} A. Okounkov, 1512.07363

\bibitem{O2}
A. Okounkov and A. Smirnov, 1602.09007

\bibitem{O3}
A. Smirnov, 1612.01048

\bibitem{DI} J. Ding and K. Iohara, 
Lett. Math. Phys. {\bf 41} (1997) 181-193, q-alg/9608002

\bibitem{Miki} K. Miki, J. Math. Phys. {\bf 48} (2007) 123520

\bibitem{DIM1} B. Feigin and A. Tsymbaliuk, Kyoto J. Math. {\bf 51} (2011) 831-854, arXiv:0904.1679

\bibitem{FFJMM} B. Feigin, E. Feigin, M. Jimbo, T. Miwa and E. Mukhin, Kyoto J. Math. {\bf 51} (2011) 337-364, arXiv:1002.3100

\bibitem{DIM2} B.~Feigin, K.~Hashizume, A.~Hoshino, J.~Shiraishi and S.~Yanagida, J.~Math.~Phys. \textbf{50} (2009) 095215, arXiv:0904.2291

\bibitem{DIM3}
B. Feigin, A. Hoshino, J. Shibahara, J. Shiraishi and S. Yanagida, arXiv:1002.2485

\bibitem{DIM4}
B. Feigin, E. Feigin, M. Jimbo, T. Miwa and E. Mukhin, Kyoto J. Math. {\bf 51} (2011) 365-392, arXiv:1002.3113

\bibitem{DIM5}
  H.~Awata, B.~Feigin, A.~Hoshino, M.~Kanai, J.~Shiraishi and S.~Yanagida,
  RIMS k\={o}ky\={u}roku {\bf 1765} (2011) 12-32;
  arXiv:1106.4088

\bibitem{DIM6}
B. Feigin, M. Jimbo, T. Miwa and E. Mukhin, Kyoto J. Math. {\bf 52} (2012) 621-659, arXiv:1110.5310

\bibitem{DIM7}
B. Feigin, M. Jimbo, T. Miwa and E. Mukhin, arXiv:1502.07194

\bibitem{AFS} H. Awata, B. Feigin and J. Shiraishi, arXiv:1112.6074

\bibitem{FJMM}
  B.~Feigin, M.~Jimbo, T.~Miwa and E.~Mukhin,
  J.Alg. {\bf 380} (2013) 78-108, arXiv:1204.5378

\bibitem{Feigin:2013fga}
  B.~Feigin, M.~Jimbo, T.~Miwa and E.~Mukhin,
  Adv.Math.\  {\bf 300} (2016)  229,
  arXiv:1309.2147

  \bibitem{FJMM1} B.~Feigin, M.~Jimbo, T.~Miwa and E.~Mukhin, arXiv:1603.02765

\bibitem{DIMl}
  B.~Feigin, M.~Jimbo and E.~Mukhin,
  arXiv:1705. 07984

\bibitem{Gleb} G. Aminov, A. Mironov and A. Morozov, JHEP, {\bf 23} (2017) 2017, arXiv:1709.04897

\bibitem{KZ} V. Knizhnik and A. Zamolodchikov, 
Nucl. Phys. {\bf B247} (1984) 83

\bibitem{EFK}
 P.I.~Etingof, I.B.~Frenkel and A.A.~Kirillov, Jr.,
 ``Lectures on Representation Theory and Knizhnik-Zamolodchikov Equations,''
 Mathematical surveys and monographs {\bf 58},  AMS (1998).

\bibitem{FR} I. Frenkel and N. Reshetikhin, 
Comm. Math. Phys. {\bf 146} (1992) 1-60

\bibitem{KZB1} D. Bernard, 
Nucl. Phys. {\bf B303} (1988) 77-93; {\it ibid.}
{\bf B309} (1988) 145-174

\bibitem{KZB2}
G. Felder and A. Varchenko, Int.Math.Res.Notices (1995) 221-233, hep-th/9502165

\bibitem{Etin1} P. Etingof and A. Varchenko, math/9907181

\bibitem{Etin2} P. Etingof and A. Varchenko, math/0302071

\bibitem{Etin3}
P. Etingof, O. Schiffmann and A. Varchenko, 
arXiv:math/0207157

\bibitem{Sun} Yi Sun, 
arXiv:1609.09038

\bibitem{SchV1} V. Schechtman and A. Varchenko, 
Lett. Math. Phys. {\bf 20} (1990) 279

\bibitem{SchV2} V. Schechtman and A. Varchenko, 
Inv. Math.  {\bf 106} (1991) 139

\bibitem{SchV3}
H. Awata, A. Tsuchiya and Y. Yamada, 
Nucl. Phys. {\bf B365} (1991) 680

\bibitem{SchV4}
H. Awata, 
Prog. Theor. Phys. Suppl. {\bf 110} (1992) 303, hep-th/9202032

\bibitem{SchV5}
H. Awata, S. Odake and J. Shiraishi,
Nagoya Repository, http://hdl.handle.net/2237/25736, 1993

\bibitem{FTV1}
G.~Felder, V.~Tarasov and A.~Varchenko,
Amer.Math.Soc.Transl. {\bf 180} (1997) 45

\bibitem{FTV2}
G.~Felder, V.~Tarasov and A.~Varchenko,
  Int.J.Math. {\bf 10} (1999) 943, q-alg/9705017

\bibitem{FV1}
G.~Felder and A.~Varchenko,
math.QA/9809139

\bibitem{Delli1} H. W. Braden, A. Marshakov, A. Mironov and A. Morozov,
  Nucl. Phys. {\bf B573} (2000) 553--572, hep-th/9906240

\bibitem{Delli2}
  A. Mironov and A. Morozov, Phys. Lett. {\bf B475} (2000) 71-76, hep-th/9912088

\bibitem{Delli3}
  A. Mironov and A. Morozov, hep-th/0001168

\bibitem{Gleb1} G. Aminov, A. Mironov and A. Morozov, A. Zotov, Phys.Lett. {\bf B726} (2013) 802-808, arXiv:1307.1465

\bibitem{Gleb2}
G. Aminov, H.W. Braden, A. Mironov, A. Morozov and A. Zotov, JHEP, {\bf 01} (2015) 033, rXiv:1410.0698

\bibitem{Gleb3}
G. Aminov, A. Mironov and A. Morozov, The European Physical Journal C, {\bf 76(8)} (2016) 1-19, arXiv:1606.05274

\bibitem{Nak1}
  H.~Nakajima,
 Duke Math.J. {\bf 76} (1994) 365-416

\bibitem{Nak2}
  H.~Nakajima,
 J.Amer.Math.Soc. {\bf 14} (2001) 145-238, math/9912158

\bibitem{D1} V. Drinfeld, Doklady Akad. Nauk SSSR {\bf 283} (1985) 1060

\bibitem{J1} M. Jimbo, 
Lett. Math. Phys. {\bf 10} (1985) 63-69

\bibitem{J2} M. Jimbo,
Lett. Math. Phys. {\bf 11} (1986) 247-252;
Commun. Math. Phys. {\bf 102} (1986) 537-547

\bibitem{D2} V.G. Drinfeld, 
Soviet Math. Doklady, {\bf 36} (1988) 212-216

\bibitem{VV2}
  M.~Varagnolo and E.~Vasserot,
  Invent.Math. {\bf 133} (1998) 133-159,
  q-alg/9612035

\bibitem{VV3}
  M.~Varagnolo and E.~Vasserot,
  Math.Res.Lett. {\bf 18} (1999) 1005-1028,
math/9902091

\bibitem{Nag1}
  K.~Nagao,
  J.Alg. {\bf 321} (2009) 3764-3789,
math/0703107

\bibitem{Nag2}
  K.~Nagao,
  Osaka J.Math. {\bf 46} (2009) 877-907,
math/0709.1767

\bibitem{Neg}
  A.~Negut,
  arXiv:1504.06525

%

\bibitem{Saito}
  Yoshihisa~Saito,
  Publ.Res.Inst.Math.Sci. {\bf 34} (1998), no. 2, 155-177,
q-alg/9611030

\bibitem{STU}
  Y.~Saito, K.~Takemura and D.~Uglov,
  Transform. Groups 3 (1998), no. 1, 75-102,
  q-alg/9702024


\bibitem{Taki:2007dh}
  M.~Taki,
JHEP {\bf 0803} (2008) 048,
arXiv:0710.1776

\bibitem{Awata:2008ed}
  H.~Awata and H.~Kanno,
  Int.J.Mod.Phys.\ {\bf A24} (2009) 2253,
arXiv:0805.0191

\bibitem{confMAMO1} A. Marshakov, A. Mironov and A. Morozov,
Phys. Lett. {\bf B265} (1991) 99

\bibitem{confMAMO2}
A. Mironov and S. Pakuliak, Theor.Math.Phys. {\bf 95} (1993) 604-625 (Teor.Mat.Fiz. {\bf 95} (1993) 317-340), hep-th/9209100

\bibitem{confMAMO3}
S. Kharchev, A. Marshakov, A. Mironov, A. Morozov and S. Pakuliak,
Nucl. Phys. {\bf B404} (1993) 717-750, hep-th/9208044

\bibitem{confMAMO4}
H.~Awata, Y.~Matsuo, S.~Odake and J.~Shiraishi,
  Phys.\ Lett.\ B {\bf 347} (1995) 49, hep-th/9411053

\bibitem{confMAMO5}
H.~Awata, Y.~Matsuo, S.~Odake and J.~Shiraishi,
Nucl.Phys. {\bf B449} (1995) 347-374, hep-th/9503043

\bibitem{confMAMO6}
H.~Awata, Y.~Matsuo, S.~Odake and J.~Shiraishi, Soryushiron Kenkyu {\bf 91} (1995) A69-A75, hep-th/9503028

\bibitem{AGTmamo1} R. Dijkgraaf and C. Vafa, arXiv:0909.2453

\bibitem{AGTmamo2}
H. Itoyama, K. Maruyoshi and T. Oota,
Prog. Theor. Phys. {\bf 123} (2010) 957-987, arXiv:0911.4244

\bibitem{AGTmamo3}
T. Eguchi and K. Maruyoshi,
arXiv:0911.4797

\bibitem{AGTmamo4}
T. Eguchi and K. Maruyoshi,
arXiv:1006.0828

\bibitem{AGTmamo5}
R. Schiappa and N. Wyllard,
arXiv:0911.5337

\bibitem{AGTmamo6}
P. Sulkowski, JHEP {\bf 04} (2010) 063, arXiv:0912.5476

\bibitem{AGTmamo7}
H. Itoyama and T. Oota, Nucl. Phys. {\bf B838} (2010) 298-330, arXiv:1003.2929

\bibitem{AGTmamo8}
A. Mironov, A. Morozov, and And. Morozov, Nucl. Phys. {\bf B843} (2011) 534, arXiv:1003.5752

\bibitem{Fucito:2004ry}
  F.~Fucito, J.~F.~Morales and R.~Poghossian,
  Nucl.Phys.\ {\bf B703} (2004) 518,
  hep-th/0406243

\bibitem{Fujii:2005dk}
  S.~Fujii and S.~Minabe,
  SIGMA {\bf 13} (2017) 052,
  math/0510455

\bibitem{TU}
K.~Takemura and D.~Uglov,
J.Phys. {\bf A 30} (1997) 3685-3717,
 solv-int/9611006

\bibitem{Uglov:1997ia}
  D.~Uglov,
T  of dynamical correlation functions in the spin Calogero-Sutherland model,''
  Comm.Math.Phys.\  {\bf 193} (1998) 663,
hep-th/9702020

%
\bibitem{Belavin:2011pp}
  V.~Belavin and B.~Feigin,
  JHEP {\bf 1107} (2011) 079,
  arXiv:1105.5800

\bibitem{Nishioka:2011jk}
  T.~Nishioka and Y.~Tachikawa,
  Phys.Rev.\ {\bf D84} (2011) 046009,
  arXiv:1106.1172

\bibitem{Belavin:2011sw}
  A.~A.~Belavin, M.~A.~Bershtein, B.~L.~Feigin, A.~V.~Litvinov and G.~M.~Tarnopolsky,
  Comm.Math.Phys.\  {\bf 319} (2013) 269,
  arXiv:1111.2803

\bibitem{Belavin:2012eg}
  A.~A.~Belavin, M.~A.~Bershtein and G.~M.~Tarnopolsky,
  JHEP {\bf 1303} (2013) 019,
  arXiv:1211.2788

\bibitem{Itoyama:2013mca}
  H.~Itoyama, T.~Oota and R.~Yoshioka,
  Nucl.Phys.\ {\bf B877} (2013) 506,
  arXiv:1308.2068

\bibitem{Itoyama:2014pca}
  H.~Itoyama, T.~Oota and R.~Yoshioka,
  Nucl.Phys.\ {\bf B889} (2014) 25,
  arXiv:1408.4216

\bibitem{eval} See a review and references, e.g., in:\\
M. Jimbo, {\sl Topics from Representations of $U_q(g)$. An Introductory Guide to Physicists},
Nankai Lectures on Mathematical Physics, 1992: World Scientific, Singapore, pp. 1-61

\bibitem{specdu1} E. Mukhin, V. Tarasov and A. Varchenko, 
math/0510364;
Adv. Math. {\bf 218} (2008) 216-265, math/0605172

\bibitem{specdu2}
A. Mironov, A. Morozov, Y. Zenkevich and A. Zotov, JETP Lett. {\bf 97} (2013) 45, arXiv:1204.0913

\bibitem{specdu3}
A. Mironov, A. Morozov, B. Runov, Y. Zenkevich and A. Zotov, Lett. Math. Phys. {\bf 103} (2013) 299,
arXiv:1206.6349

\bibitem{specdu4}
A. Mironov, A. Morozov, B. Runov, Y. Zenkevich and A. Zotov, JHEP {\bf 1312} (2013) 034, arXiv:1307.1502

\bibitem{specdu5}
L. Bao, E. Pomoni, M. Taki and F. Yagi, JHEP {\bf 1204} (2012) 105, arXiv:1112.5228

\bibitem{Sham1} M.~Aganagic, N.~Haouzi, C.~Kozcaz and
  S.~Shakirov, 
  arXiv:1309.1687

\bibitem{Sham2}
M.~Aganagic, N.~Haouzi and S.~Shakirov, 
  arXiv:1403.3657

\bibitem{Sham3}
M.~Aganagic and N.~Haouzi, arXiv:1506.04183

\bibitem{SWqint1} N. Nekrasov and S. Shatashvili, arXiv:0908.4052

\bibitem{SWqint2}
A. Mironov and A. Morozov, 
JHEP {\bf 04} (2010) 040, arXiv:0910.5670

\bibitem{SWqint3}
A. Mironov and A. Morozov,
J. Phys. {\bf A43} (2010) 195401, arXiv:0911.2396

\bibitem{SWqint4}
A. Marshakov, A. Mironov and A. Morozov, J. Geom. Phys. {\bf 61} (2011) 1203-1222, arXiv:1011.4491

\end{thebibliography}
\end{document}